\documentclass[twocolumn]{svjour3} 

\usepackage[american]{babel}
\usepackage[utf8]{inputenc}
\usepackage{ae,aecompl}
\usepackage[numbers]{natbib}
\usepackage{graphicx}

\usepackage{bm}
\usepackage{url}
\usepackage{color}
\usepackage{booktabs}
\usepackage{subfigure}

\usepackage{amsmath} 
\usepackage{amssymb}
\usepackage{commath}
\usepackage{dsfont}

\newcommand{\R}{\ensuremath{\mathbb{R}}}    

\journalname{Meccanica} %

\begin{document}

\title{On the reduction of nonlinear electromechanical systems}

\author{Americo Cunha~Jr \and
Marcelo Pereira \and
Rafael Avan\c{c}o \and
\mbox{Angelo Marcelo Tusset} \and 
Jos\'{e} Manoel Balthazar
}

\institute{A. Cunha~Jr \at
              Rio de Janeiro State University, Brazil\\
		     ORCID: 0000-0002-8342-0363\\
              \email{americo.cunha@uerj.br}
           \and 
           M. Pereira \at 
           SENAI Innovation Institute, Brazil\\
           Rio de Janeiro State University, Brazil\\
           ORCID: 0000-0003-1655-7092\\
           \email{mardapereira@firjan.com.br}
           \and 
           R. Avan\c{c}o \at 
           Federal University of Maranhão, Brazil\\
           ORCID: 0000-0003-2276-0230\\
           \email{avancorafael@gmail.com}
           \and
           A. M. Tusset \at 
           Federal University of Technology – Paran\'{a}, Brazil\\
           ORCID: 0000-0003-3144-0407\\
           \email{a.m.tusset@gmail.com}
           \and
           J. M. Balthazar \at 
           S\~{a}o Paulo State University, Brazil\\
           Federal University of Technology – Paran\'{a}, Brazil\\
           ORCID: 0000-0002-9796-3609\\
           \email{jmbaltha@gmail.com}
}

\date{Received: date / Accepted: date}

\maketitle

\begin{abstract}
The present work revisits the reduction of the nonlinear dynamics of an electromechanical system through a quasi-steady state hypothesis, discussing the fundamental aspects of this type of approach and clarifying some confusing points found in the literature. Expressions for the characteristic time scales of dynamics are deduced from a physical analysis that establishes an analogy between electromechanical dynamics and the kinetics of a chemical reaction. It provides a physical justification, supplemented by non-dimensionalization and scaling of the equations, to reduce the dynamics of interest by assuming a quasi-steady state for the electrical subsystem, eliminating the inductive term from the electrical equation. Numerical experiments help to illustrate the typical behavior of the electromechanical system, a boundary layer phenomenon near the initial dynamic state, and the validity limits of the electromechanical quasi-steady-state assumption discussed here.

\keywords{nonlinear dynamics \and electromechanical dynamics \and DC~motor-cart system \and model-order reduction \and quasi-steady-state assumption}
\end{abstract}

\section{Introduction}
\label{intro}

The interest in the dynamic behavior of electromechanical nonlinear systems is not a new, being explored in open literature \cite{avanco2018p23,Balthazar2004,Chattopadhyay1975p809,Gonzalez-Carbajal2017p1377,Jiang2021p1205,Kazmierczak2012,Pham2018p385,Shvets2008,Szmit2016p953,Trimmer1989p17,vonWagner2002p861} and classical books \cite{Alifov1990,Kononenko1969, Moon2002, nayfeh1979} for decades. Despite that, many modern applications of engineering and science are based on the interaction between an electric system with a mechanical counterpart, such as energy harvesting technologies \cite{clementino2014p283,cottone2009p080601,cunhajr2021p137,erturk2009p254102,adhikari2012p1505,cunhajr_belhaq2019,cunhajr2018p01001,Balthazar2017p2583,rocha2018p3684}, micro/nano electromechanical resonators \cite{Kazmi2018p113,Ruzziconi2013p1761,Younis2003p91,Younis2003p672,Zehnder2018p92}, drill-strings \cite{cunhajr2015p849,cayres2018p16009} etc. In this way, the nonlinear dynamics of electromechanical systems continues to be a current and important topic of research.

Due to the complexity (or high computational cost) associated with the simulation of these dynamical systems, approaches that seek to reduce the order (dimension) of the underlying mathematical model are common in the literature \cite{avanco2018p23,Balthazar2003p613,Balthazar2018p19,belato2001p1699,Goncalves2016p2203,goncalves2014p5115,rocha2018p3684,Rocha2018}. In general, these reduction approaches seek to decrease the model-order by considering a restriction of the original dynamical system to a structure-preserving low-dimensional\footnote{Low in this context means small compared with the phase-space dimension.} manifold, i.e., a manifold that preserves the main characteristics of the nonlinear dynamical system \cite{Holmes1996,mass1992p239,Reis2008} or by projecting the original dynamics into low-dimensional subspace \cite{Benner2015p483,Chinesta2011p395,Rathinam2003p1983}.

A simple approach to reducing the order of an electromechanical system involves the elimination of the inductive term from the model equations, which is done when the electrical dynamics characteristic time is much faster than the mechanical time-scale of the problem. The idea behind this simplification is that the electrical and mechanical dynamics are ``competing'', and once the electric time-scale is much faster than its mechanical correspondent (typically, some orders of magnitude), the mechanical system only show significant variations in its behavior a long time after the beginning of the electric dynamics, that is already in dynamic equilibrium \cite{Blekhman2008p21,Cveticanin2018,Evan-Iwanowski1976,Kononenko1969,nayfeh1979}. The same idea is widely used in reducing chemical kinetics mechanisms. The chemical species that react most rapidly are assumed to be in equilibrium so that the corresponding differential equations become simple algebraic relations, reducing the model dimension. This procedure is known in the chemical literature as \emph{quasi-steady-state assumption} \cite{Goeke2015,Segel1989p446}.

In some recent publications, Lima and Sampaio \cite{lima2018_1}, and Lima et al. \cite{Lima2019p552} put in check the validity of this \emph{electromechanical quasi-steady-state assumption}, claiming that, as this simplification decouples the electrical system from its mechanical counterpart, it loses the main features of the system's physics, leading to erroneous predictions. Using the electromechanical system presented in \cite{Lima2016p931} as an example, they give and discuss a limit case in which this model-order reduction approach leads to erroneous predictions.

Although this theme is not new, the discussion raised in \cite{lima2018_1,Lima2019p552} is legitimate, mainly because of the exciting observations the authors make about the essence of electromechanical coupling. However, the authors of this work do not agree with all of their arguments, especially with the conclusion to generally invalidate the use of the quasi-steady-state strategy for model-order reduction of electromechanical systems. Indeed, if the limit of validity of the reduction hypothesis is respected, the simulations presented in this paper show that the qualitative predictions of the reduced-order model are excellent.

Despite this model-order reduction procedure being classic, the discussion raised by Lima and co-authors \cite{lima2018_1,Lima2019p552} show that the subject is still not well understood by many researchers and is worthy of a pedagogical discussion. 

In this sense, this work aims to present a didactic discussion about the electromechanical quasi-steady-state assumption used for model-order reduction in nonlinear electromechanical systems to clarify its fundamentals and limits of applicability. A formal justification for the model-order reduction procedure is presented from two different points of view (physical and mathematical) and a discussion on the system's qualitative behavior. The manuscript also quantifies how shorter the electric time must be for the simplifying hypothesis to be valid. 

To the best of the authors' knowledge, the physical justification for reducing the electromechanical system provided in this work, in analogy with the simplification of chemical kinetics mechanisms, is an original interpretation of this problem, not available in the literature. So, the paper also contributes to understanding the similarities between electromechanical systems with chemical reactions. 

The rest of this paper is organized as follows. The electromechanical system of interest is introduced in section~\ref{full_dyn}. A detailed analysis of the time scales intrinsically to this dynamical system can be seen in section~\ref{time-scale_analysis}. A dimensionless formulation for the dynamical system equations is presented in section~\ref{dimless_form}. The reduced-order model for the electromechanical system is deduced in section~\ref{red_order_dyn_model}. Numerical experiments to illustrate some general characteristics and peculiarities of the reduced-order model are reported in the sequence in section~\ref{num_experiments}. Finally, in section~\ref{concl_remaks}, the manuscript is closed with the final considerations.

\begin{figure}[h]
	\centering
	\includegraphics[scale=0.2]{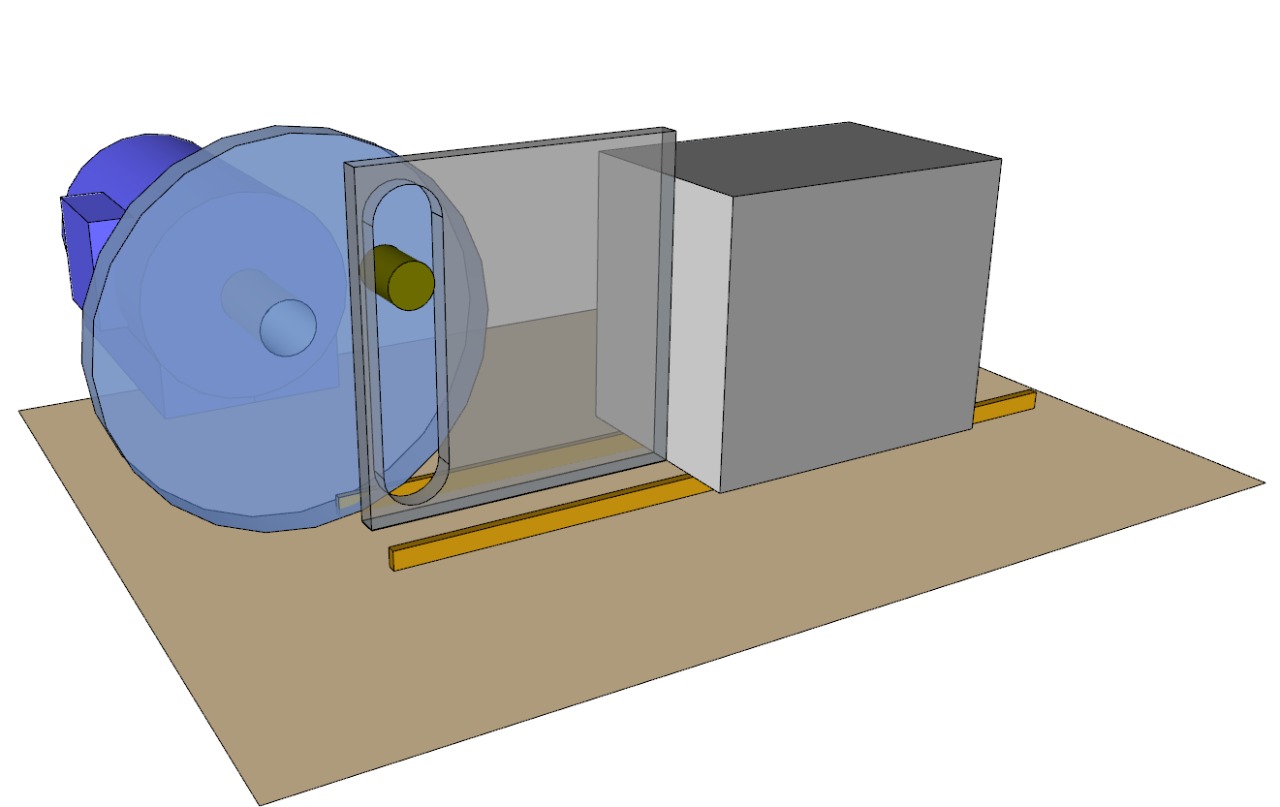}
	\caption{Schematic representation of an electromechanical composed by a cart in horizontal translation coupled to a DC motor via a slotted link mechanism.}
\label{eletromech_system_fig}
\end{figure}

\section{Full-order dynamic model}
\label{full_dyn}

Consider the electromechanical system analyzed in references \cite{Lima2016p931,lima2018_1}, which is presented in Figure~\ref{eletromech_system_fig}, in which a video animation of the typical dynamic behavior can be seen in the Supplementary Material 01 and in reference \cite{eletromech_video1}. This system consists of a cart that undergoes linear horizontal translation movement, which is coupled via a slotted link mechanism to a DC motor.

\begin{figure}[h]
	\centering
	\includegraphics[scale=0.25]{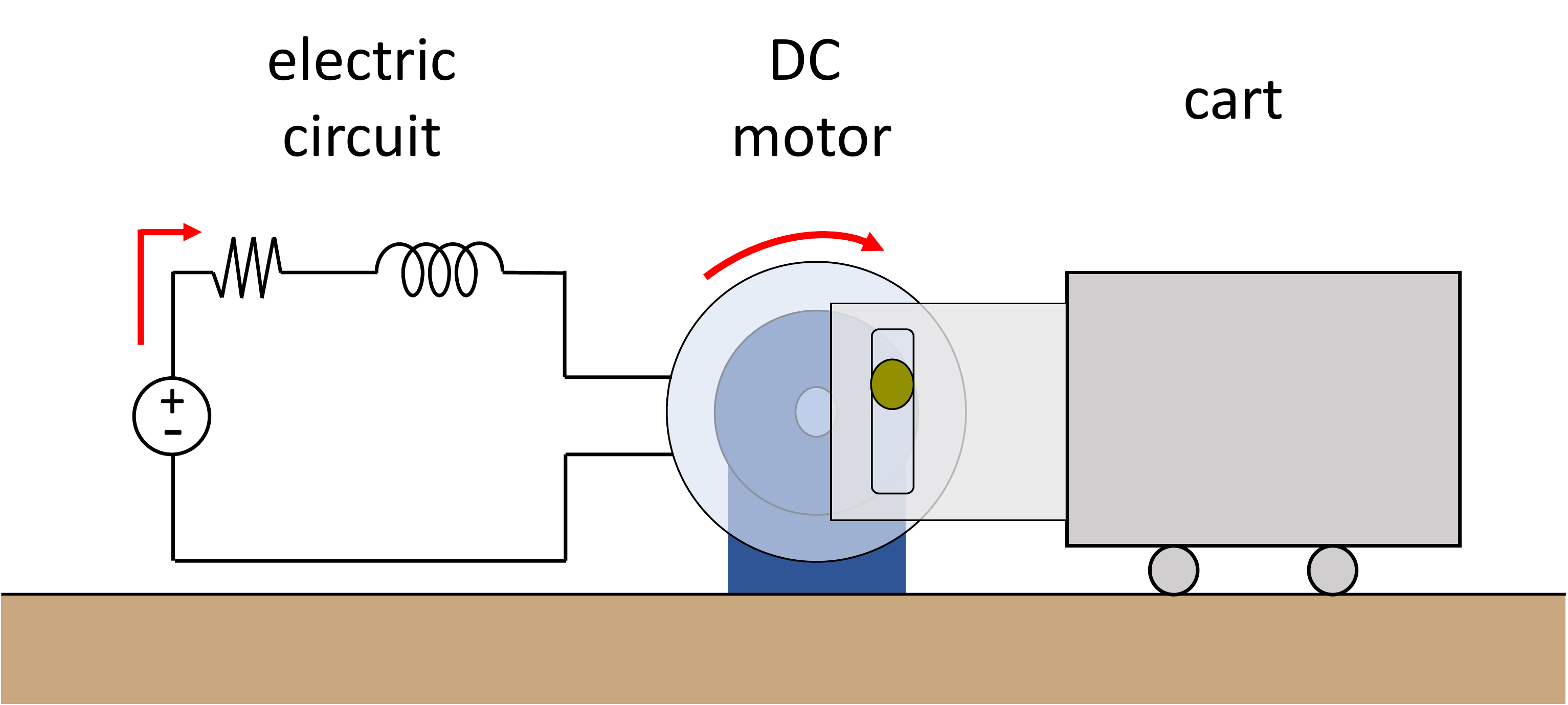}
	\caption{Idealization of the electromechanical system: a horizontally sliding mass coupled to a DC motor that is represented by an equivalent electrical circuit.}
\label{eletromech_system_fig2}
\end{figure}

In terms of physical modeling, this electromechanical system can be idealized as in Figure~\ref{eletromech_system_fig2}, where the cart is represented by a lumped mass, horizontally sliding without friction, coupled via a slotted link mechanism to a DC motor, which is represented by an equivalent electrical circuit composed of a voltage source and resistive and inductive elements.

This type of electromechanical system has its dynamical behavior evolving according to
\begin{equation}
	L \, Q^{\, ''} + R \, Q^{\, '} + G \, \Theta^{\, '}  = \mathcal{V},
\label{electric_eq}
\end{equation}
\begin{equation}
	J \, \Theta^{\, ''} + B \, \Theta^{\, '} - G \, Q^{\, '}= \mathcal{T},
\label{mech_eq}
\end{equation}
where $Q^{\, '} = Q^{\, '}(T)$ and $\Theta = \Theta(T)$ respectively denote the electrical current and angular displacement in time $T$; the upper prime is an abbreviations for time derivative, i.e., $\square^{\, '} = d \, \square / dT$; $L$ accounts an electrical inductance; $R$ represents an electrical resistance; $J$ a rotational inertia; $B$ describes a damping coefficient; while $G$ is an electromechanical coupling coefficient. The voltage source $\mathcal{V} = \mathcal{V}(T)$ and the external torque $\mathcal{T} = \mathcal{T}(T)$ correspond to (possibly) time-dependent external excitations. The torque may also be a function of the electromechanical system coordinates and their derivatives, i.e., $\mathcal{T} = \mathcal{T}(\Theta,\Theta^{\, '},\Theta^{\, ''},T)$.

Besides that, the link mechanism shown in Figure~\ref{eletromech_system_fig} imposes to the mechanical subsystem a nonlinear kinematical constraint  of the form
\begin{equation}
	X = D \, \cos{\Theta},
\label{kinematic_eq}
\end{equation}
which relates the rotational and translational motions, respectively defined by angle $\Theta$ and cart horizontal displacement $X$, and with $D$ denoting the slotted link pin eccentricity.

Regarding the torque exerted by the motor on the shaft, it can be shown that it is given by
\begin{equation}
	\mathcal{T} = F \, D \, \sin{\Theta},
\label{torque_eq0}
\end{equation}
where $F$ is the horizontal force exerted by the motor on the cart, which can be obtained from Newton's second law
\begin{equation}
	M \, X^{\, ''} = F,
\label{newton_eq}
\end{equation}
where $M$ is the cart mass. 

Thus, combining Eqs.(\ref{kinematic_eq}), (\ref{torque_eq0}) 
and (\ref{newton_eq}), one has an alternative representation for the torque
\begin{equation}
	\mathcal{T} = - M \, D^2 \, \sin{\Theta} \, \left( \sin{\Theta} \, \Theta^{\, ''}  + \cos{\Theta} \, \Theta^{\, ' \,2}  \right).
\label{torque_eq}
\end{equation}

Once this electromechanical system departs from the 
initial state
\begin{equation}
	Q^{\, '}(0) = Q^{\, '}_0, ~~ \Theta^{\, '} (0) = \Theta^{\, '}_0, ~~ \mbox{and} ~~ \Theta(0) = \Theta_0,
\label{model_ic_eq}
\end{equation}
its dynamic behavior is completely characterized by the \emph{full-order dynamic model} defined by Eqs.(\ref{electric_eq}), (\ref{mech_eq}), (\ref{torque_eq}) and (\ref{model_ic_eq}).

\section{Time-scale analysis}
\label{time-scale_analysis}

The dynamic system of interest in this work has two characteristic time scales, one intrinsic to the DC motor equivalent electrical circuit and another related to the mechanical oscillator. Electrical oscillations occur much faster than their mechanical counterparts, and typically they differ 2 or 3 orders of magnitude \cite{Chapman2012}, so the electromechanical system dynamics presents two distinct behaviors along with its temporal evolution: (i) a short transient regime, where the dynamics is fast and influenced almost exclusively by the electric circuit; and (ii) a long quasi-stationary regime (slow drift), in which the mechanical oscillator drives the nonlinear system \cite{nayfeh1979,Segel1989p446,Shoffner2017p122}. 

Intuitively one can think that the observed dynamic behavior results from a ``competition''  between these two processes. In the first moment, the transient wins the ``competition'' since it is faster than the drift. Nevertheless, once this transient is also short, after some time, it is ``overcome'' by the slow (but long duration) quasi-stationary regime with relaxation oscillations.

Therefore, estimating the value of each of these time-scales can be extremely useful in understanding the electromechanical oscillator dynamic behavior better.

The characteristic time of the electric subsystem can be estimated from Eq.(\ref{electric_eq}) assuming $\Theta^{\, '} \approx \Theta^{\, '}_0$, since the electric dynamics is faster than the mechanical one, the electrical current $Q^{\, '}$ varies considerably before the angular velocity $\Theta^{\, '}$ differs significantly from its initial value. Consequently, Eq.(\ref{electric_eq}) becomes
\begin{equation}
	L \, Q^{\, ''} + R \, Q^{\, '}  + G \, \Theta^{\, '}_0 = \mathcal{V},
\label{time-scale_eq1}
\end{equation}
from which one obtains
\begin{equation}
\begin{split}
	Q^{\, '}(T) =  - \frac{G}{R} \, \Theta^{\, '}_0 ~+~ \frac{\mathcal{V}}{R} ~+~ ~~~~~~~~~~~~~~~~~~~~~~~~~~~~~\\
	~~~~~~~~~~~~~~~~~~~~ \left(Q^{\, '}_0 - \frac{\mathcal{V}}{R} + \frac{G}{R} \, \Theta^{\, '}_0 \right) \, \exp{\left( - \frac{T}{L/R} \right)}.
\end{split}
\label{time-scale_eq2}
\end{equation}

For typical initial conditions ($Q^{\, '}_0 + \frac{G}{R} \, \Theta^{\, '}_0 \neq \frac{\mathcal{V}}{R}$), the electric current initially exhibits an exponential decaying behavior toward a constant value
\begin{equation}
	Q^{\, '}_{\infty} = - \frac{G}{R} \, \Theta^{\, '}_0 ~+~ \frac{\mathcal{V}}{R}.
\end{equation}

The time constant of this decay, $L/R$, defines a time scale for the electrical subsystem 
\begin{equation}
	T_{Q} = \frac{L}{R},
\label{time-scale_eq3}
\end{equation}
which is called \emph{electrical characteristic time}.

One possible way to characterize a mechanical time-scale $T_{\Theta}$ is through the ratio
\begin{equation}
	T_{\Theta} = \frac{\Theta^{\, '}_{max} - \Theta^{\, '}_{min}}{|\Theta^{\, ''}|_{max}},
\label{time-scale_eq4}
\end{equation}
where $\Theta^{\, '}_{max}$ and $\Theta^{\, '}_{min}$ denote largest and smallest values assumed by the angular velocity $\Theta^{\, '}$, respectively; $|\Theta^{\, ''}|_{max}$ is the largest angular acceleration, in absolute value, that the electromechanical system undergoes during its operation. This ratio gives a global time-scale for the variable $\Theta$.

It is trivial to see that when the DC motor starts to operate from the rest (or near rest - some movement remains after shutdown due to inertia), the initial angular velocity $\Theta^{\, '}_0$ is zero (or very close to zero). As the electromechanical system of Figure~\ref{eletromech_system_fig} cannot pump energy from the motor, there is no possibility of reversion in its rotation \cite{Lima2016p931}. Thus, when one starts from rest (or near rest), the lowest possible angular velocity is the initial one, so that $\Theta^{\, '}_{min} = 0$ (or $\Theta^{\, '}_{min} \approx 0$).

\begin{figure}[ht!]
	\centering
	\includegraphics[scale=0.23]{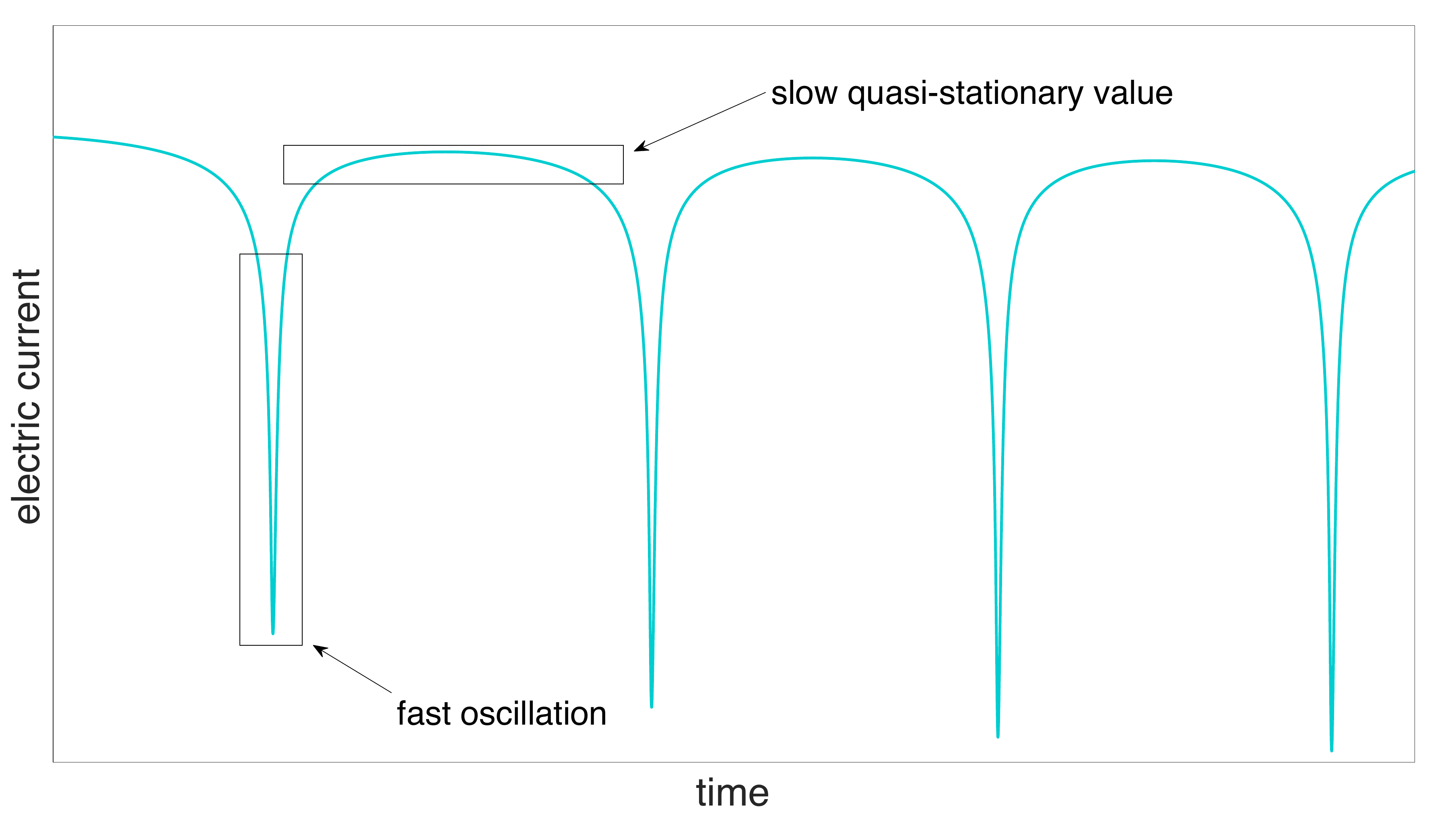}
	\caption{Illustration of the typical electric current time series, after the transient, for the electromechanical system. One can observe a fast relaxation oscillation at each period followed by a slow drift with an almost constant value.}
\label{fig_relaxation_oscillation}
\end{figure}

The instant where the maximum angular velocity occurs is more difficult to be identified directly, but $\Theta^{\, '}_{max}$ can be estimated with the aid of Figure~\ref{fig_relaxation_oscillation}, which shows the typical behavior of the electromechanical system electric current after the transient. Note that each cycle is composed of two processes, a fast relaxation oscillation followed by a slow drift with an almost constant value.

Strictly speaking, this stage of the dynamics is not a steady-state\footnote{It is crucial not to confuse the notion of steady-state (time derivative equal to zero) with the statistical idea of stationarity, which concerns the invariance of the statistics of the temporal signal.} regime, as there is temporal variation. However, one can note that the electric current value is almost constant most of the time since the oscillatory relaxations are very fast. This regime is what is called here the \emph{quasi-steady-state}.

Thus, as the electric current is almost constant\footnote{Except for short periods of relaxation oscillations, as can be see in Figure~\ref{fig_relaxation_oscillation}.} ($Q^{\, '} \approx Q^{\, '}_{\infty}$  for almost every instant), it is reasonable to think that its derivative is approximately zero, i.e., $Q^{\, ''} \approx 0$ at all times. This hypothesis applied into Eq.(\ref{electric_eq}) imply that
\begin{equation}
	 \Theta^{\, '} = \frac{\mathcal{V}}{G} - \frac{R}{G} \, Q^{\, '},
\label{time-scale_eq5}
\end{equation}
from where one notices that
\begin{equation}
	 \Theta^{\, '}_{max} = \frac{\mathcal{V}}{G},
\label{time-scale_eq6}
\end{equation}
which occurs when $Q^{\, '} = 0$.

In addition, by substituting (\ref{time-scale_eq5}) in (\ref{mech_eq}) and doing some simple algebraic manipulations, one obtains
\begin{equation}
	\Theta^{\, ''} + \left( \frac{B}{J} + \frac{G^{\, 2}}{J \, R} \right) \, \Theta^{\, '} - \frac{G \, \mathcal{V}}{J \, R} = \frac{\mathcal{T}}{J}.
\label{time-scale_eq7}
\end{equation}

At the beginning of the quasi-stationary regime, the DC motor shaft is near a rest state. In this way, the angular acceleration must be high in the early quasi-stationary regime, to be able to move the DC motor shaft (and consequently the cart). Therefore, when $\Theta^{\, ''}$ is maximum one has $\Theta^{\, '} \approx 0$ and $\Theta \approx 0$, in a way that $\mathcal{T} \approx 0$, and thus, with aid of Eq.(\ref{time-scale_eq7}), it is possible to obtain
\begin{equation}
	|\Theta^{\, ''}|_{max} = \frac{G \, \mathcal{V}}{J \, R}.
\label{time-scale_eq8}
\end{equation}

Now the estimates for $\Theta^{\, '}_{max}$, $\Theta^{\, '}_{min}$ and $|\Theta^{\, ''}|_{max}$, obtained above, can be substituted in Eq.(\ref{time-scale_eq4}) to obtain the \emph{mechanical characteristic time}, which is given by
\begin{equation}
	T_{\Theta} = \frac{J \, R}{G^{\,2}}.
\label{time-scale_eq9}
\end{equation}

Although informal from the mathematical point of view (a formal justification is given in section~\ref{red_order_dyn_model}), the time-scale analysis developed in this section has a strong physical appeal. It suggests a procedure to reduce the electromechanical system order. The main idea is that, as the electrical dynamics occur much faster than its mechanical counterpart, when the latter starts, the former is in ``equilibrium'' and its transient term $Q^{\, ''}$ can be discarded, so that Eq.(\ref{electric_eq}) is reduced to Eq.(\ref{time-scale_eq5}), an affine manifold on the system phase-space. In this way, the system dynamics can be obtained from the integration of Eq.(\ref{time-scale_eq7}), then the electrical current can be computed with aid of Eq.(\ref{time-scale_eq5}). This approach, valid when $T_{Q} \ll T_{\Theta}$, is called \emph{quasi-steady-state assumption}, being also used in chemical kinetics \cite{Eilertsen2020p108339,frank1940p695,turanyi1992p903,turanyi1993p163}, where a differential equation representing the concentration rate of change of a certain intermediate chemical species is exchanged for an algebraic equation, obtained by equating its right side to zero. Such simplification is usually adopted if this intermediate species reacts much faster than other chemical species. The faster reactions are practically in dynamic equilibrium when the other reactions start. Here, the electrical subsystem plays the role of the fast chemical species, while the mechanical part emulates a slow species.

\section{Dimensionless formulation}
\label{dimless_form}

The full-order dynamic model presented in section~\ref{full_dyn} clearly show that the electromechanical system dynamics depends on the dimensionless angle $\Theta$ and other 11 dimensional quantities: $T$, $\Theta^{\, '}$, $Q^{\, '}$, $L$, $R$, $G$, $\mathcal{V}$, $J$, $B$, $M$, and $D$. Thus, there is an vector mapping $\mathcal{F}: \R^{12} \to \R^2$ relating all these quantities through a vector equation
\begin{equation}
	\mathcal{F}( T, \Theta, \Theta^{'}, Q^{'}, L, R, G, \mathcal{V}, J, B, M, D) = 0.
	\label{da_eq1}
\end{equation}

These 12 parameters depend on 4 base quantities (mass, length, time and electrical current), in such a way that it is possible to rewrite the vector relationship from Eq.(\ref{da_eq1}) in terms of 8 dimensionless groups, i.e.,
\begin{equation}
	f( t, \theta, \dot{\theta}, \dot{q}, \ell, \nu, b, d) = 0,
	\label{da_eq2}
\end{equation}
where $f: \R^{8} \to \R^2$ is another vector mapping, and the dimensionless quantities are
\begin{equation}
\begin{split}
	t = \frac{T}{JR/G^2} \,, ~~ 
	\theta = \Theta \,, ~~
	\dot{\theta} = \frac{\Theta^{\, '}}{G^2/JR} \,, ~~ \\
	\dot{q} = \frac{Q^{\, '}}{G^3/JR^2} \,, ~~
	\ell = \frac{L}{J R^2/G^2} \,, ~~~~~~~~~~ \\
	\nu = \frac{\mathcal{V}}{G^3/JR} \,, ~~
	b = \frac{B}{G^2/R} \,, ~~ 
	d = \frac{D}{\sqrt{J/M}} \,.
\end{split}
	\label{da_eq3}
\end{equation}

It follows that the dimensionless formulation of the full-order dynamic model from Eqs.(\ref{electric_eq}) to (\ref{model_ic_eq}), which is obtained with aid of the dimensionless groups defined in (\ref{da_eq3}), is given by
\begin{equation}
	\ell \, \ddot{q} + \dot{q} + \dot{\theta} = \nu,
\label{electric_eq2}
\end{equation}
\begin{equation}
	\ddot{\theta} + b \, \dot{\theta} - \dot{q}= \tau,
\label{mech_eq2}
\end{equation}
with 
\begin{equation}
	\tau = - d^{\, 2} \, \sin{\theta} \, \left( \sin{\theta} \, \ddot{\theta} + \cos{\theta} \, \dot{\theta}^{\, 2} \right),
\label{torque_eq2}
\end{equation}
and
\begin{equation}
	\dot{q}(0) = \dot{q}_0, ~~ \dot{\theta} (0) = \dot{\theta}_0, ~~ \mbox{and} ~~ \theta(0) = \theta_0.
\label{model_ic_eq2}
\end{equation}

Note that dimensionless time-derivative is related to the dimensional one through the relationship $\dot{\square} = G^{\,2}/JR \, \square^{\, '}$. Also, the dimensionless groups are written in a way that the physical meaning of each one is very clear. The dimensionless time is $t$; the angular displacement as function of $t$ is $\theta$; $\dot{\theta}$ is the dimensionless angular velocity, which also depends on $t$; the other non-dimensional parameters represent: $\dot{q}$ the electrical current; $\ell$ the inductance; $\nu$ the voltage; $b$ the damping coefficient; and $d$ the pin eccentricity.

\section{Reduced-order dynamic model}
\label{red_order_dyn_model}

In a realistic electromechanical system, the dimensional parameters that appear in Eq.(\ref{da_eq1}) assume typical values\footnote{Some electromechanical parameters assume values in ranges that span a few orders of magnitude. The magnitudes shown in Table~\ref{table1} represent a possible choice within these ranges.}  such as those in which the orders of magnitude (units in SI system) appear in Table~\ref{table1}. The magnitude of these parameters corresponds to middle to low power motors (around 500~watts), which are fairly easy to find in industrial applications and experimental test rigs in laboratories \cite{Chapman2012,Gao2012p189757,Nayak2019p83,Petrovas2018p415,Saab2001p957638,Sendrescu2012,Toliyat2004,Yeadon2001}.

\begin{table}[h!]
	\centering
	\caption{Orders of magnitude for the dimensional parameters of a typical electromechanical system.}
	\vspace{5mm}
	\begin{tabular}{cc}
		\toprule
		System   & Typical Value\\
        Parameter & (order of magnitude)\\
		\midrule
		$D$ & $\sim$ $10^{-1}$\\
		$J$ & $\sim$ $10^{-4}$\\
		$L$ & $\sim$ $10^{-3}$\\
		$M$ & $\sim$ $10^{0}~~$\\
		$G$ & $\sim$ $10^{-1}$\\
		$\mathcal{V}$ & $\sim$ $10^{1}~~$\\
		$~~Q^{\, '}$ & $\sim$ $10^{2}~~$\\
		$R$ & $\sim$ $10^{0}~~$\\
		$T$ & $\sim$ $10^{-2}$\\
		$~~\Theta^{\, '}$ & $\sim$ $10^{2}~~$\\
		$B$ & $\sim$ $10^{-2}$\\
		\bottomrule
	\end{tabular}
	\label{table1}
\end{table}

From the values in Table~\ref{table1}, recalling the definitions of dimensionless parameters in (\ref{da_eq3}), it may be noted that $\ddot{q}$, $\dot{q}$, $\dot{\theta}$ and $\nu$ are all quantities whose order of magnitude is 1, i.e., $\ddot{q} \sim \mathcal{O}(1)$, $\dot{q} \sim \mathcal{O}(1)$, $\dot{\theta} \sim \mathcal{O}(1)$ and $\nu \sim \mathcal{O}(1)$. Consequently, as $\ell \sim \mathcal{O}(10^{-1}) \ll 1$, the inductive term in Eq.(\ref{electric_eq2}) can be safely discarded so that
\begin{equation}
	\dot{q} + \dot{\theta} = \nu,
\label{red_eq1}
\end{equation}
which, when replaced in Eq.(\ref{mech_eq2}), results in
\begin{equation}
	\ddot{\theta} + (b+1) \, \dot{\theta} - \nu = \tau.
\label{red_eq2}
\end{equation}

It is noteworthy that the inductive term is neglected because the dimensionless parameter $\ell \ll 1$ is small compared to other terms in the electrical equation. There is no ad hoc hypothesis such as $\ddot{q} \approx 0$ in this approach.

The dimensionless \emph{reduced-order dynamic model} that results from this simplification is defined by Eqs.(\ref{torque_eq2}), (\ref{red_eq1}) and (\ref{red_eq2}), as well as the initial conditions $\dot{\theta} (0) = \dot{\theta}_0$, and $\theta(0) = \theta_0$. The quasi-steady-state approach simplifies the dynamic model from a three-dimensional first-order system of differential equations to a first-order differential-algebraic system of dimension two.

The reader may observe that, as Eq.(\ref{red_eq2}) depends only on the mechanical variables $\theta$ and $\dot{\theta}$, i.e., there is no dependence with electric current $\dot{q}$, the mechanical dynamics is decoupled from the electrical one, a fact that is well observed by Lima et al. \cite{lima2018_1,Lima2019p552}. Meanwhile, the electrical system is still influenced by the mechanical one, once electric current $\dot{q}$ and angular velocity $\dot{\theta}$ are linked through Eq.(\ref{red_eq1}), that defines an affine manifold on the phase-space $(\theta,\dot{\theta},\dot{q})$, a fact that these authors do not comment.

In practical terms, the mechanical dynamics is obtained from the integration of (\ref{red_eq2}), and then the electric current is calculated using the algebraic constraint (\ref{red_eq1}). However, this procedure apparently leads to a paradox. The initial current in the reduced dynamics is given by $\dot{q}(0) = \nu - \dot{\theta}_0$, because it comes from the algebraic constraint (\ref{red_eq1}). Suppose the full dynamic has $\dot{q}_0 \neq \nu - \dot{\theta}_0$. In that case, the reduced-order model cannot satisfy this initial condition, i.e., accurately represents the system's initial state. This fact is one of the main criticisms of Lima et al. \cite{lima2018_1,Lima2019p552} against this type of model-order reduction technique, arguing that such simplification, by missing initial information of the dynamics, loses its essential (main) characteristics. Nevertheless, this apparent paradox is overcome by the explanation presented below, which is inspired by the development presented by S. Strogatz \cite[section 3.5]{strogatz2014}.

The dynamical system state at time $t$ can be interpreted as the position of a particle, moving along a trajectory that starts in the initial condition (\ref{model_ic_eq2}), whose velocity is determined by the vector field
\begin{equation}
		\left( \begin{array}{c}
			 \dot{\theta}\\
			 \ddot{\theta}\\
			 \ddot{q}\\
		\end{array} \right)
		=
		\left( \begin{array}{c}
			 \dot{\theta}\\
			 \displaystyle \frac{\dot{q} - d^{\,2} \, \sin \theta \, \cos \theta \, \dot{\theta}^{\,2} -b \, \dot{\theta}}{1 + d^{\,2} \, \sin^{\,2} \theta}\\
			 \displaystyle \frac{\nu - \dot{q} - \dot{\theta}}{\ell}\\
		\end{array} \right).
		\label{vector_field_eq}
\end{equation}

According to the Table~\ref{table1}, one has $b \sim \mathcal{O}(10^{-2})$ and $d \sim \mathcal{O}(1)$. In addition, in average terms $\sin^2 \theta \sim 1/2$ and $\sin \theta \, \cos \theta \sim 0$. Considering also a typical set of (non-zero) initial condition, assumed without any loss of generality as being above the affine manifold (\ref{red_eq1}), i.e., $\dot{q}_0 + \dot{\theta}_0 > \nu$, one can show that $\ddot{\theta} \sim \mathcal{O}(1)$ and $\ddot{q} \sim \mathcal{O}(1 / \ell) \gg 1$. 

Since the third component of the velocity vector (\ref{vector_field_eq}) is negative and much more prominent in magnitude than the other two, the dynamical system state goes rapidly downwards towards a region in the neighborhood of the affine manifold (\ref{red_eq1}), whose thickness is $\mathcal{O}(\ell)$. This region merges with the affine manifold as $\ell \to 0$. In what follows, the full-order dynamics evolve (approximately) according to the reduced-order model equations.

Therefore, the singular limit $\ell \to 0$ is not a serious deficiency of the model-order reduction method. It only induces the formation of a temporal boundary layer at $t=0$, so that the dynamics is composed of two parts. A very fast initial transient, where the approximation is not valid, followed by a slow drift around the affine manifold (\ref{red_eq1}), where the reduced-order model represents very well the original system dynamics. Despite being a well know and crucial fact in perturbation theory \cite{Awrejcewicz2012,nayfeh2000,Verhulst2007p747}, it seems this issue was unnoticed by Lima et al. \cite{lima2018_1,Lima2019p552}.

The validity of the electromechanical quasi-steady-state approach requires $\ell \ll 1$, which is equivalent to
\begin{equation}
    \frac{L}{J R^2/G^2} = \frac{L/R}{J R/G^2} = \frac{T_{Q}}{T_{\Theta}} \ll 1,
\end{equation}
i.e., $T_{Q} \ll T_{\Theta}$ (electric time much smaller than mechanical time). The mathematical development of this section provides a \emph{formal justification} for the physical argumentation presented in section~\ref{time-scale_analysis}.

\section{Results and discussion}
\label{num_experiments}

This section presents some numerical experiments, conducted with the Matlab code \textbf{ElectroM} \cite{ElectroM}, that illustrate general characteristics, and some peculiarities, of the reduced-order model. The dimensionless parameters adopted for the full-order dynamic model, corresponding to the typical values of Table~\ref{table1}, are the following: $\ell = 0.05$, $b=1$, $\nu=1$ and $d=10$. Unless something is said to the contrary, both dynamical systems (full and reduced) are integrated for 200 units of dimensionless time, from the initial condition $(\theta_0, \dot{\theta}_0, \dot{q}_0) = (0,0,3\, \nu)$, which is typical ($\dot{\theta}_0+\dot{q}_0 \neq \nu$). The reduced model considers the same parameters, but does not have the inductive term. 

\subsection{The typical electromechanical dynamic behavior and the reduced-order model approximation}

\begin{figure}[ht!]
	\centering
	\includegraphics[scale=0.4]{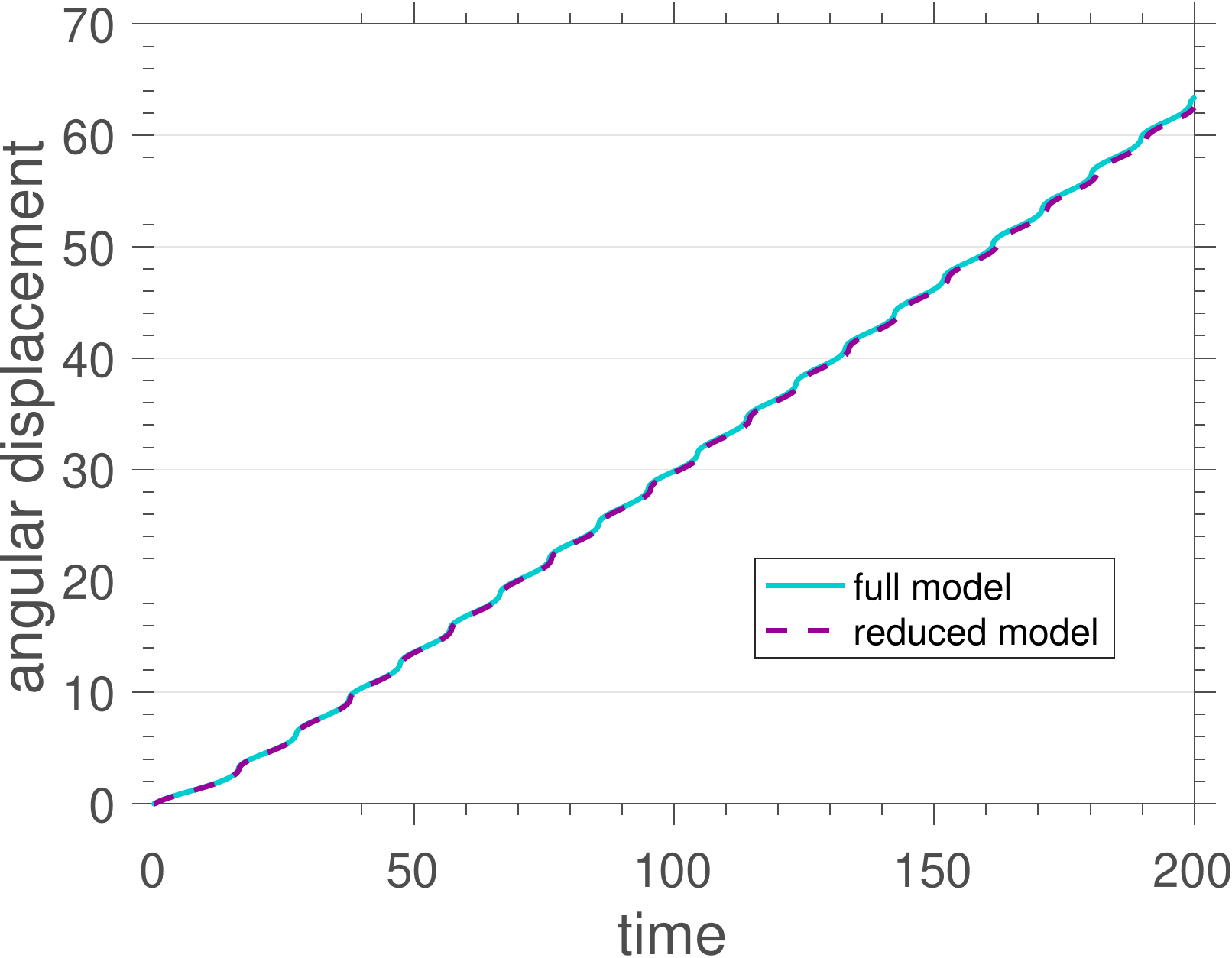}
	\includegraphics[scale=0.4]{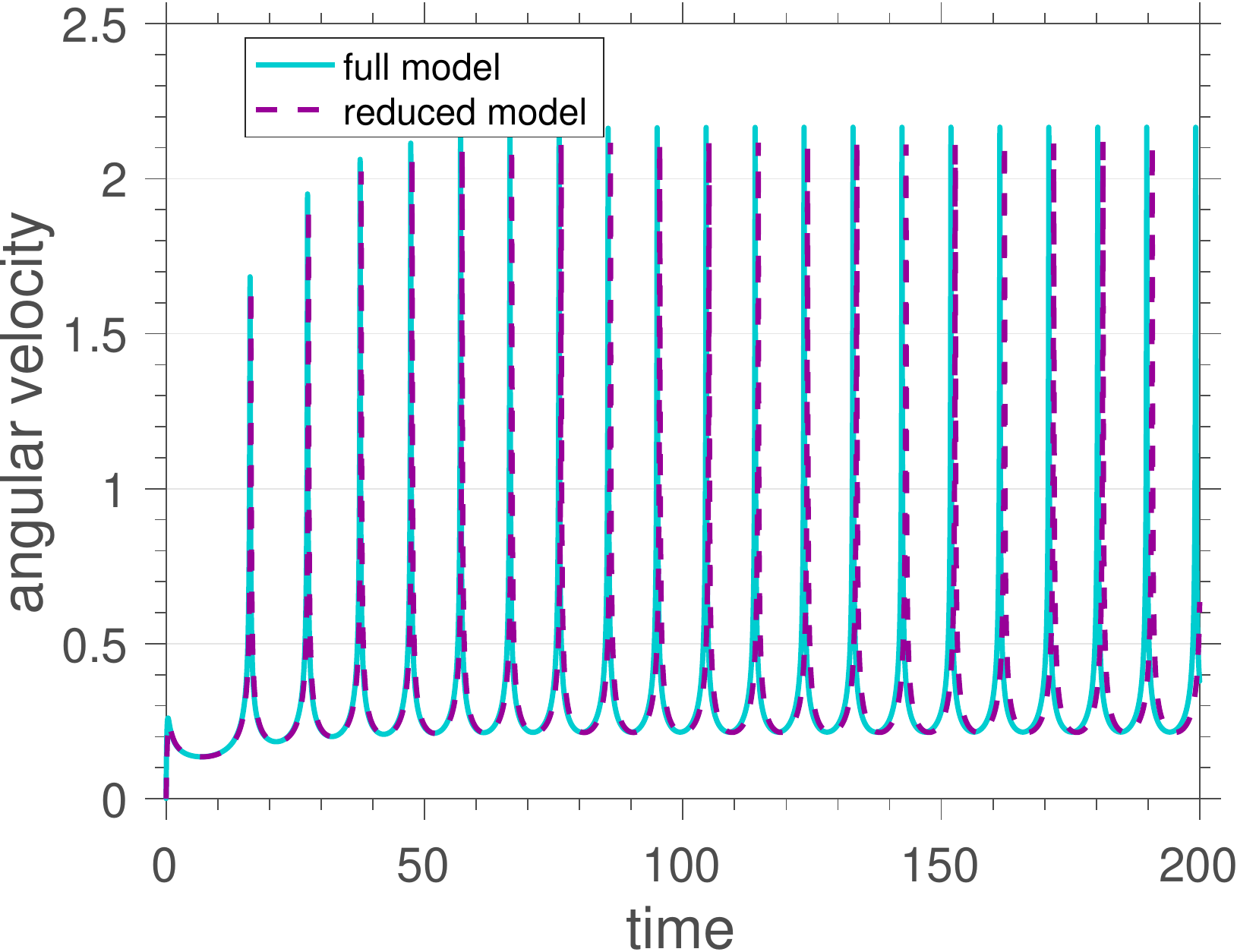}
	\includegraphics[scale=0.4]{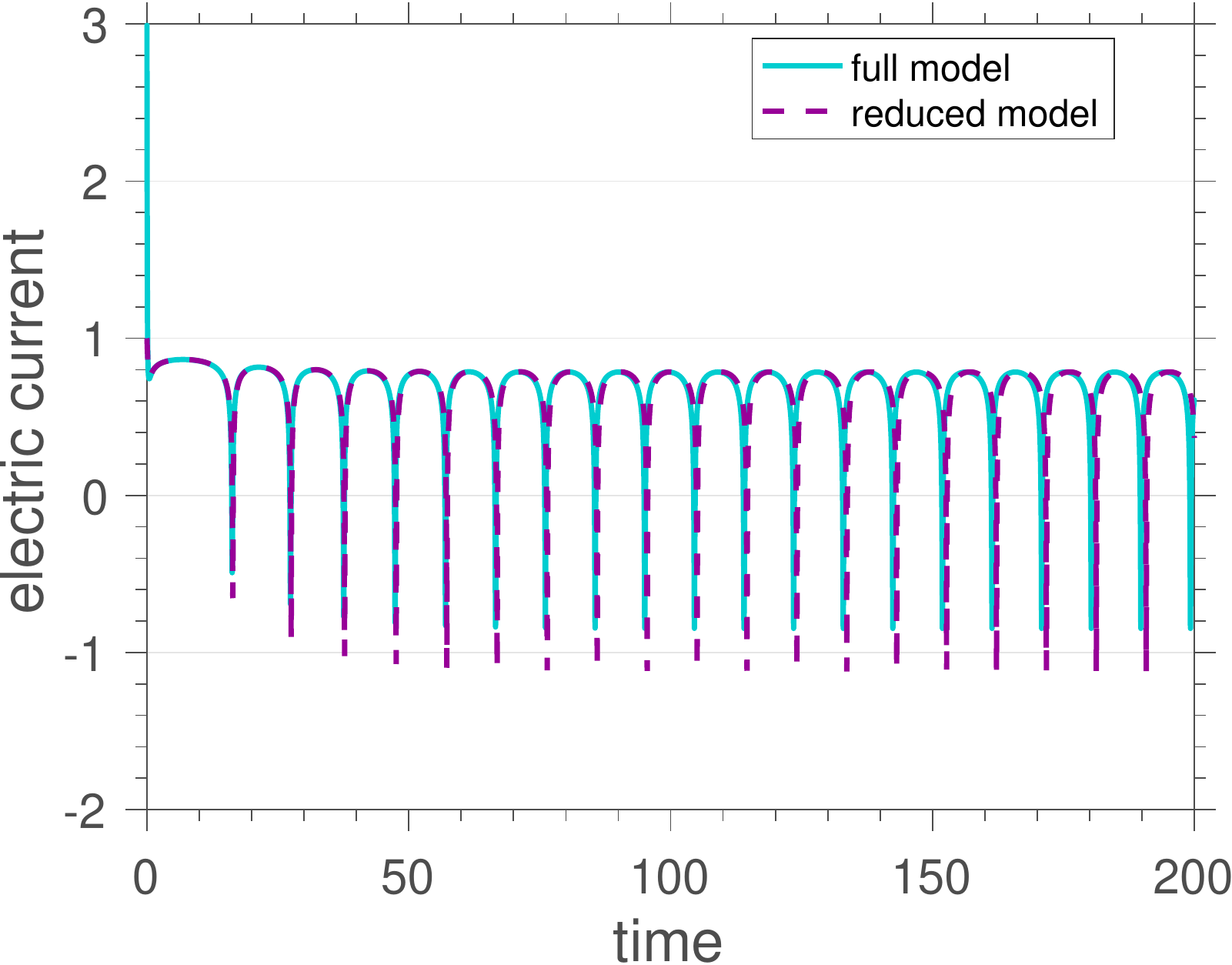}
	\caption{Comparison between the time series of the full model with their counterparts in the reduced-order model: angular displacement (top); angular velocity (middle); and  electric current (bottom). Dimensionless parameters: $\ell = 0.05$, $b=1$, $\nu=1$, $d=10$, $(\theta_0, \dot{\theta}_0, \dot{q}_0) = (0,0,3 \, \nu)$.}
\label{fig_time-series1}
\end{figure}

\begin{figure}[ht!]
	\centering
	\includegraphics[scale=0.4]{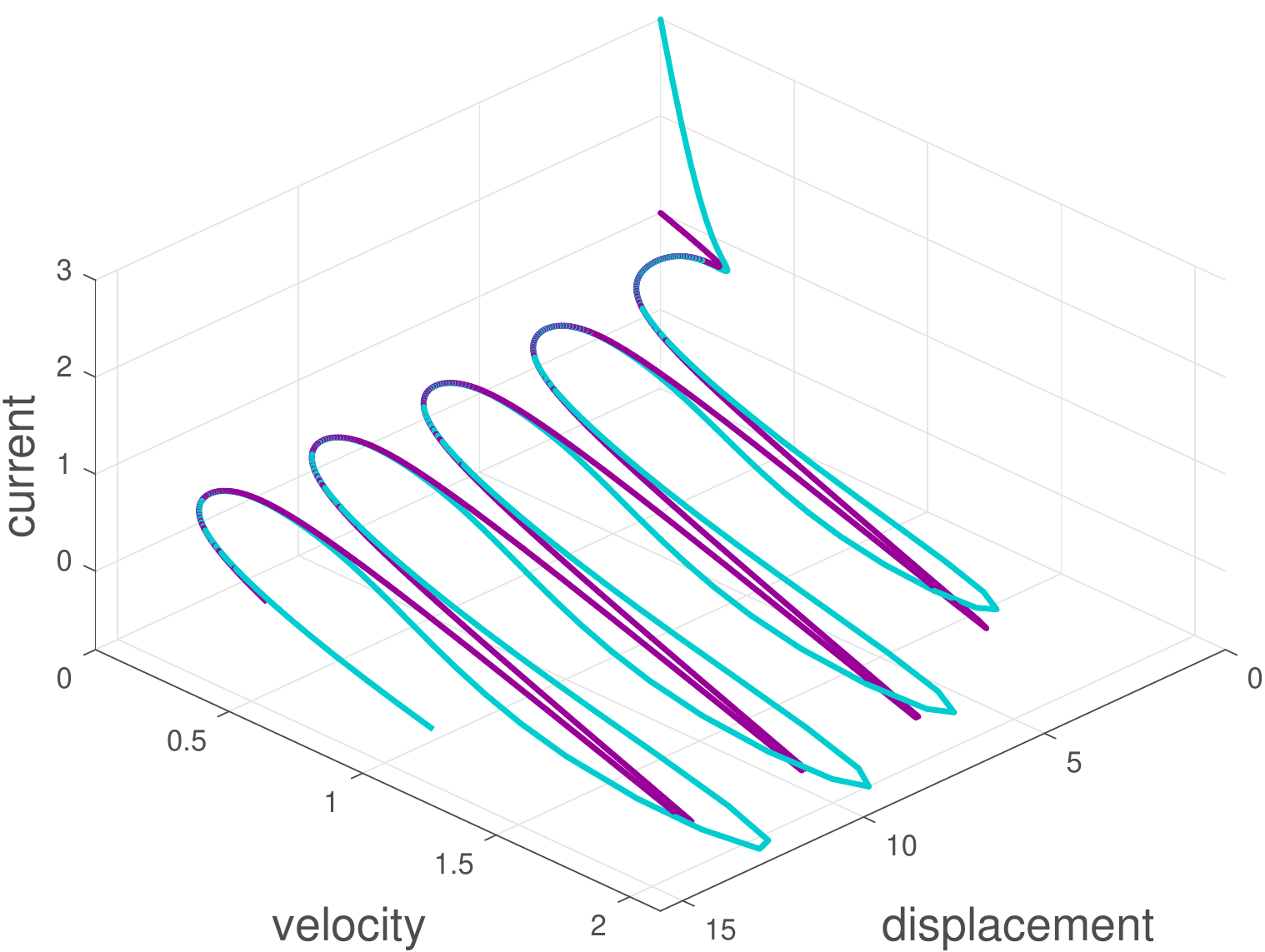}
    \includegraphics[scale=0.4]{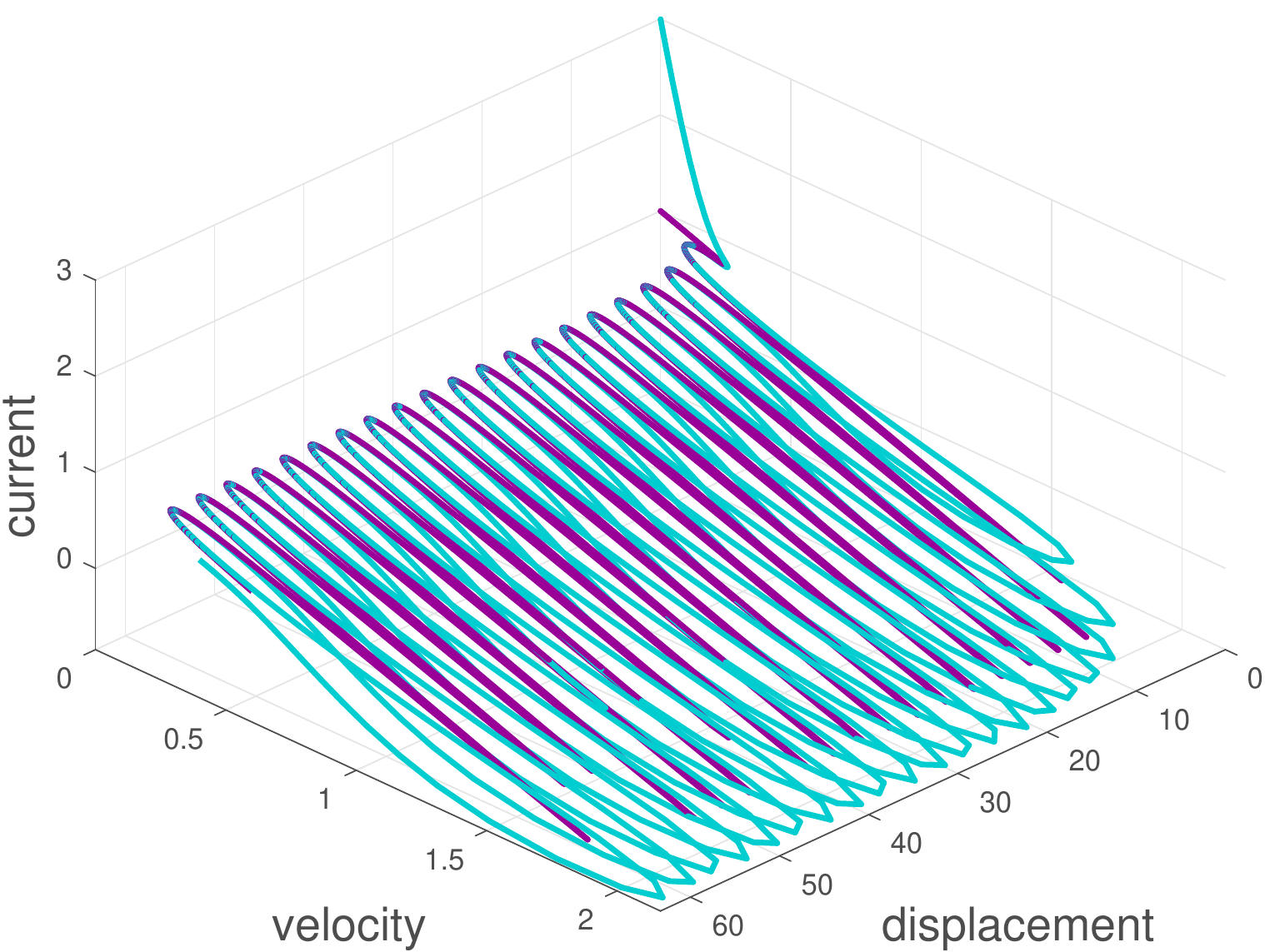}
	\includegraphics[scale=0.4]{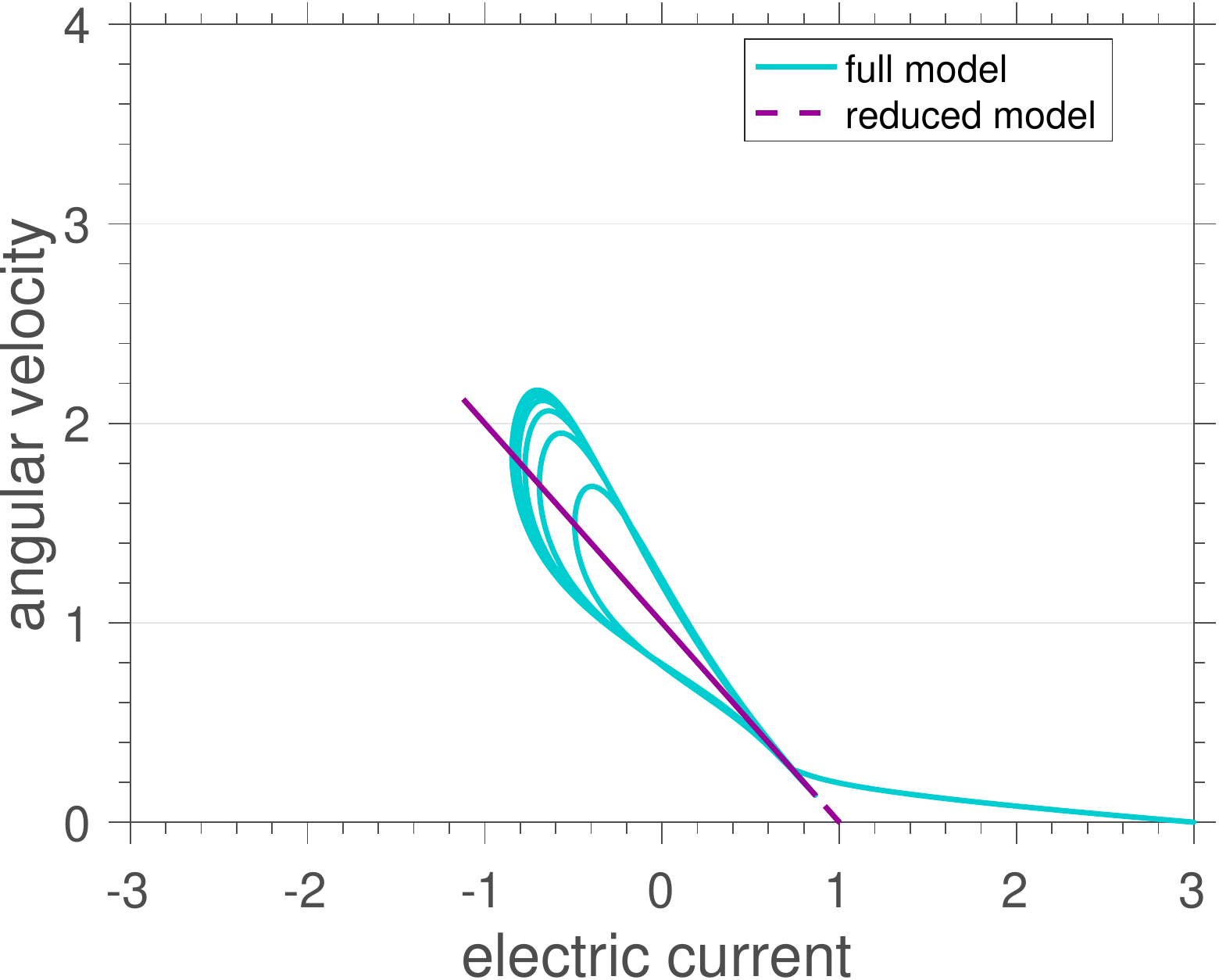}
	\caption{Comparison between the phase-space trajectories, for both models, from two different perspectives (top and middle), and their projections in the $\dot{\theta} \times \dot{q}$ plane (bottom).  The blue (external) curve corresponds to the full model, while the pink (internal) curve represents the dynamics of the reduced model. Dimensionless parameters: $\ell = 0.05$, $b=1$, $\nu=1$, $d=10$, $(\theta_0, \dot{\theta}_0, \dot{q}_0) = (0,0,3 \, \nu)$.}
\label{fig_phase-space1}
\end{figure}

The time series for angular displacement $\theta(t)$, angular velocity $\dot{\theta}(t)$, and electric current $\dot{q}(t)$, for both models (full and reduced), are shown in Figure~\ref{fig_time-series1}. The trajectories of these two systems in phase-space, as well as their projections in $\dot{\theta} \times \dot{q}$ plane, can be seen in Figure~\ref{fig_phase-space1}. 

Observing the results of Figure~\ref{fig_time-series1} one can note that, for the adopted set of parameters, the reduced-order model can capture very well the full model qualitative behavior, since there is a high correlation between the time series of the two models.

However, some readers might argue that while there is a good correlation between the time series, the two dynamics accumulate into different attractors, which are pretty different at first sight (see Figure~\ref{fig_phase-space1}), and, because of this fact, the reduced model does not produce a good approximation. While this is a valid concern and quite natural for beginners in nonlinear dynamics, a more mature understanding of a nonlinear system's geometry reveals that this is a very naive overview of the reduced-order dynamic model.

The modern (geometric) theory of nonlinear dynamical systems \cite{arnold1992,hirsch2012,perko2006,Verhulst2012b}, which dates back to the pioneering works of Poincaré in the early 20th-century \cite{Verhulst2012}, study the qualitative behavior of the underlying solutions. This new paradigm of dynamics is closely related to the fact that for the vast majority of dynamical systems is simply impossible to obtain a closed-form solution, so understanding the system's geometric aspects becomes essential to say something about the underlying solutions. Although, nowadays, the lack of a formula is not a barrier to quantitative analysis, as there are consolidated numerical methods and great computing power, the main lesson that comes from the geometric theory of dynamical systems is that the most interesting questions\footnote{Some interesting questions: What is the solution asymptotic behavior? How is the solution stability affected by varying a system parameter? How sensitive is a solution to perturbations on the initial conditions? etc.} to be asked are qualitative.

Based on this (qualitative) dynamic perspective, what matters in a reduced-order model is its intrinsic ability to reproduce the most important qualitative aspects of the original system's behavior. Quantitative differences are secondary for most purposes of interest. In this sense, the reduced model shown above proved to be an excellent approximation for the original system dynamics since it reproduces the key features of the original time series very well. 

\begin{figure}[ht!]
	\centering
    \includegraphics[scale=0.4]{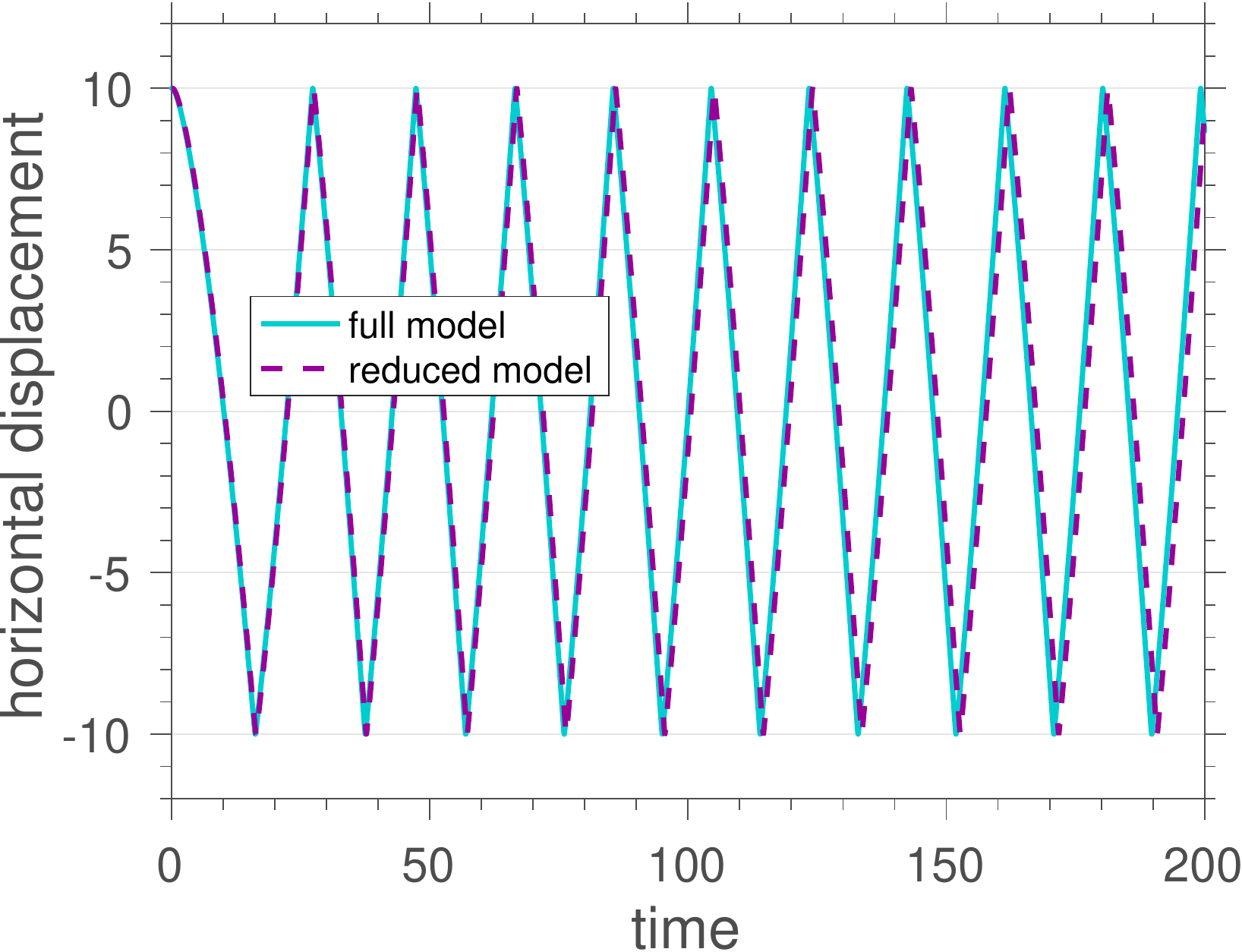}
	\includegraphics[scale=0.4]{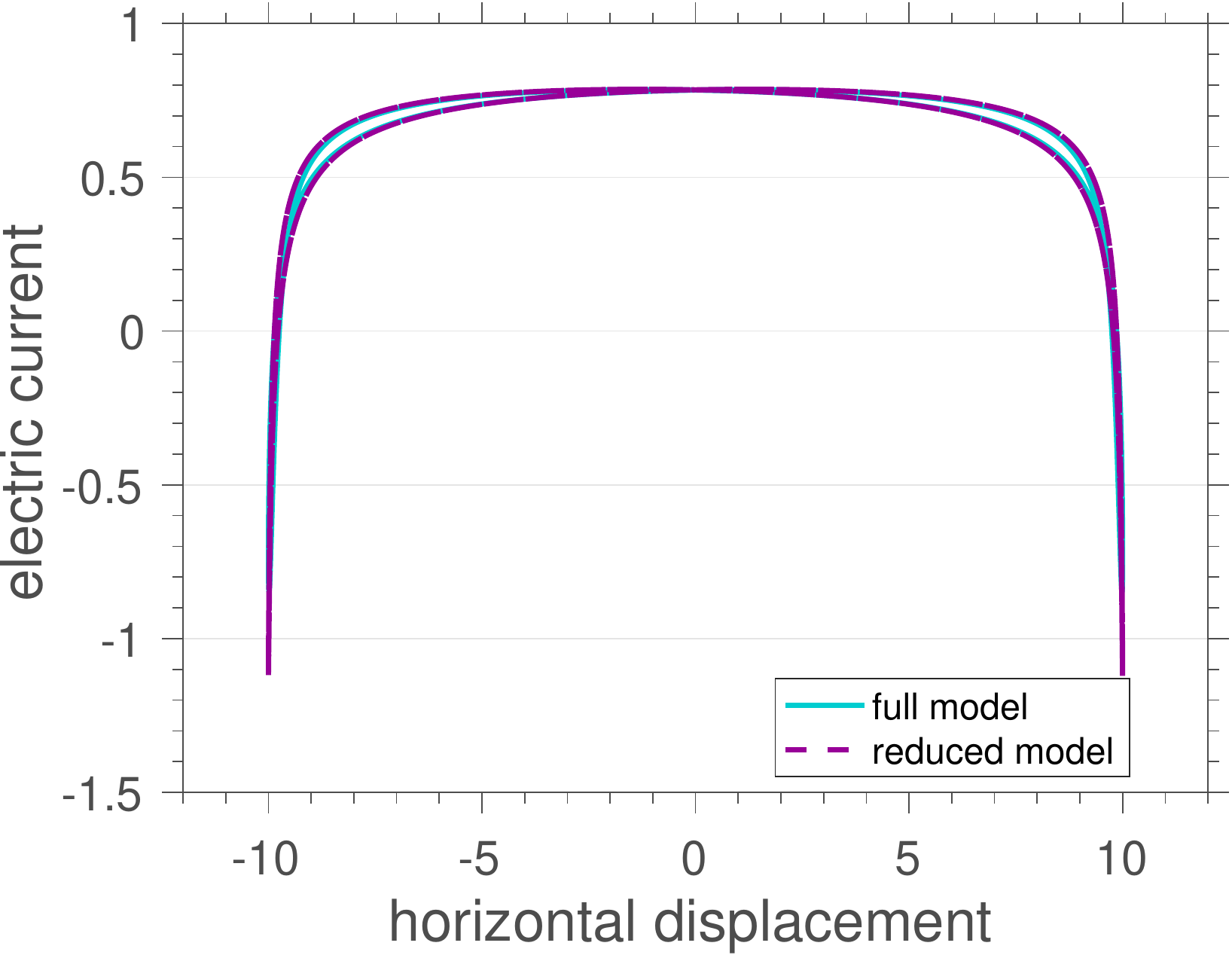}
	\caption{Comparison, for both models, between the horizontal displacement time series (top); and the dynamics projection in the $\dot{q} \times x$ plane (bottom). Dimensionless parameters: $\ell = 0.05$, $b=1$, $\nu=1$, $d=10$, $(\theta_0, \dot{\theta}_0, \dot{q}_0) = (0,0,3 \, \nu)$.}
\label{fig_phase-space_projections1}
\end{figure}

Note in Figure~\ref{fig_time-series1} that, for both $\dot{\theta}(t)$ and $\dot{q}(t)$, the time series of the reduced model present the oscillatory pattern of fast growth and decay (or vice versa) which is characteristic of full dynamics. This relaxation oscillation behavior is a consequence of the kinematic constraint defined by Eq.(\ref{kinematic_eq}), which enforces an inversion of direction in the cart horizontal displacement $x(t)$ when it arrives at a certain limit value, as can be seen in the Figure~\ref{fig_phase-space_projections1}, which displays the dynamic evolution of $x(t)$ and the projection of the system dynamics in the $\dot{q} \times x$ plane. From Figure~\ref{fig_phase-space_projections1} it is also clear that the nonlinear relationship between $q(t)$ and $x(t)$, a consequence of the electromechanical coupling in Eqs.(\ref{electric_eq}) and (\ref{mech_eq}), is also very well represented by the reduced model from Eqs.(\ref{red_eq1}) and (\ref{red_eq2}), although in this case the mechanical equation is decoupled of the electrical equation (but not the contrary).

\begin{figure}[ht!]
	\centering
	\includegraphics[scale=0.4]{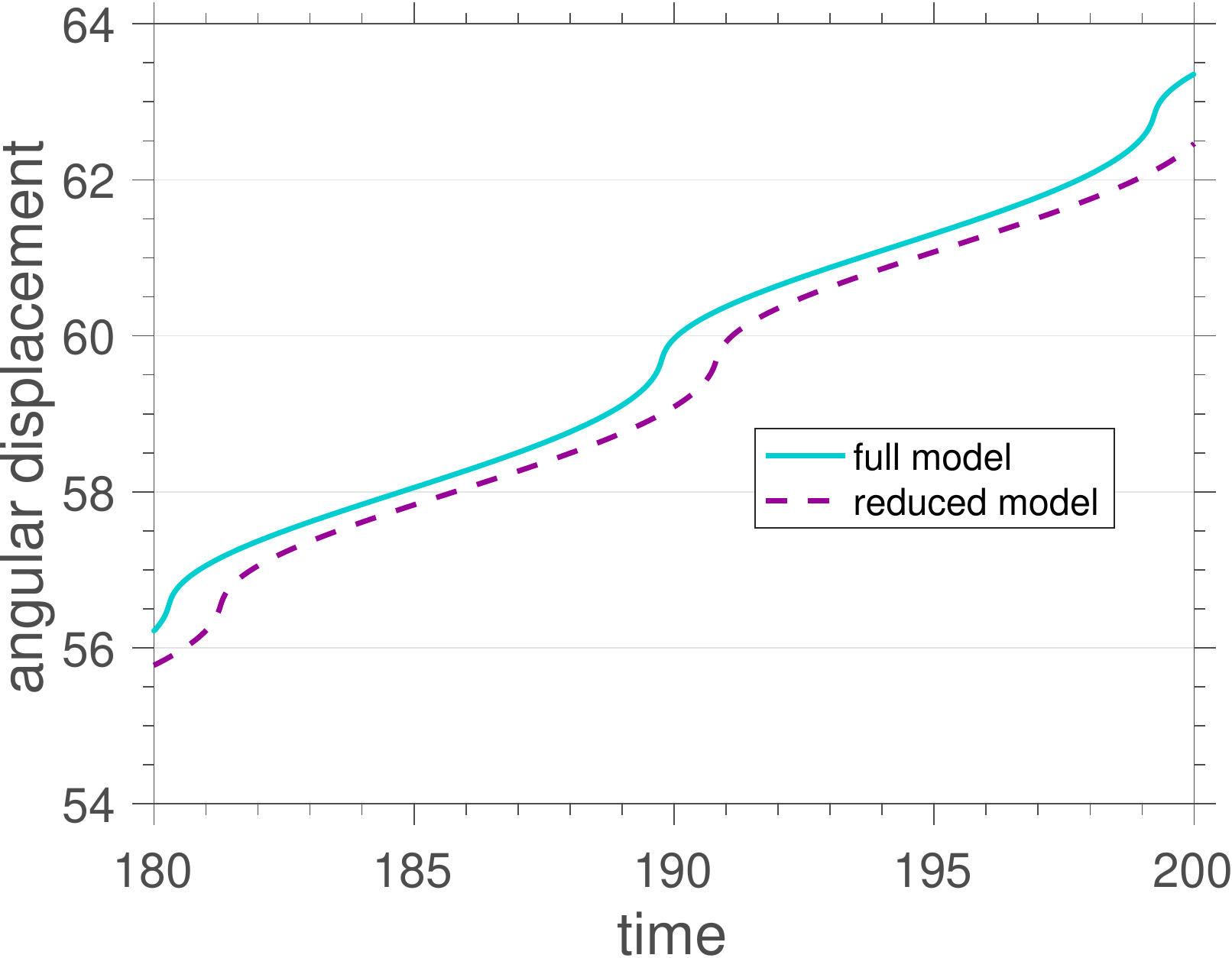}
	\includegraphics[scale=0.4]{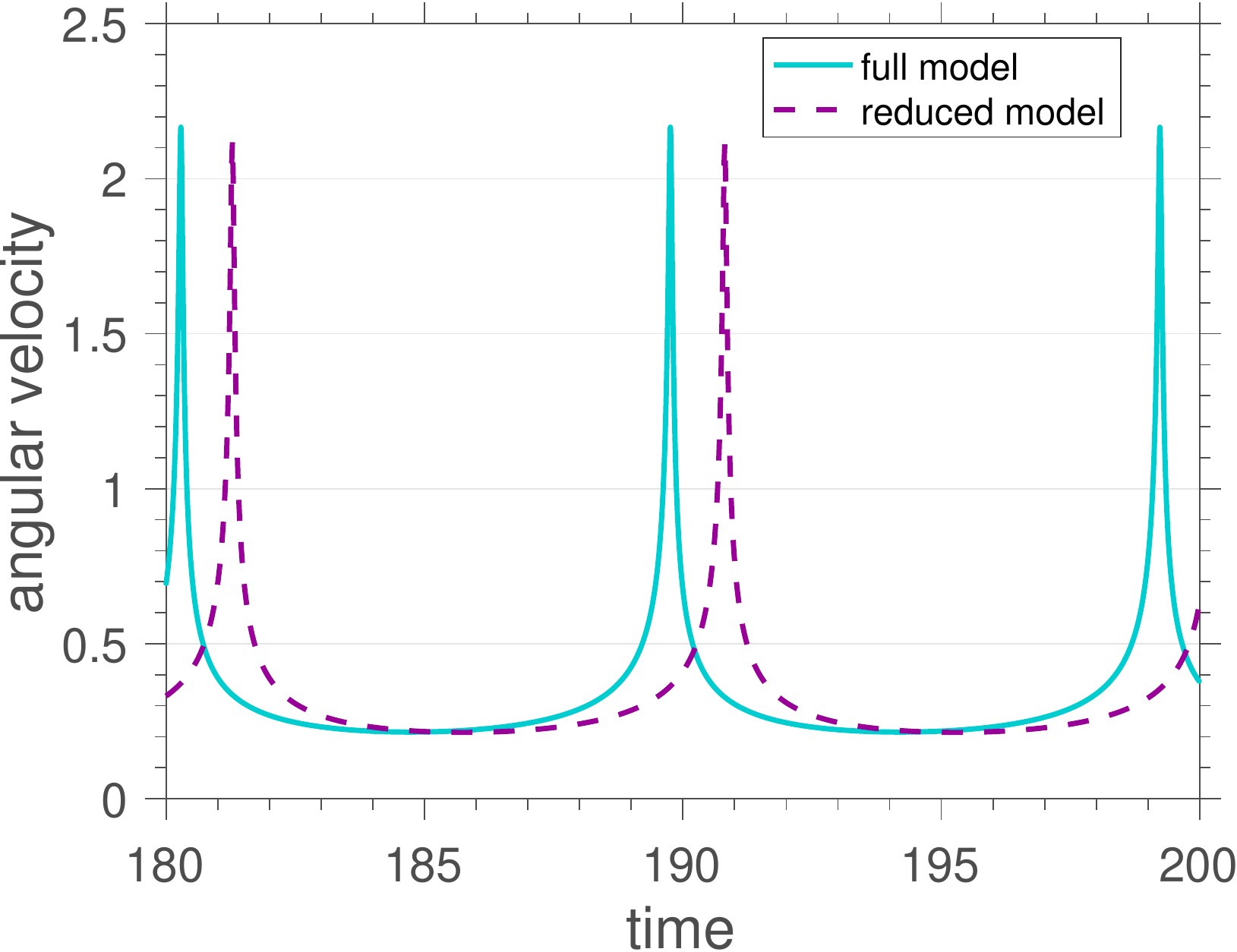}
	\includegraphics[scale=0.4]{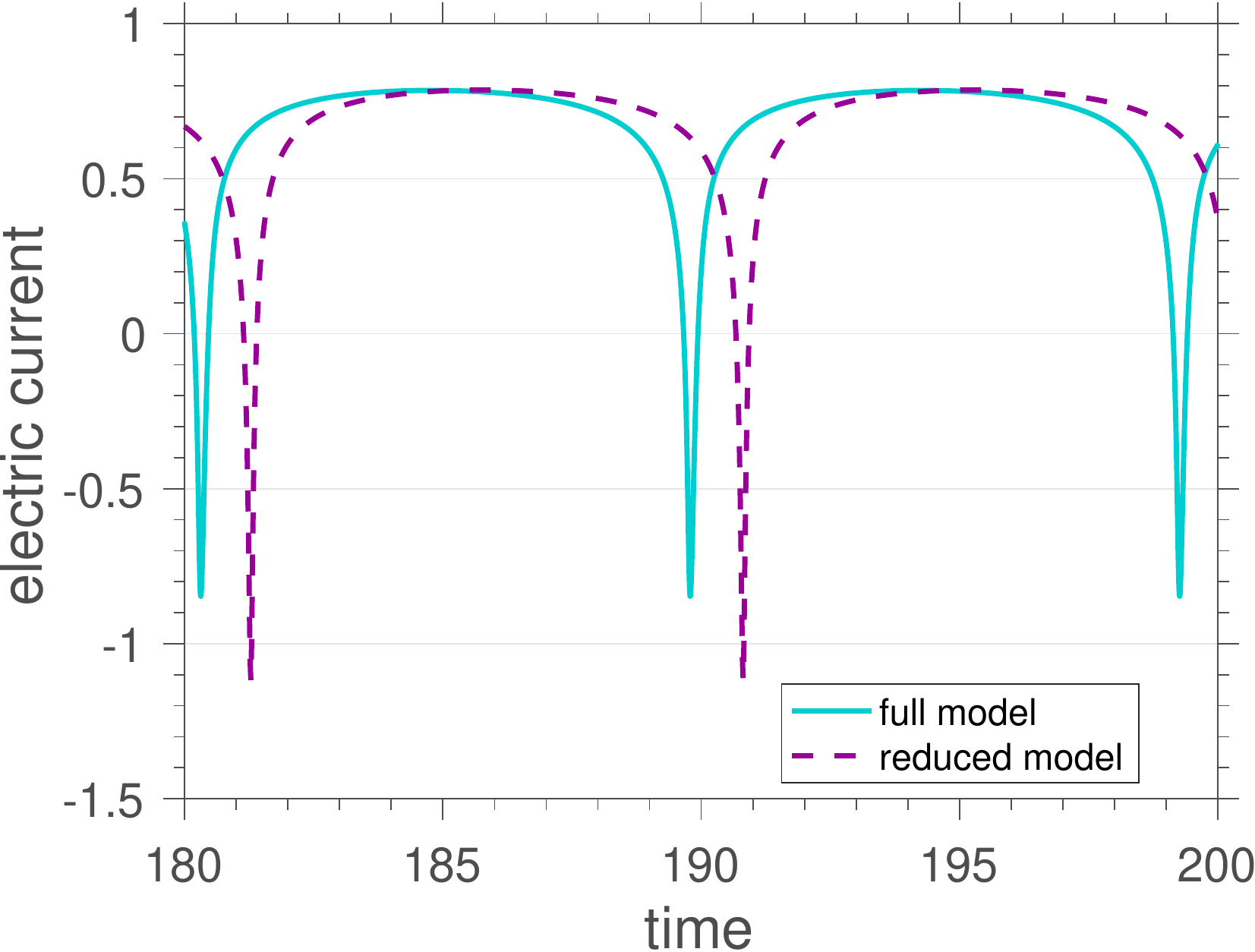}
	\includegraphics[scale=0.4]{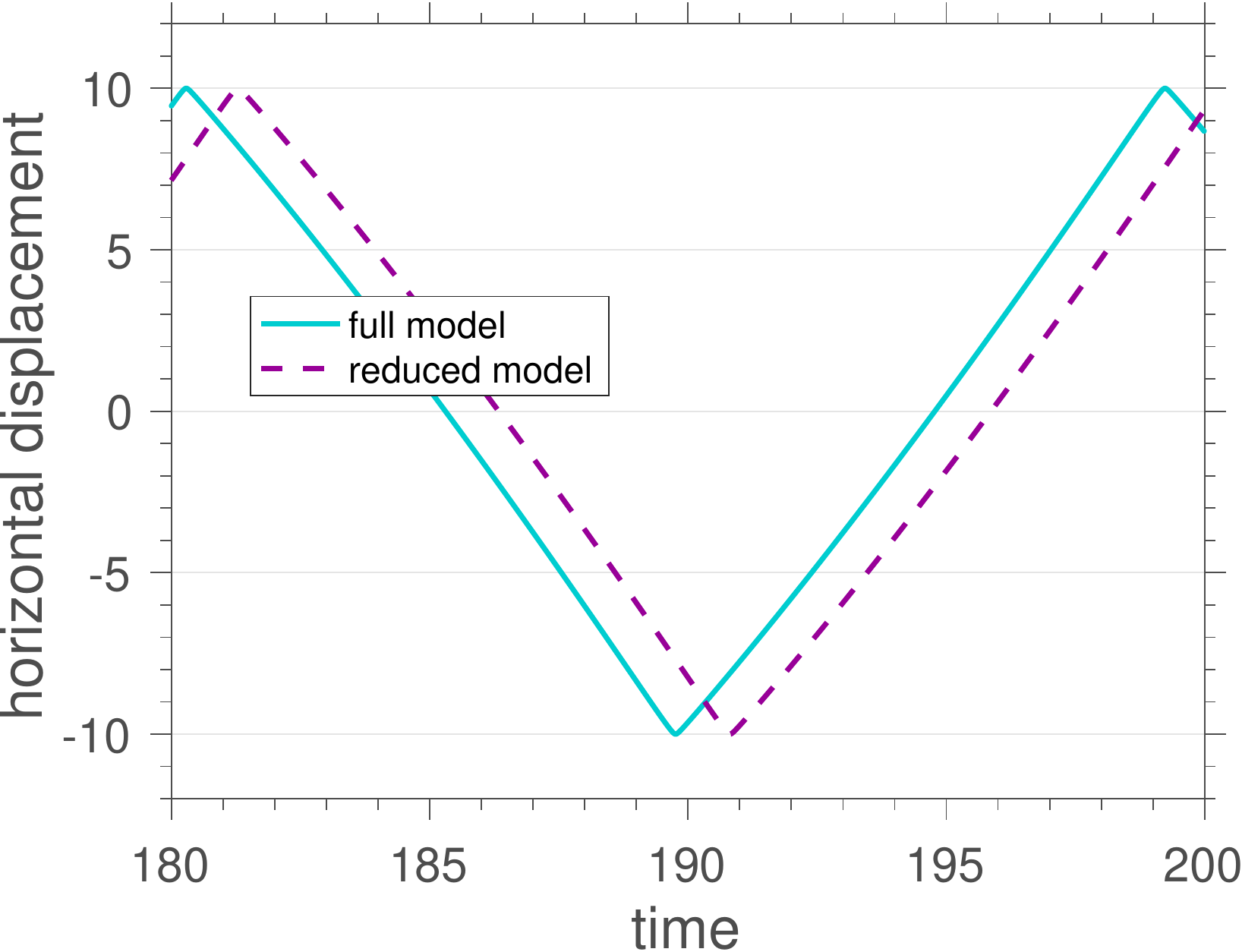}
	\caption{Enlarged view of the final part of the time series of $\theta(t)$, $\dot{\theta}(t)$, $\dot{q}(t)$ and $x(t)$, respectively, for the full and the reduced-order models. There is a discrepancy between the two dynamics. Dimensionless parameters: $\ell = 0.05$, $b=1$, $\nu=1$, $d=10$, $(\theta_0, \dot{\theta}_0, \dot{q}_0) = (0,0,3 \, \nu)$.}
\label{fig_timeseries1_zoom}
\end{figure}

Although the curves shown in Figure~\ref{fig_time-series1} suggest that the quantitative agreement between the time series is excellent, this is not the case, as can be seen in Figure~\ref{fig_timeseries1_zoom}, which shows in more detail the final part of the time series of $\theta(t)$, $\dot{\theta}(t)$, $\dot{q}(t)$ and $x(t)$, for the full and the reduced-order models. It is possible to see that the reduced-order model time series are delayed concerning the full-order model response. This quantitative discrepancy can be relatively small if the time interval considered in the analysis is very short. In this case, the approximation may also be good from a quantitative perspective. However, more often, the reduced dynamics may become very delayed to the full-order one for long time intervals. This delay is a consequence of the under-prediction of the first natural frequency of the system response by the reduced-order model, as shown in section~\ref{sect_val_limity}.

To better visualize this delay, as well as its consequences, the reader is invited to watch the video animation of the system trajectory in phase-space available in Supplementary Material 02 or in \cite{eletromech_video2}. It is observed that despite the trajectory of the reduced model being (always) delayed to the original system, there is a good qualitative agreement between the two dynamics. Indeed, the (zig-zag) relaxation oscillatory pattern inherent in the full model is also observed in the reduced dynamics. In the $\theta$ direction, where the system attractor is unlimited (see the upper part of the Figure~\ref{fig_phase-space1}), the difference between the two dynamics becomes arbitrarily large as $t \to \infty$. However, in the time series of $\dot{\theta}(t)$ and $\dot{q}(t)$ the delay is not able to induce an arbitrarily large divergence between the two dynamics, since the cross-section $\dot{\theta} \times \dot{q}$ of the system attractor (bottom part of the Figure~\ref{fig_phase-space1}) is limited, which prevents the distance between full and reduced time series from becoming too large. Furthermore, due to the periodic behavior of these series, after some time, the two dynamics in $\dot{\theta} \times \dot{q}$ plane briefly return to phase, then move away again, and continue in this pattern indefinitely. In the case of the electromechanical system analyzed in this paper, this limited distance, with a periodic return to the same phase of the original dynamics, is what guarantees the consistency of the employed quasi-steady state approach, since it ensures that the qualitative behavior of the two systems will never be very different.

\subsection{The quasi-stationary nature of electrical dynamics}

At this point, the reader may be confused about the hypothesis discussed in section~\ref{time-scale_analysis}, where electrical dynamics is assumed to be in a quasi-stationary regime, as the electric current time series shown in Figure~\ref{fig_time-series1} presents large fluctuations, which in principle eliminates any notion of stationarity. This is a very natural question whose understanding requires thinking about the dynamics in the following way.

Current fluctuations occur according to a mechanism of relaxation oscillations, whose abrupt changes in value occur for very short periods so that for the vast majority of the analyzed time interval, the time series assumes an approximately constant value. Hence, the notion of quasi-stationarity is employed rather than the strict notion of stationarity.

\subsection{The temporal boundary layer resulting from non-compatible initial conditions}

As pointed out in references \cite{lima2018_1,Lima2019p552}, and explained in detail in section~\ref{red_order_dyn_model}, the two dynamics are very discrepant in the first instants of evolution due to an incompatibility between the initial conditions, which is a result of dropping out the inductive term (electric inertia) during the model reduction procedure. This topic is addressed in this section, with a discussion that complements the theoretical explanation in section 5, illustrating that the period where the two dynamics are very dissimilar is transitory and very fast, not being, therefore, a strong limitation for the use of the reduced-order model in long-term qualitative analyses. 

\begin{figure}[h!]
	\centering
	\includegraphics[scale=0.4]{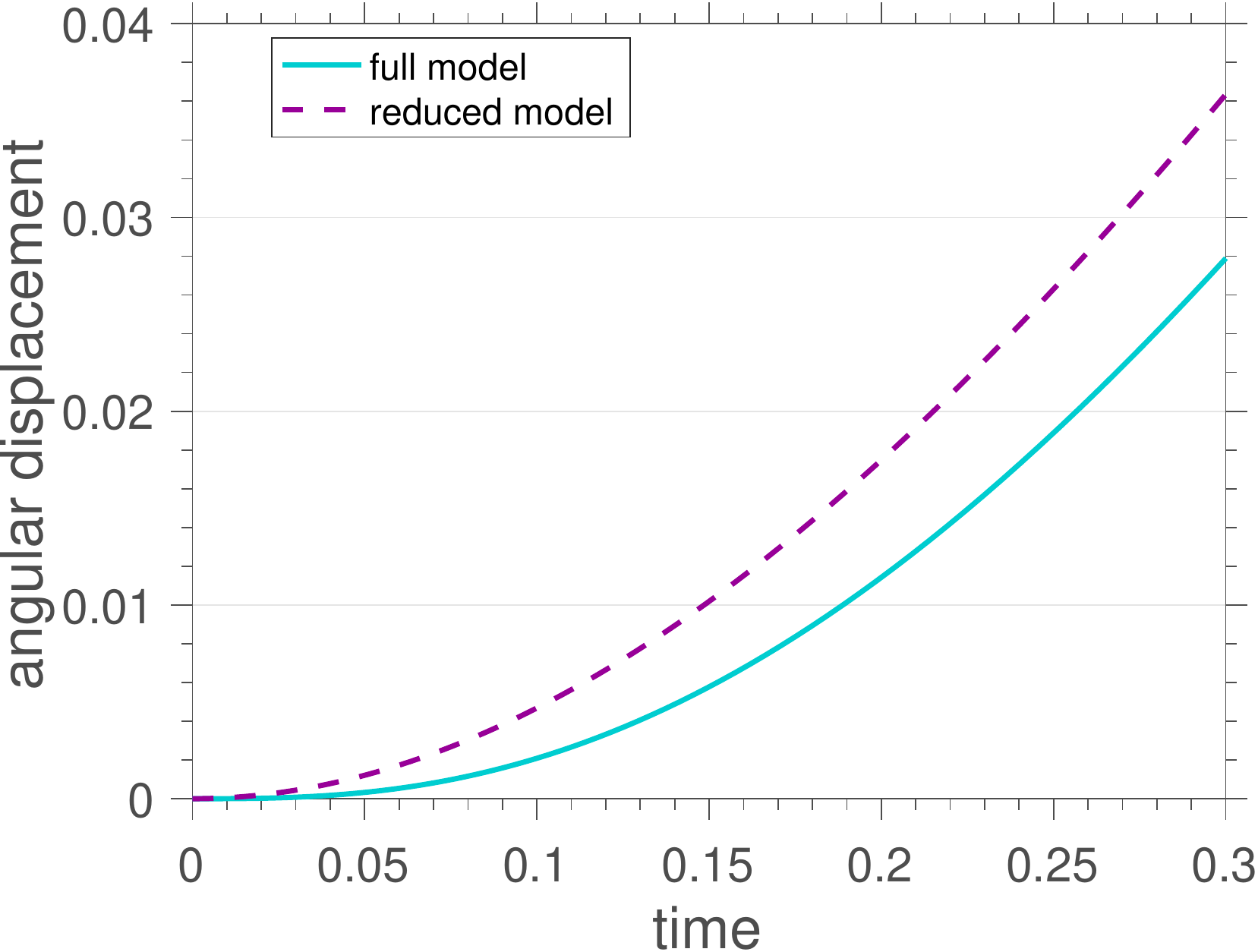}
	\includegraphics[scale=0.4]{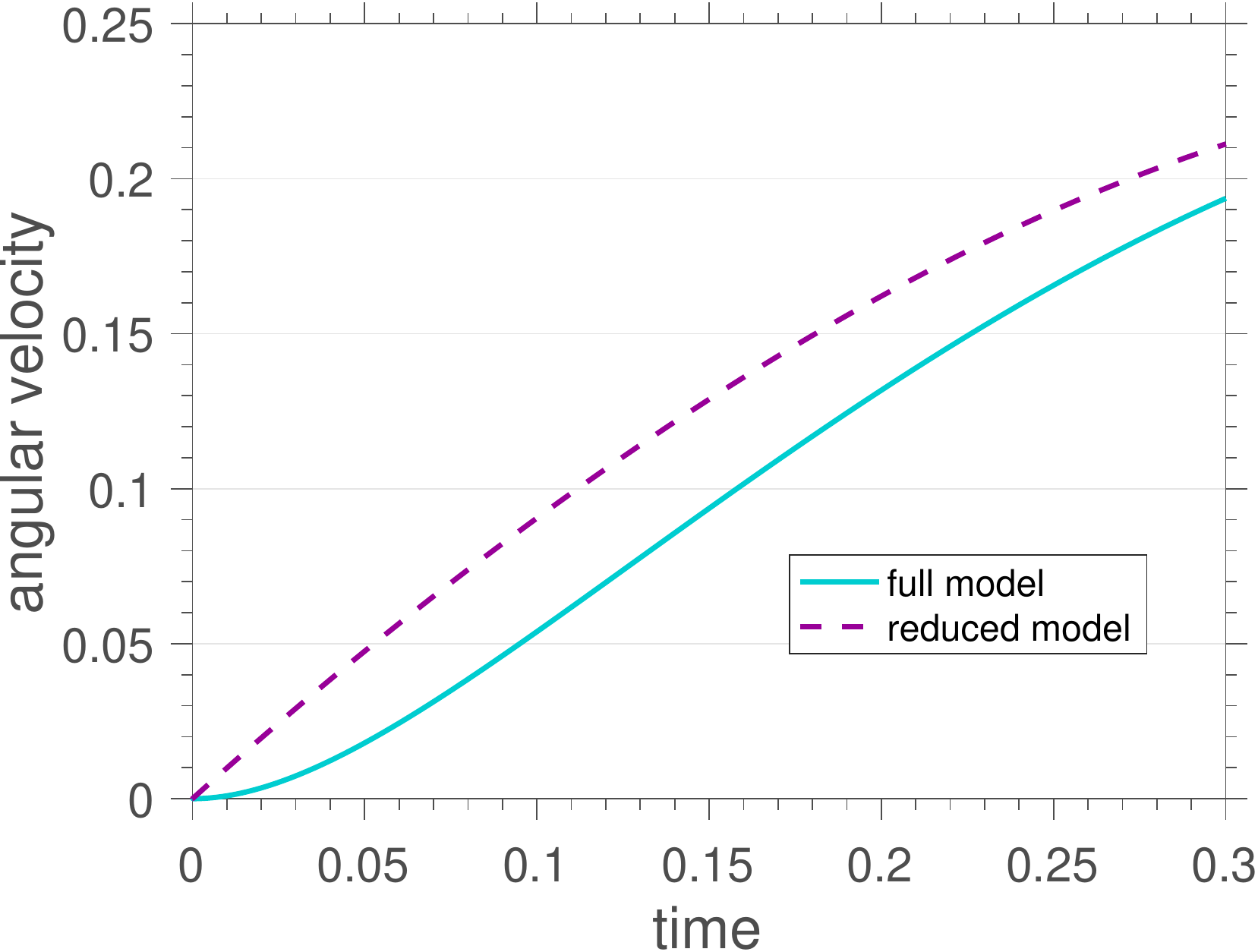}\\
	\includegraphics[scale=0.4]{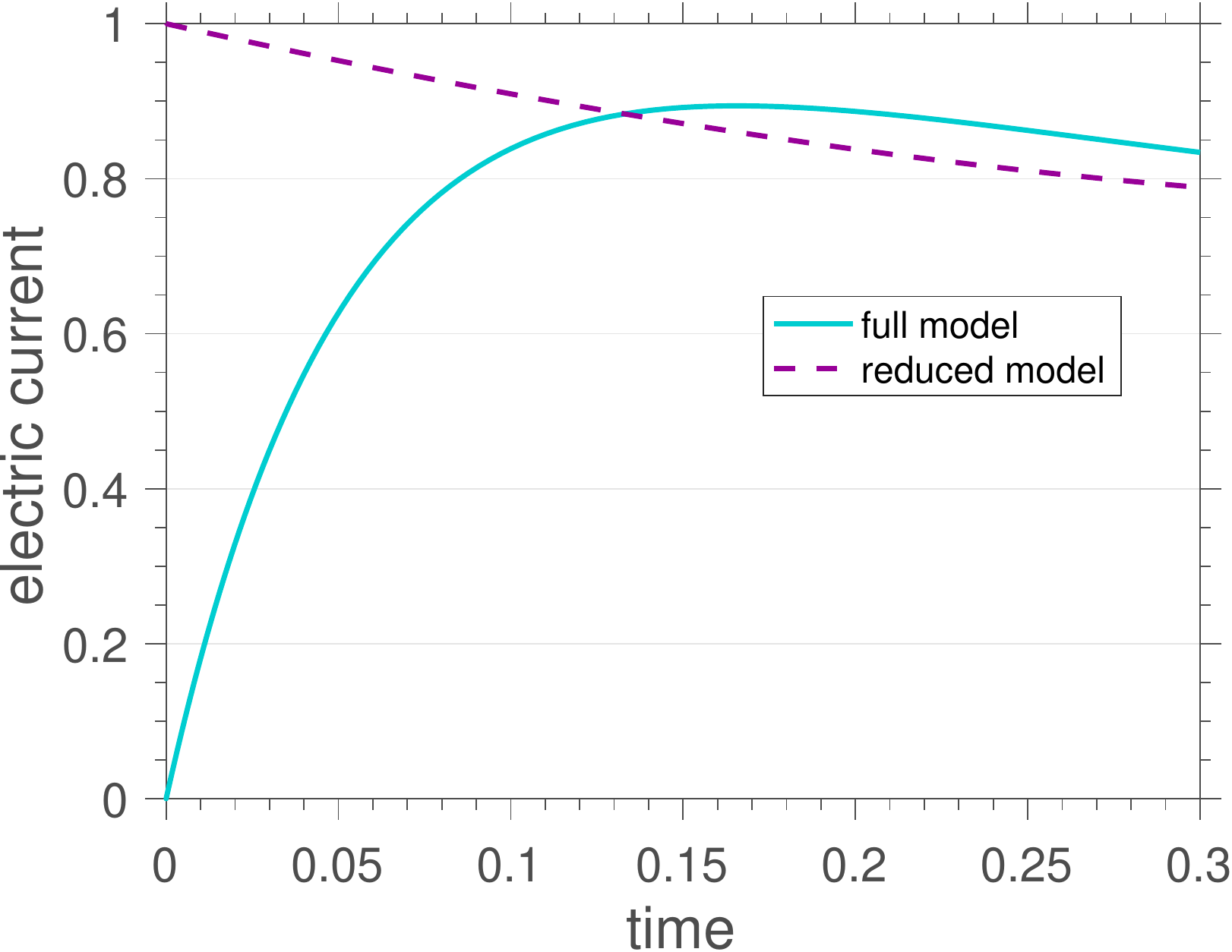}
	\includegraphics[scale=0.4]{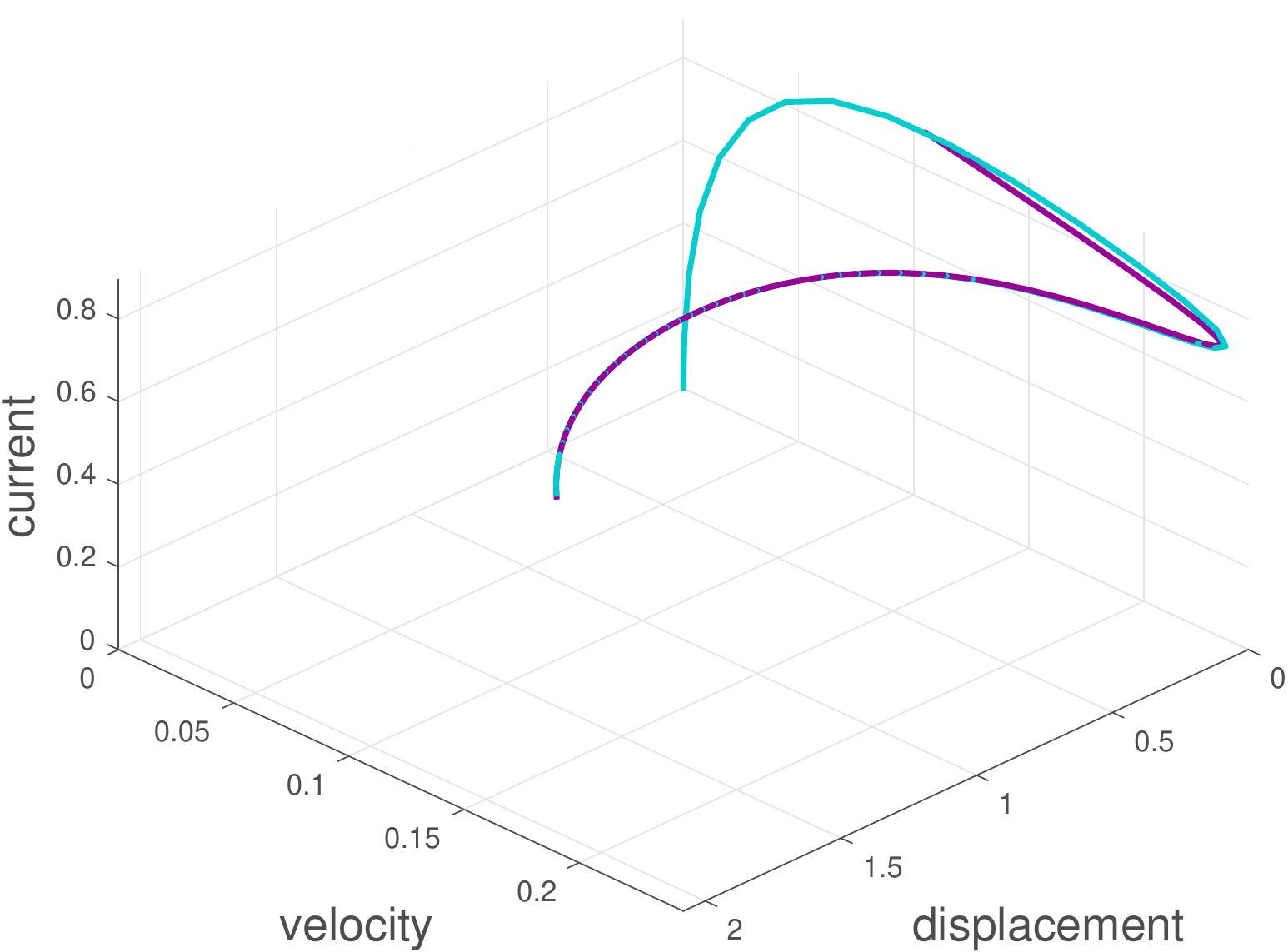}
	\caption{Comparison, for both dynamics, between $\theta(t)$, $\dot{\theta}(t)$, $\dot{q}(t)$ time series and the phase-space trajectory at the beginning of the time evolution. Dimensionless parameters: $\ell = 0.05$, $b=1$, $\nu=1$, $d=10$, $(\theta_0, \dot{\theta}_0, \dot{q}_0) = (0,0,0)$.}
\label{timeseries2_fig}
\end{figure}

The reader is invited to observe Figure~\ref{timeseries2_fig}, which compares (for both models) $\theta(t)$, $\dot{\theta}(t)$, $\dot{q}(t)$ time series, and the phase-space trajectory, with the full model starting from the initial condition $(\theta_0, \dot{\theta}_0, \dot{q}_0) = (0,0,0)$, that is typical in the sense $\dot{\theta}_0+\dot{q}_0 \neq \nu$. Note that, despite the absence of a current derivative in the electric equation, the initial values $\theta_0$ and $\dot{\theta}_0$ coincide in both models, being $\dot{q}_0$ the only initial condition that the reduced-order model does not meet. Such ``defect'', i.e., the inability of the reduced-order model to capture a good estimate of the electric current in the very first moments, is eliminated very quickly -- less than $0.05$ units of dimensionless time -- since the $\dot{q}(t)$ time series associated with the full-order model quickly moves towards its reduced-order model counterpart (third graph in Figure~\ref{timeseries2_fig}), starting to ``guide'' the reduced dynamics indefinitely. In this short period, where a temporal boundary layer forms in the vicinity of $t=0$, the effects of the inductive term are important, and the two dynamics differ significantly as the reduced one does not feel its inertial effects. A slow steady-state drift follows, where inertial effects no longer affect the dynamic behavior. In this second stage of the dynamics, the distance between the two trajectories is $\mathcal{O}(\ell)$.

This two-phase behavior becomes clearer when you appreciate the fourth graph in Figure~\ref{timeseries2_fig}, where the full dynamics leap towards the reduced dynamics (see the small stretch of the curve in pure cyan color). Although the bottom graph in Figure~\ref{timeseries2_fig} shows only a small (initial) portion of the dynamic evolution in phase-space, the attractor where the full dynamics accumulates (after the transient) has the same ``airfoil'' shape shown in Figure~\ref{fig_phase-space1}, with an embedded affine manifold that represents the reduced-order system attractor. Thus, full dynamics steadily evolves in the ``airfoil'', while reduced dynamics pursues the last (with some delay), but excellent correlation, within the affine manifold $\dot{\theta}+\dot{q} = \nu$ which is $\mathcal{O}(\ell)$ inside the ``airfoil''. This is typical behavior when initial conditions are such that $\dot{\theta}_0+\dot{q}_0 \neq \nu$. In the particular (and rare) case where $\dot{\theta}_0+\dot{q}_0 = \nu$ the reduced-order dynamics satisfies all initial conditions and the boundary layer does not exist, the two dynamics evolve $\mathcal{O}(\ell)$ apart for all $t>0$.

\subsection{When full-order and reduced-order dynamics are pretty different and the validity limit of the approximation
}
\label{sect_val_limity}

\begin{figure*}
\centering
\includegraphics[scale=0.25]{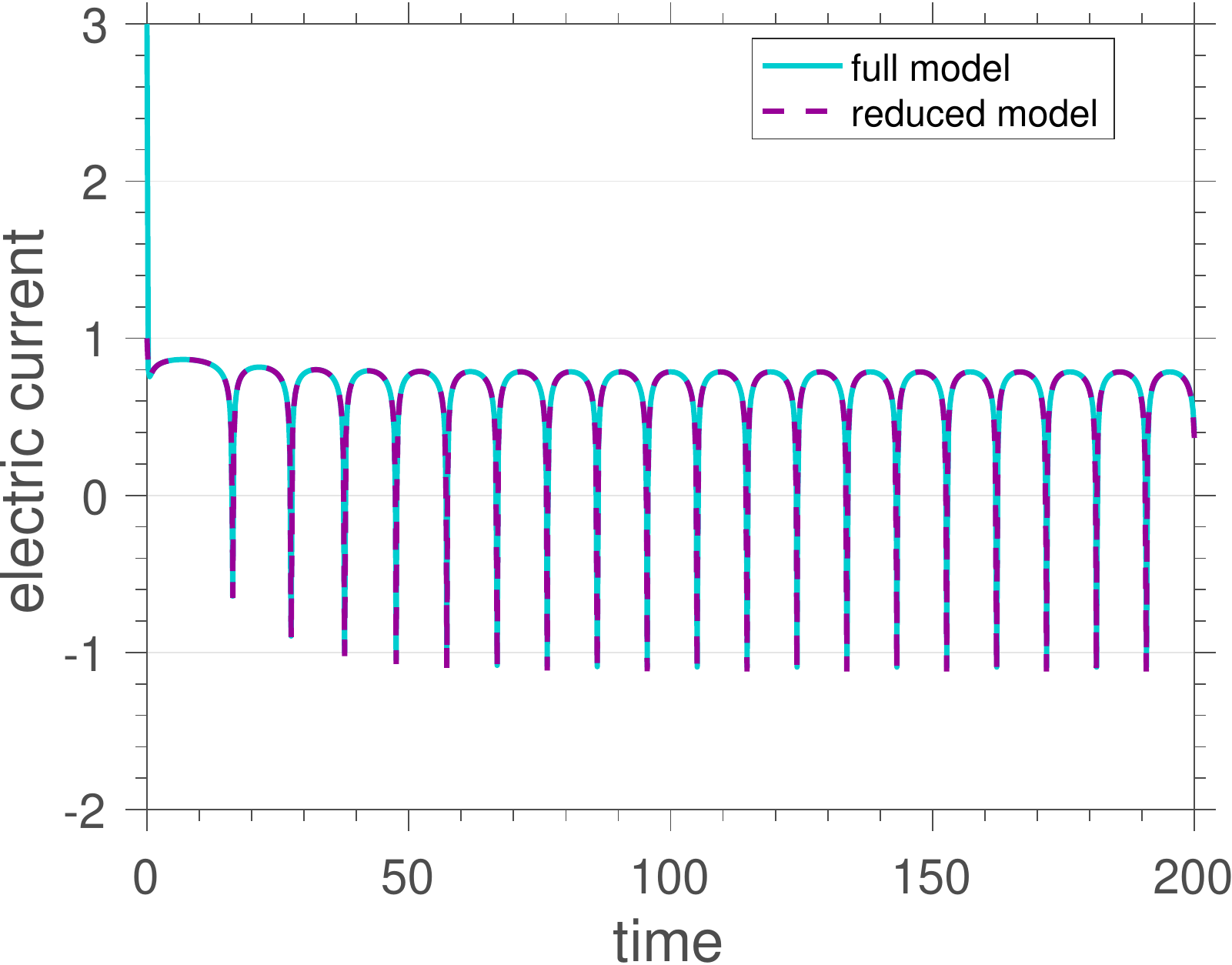}
\includegraphics[scale=0.25]{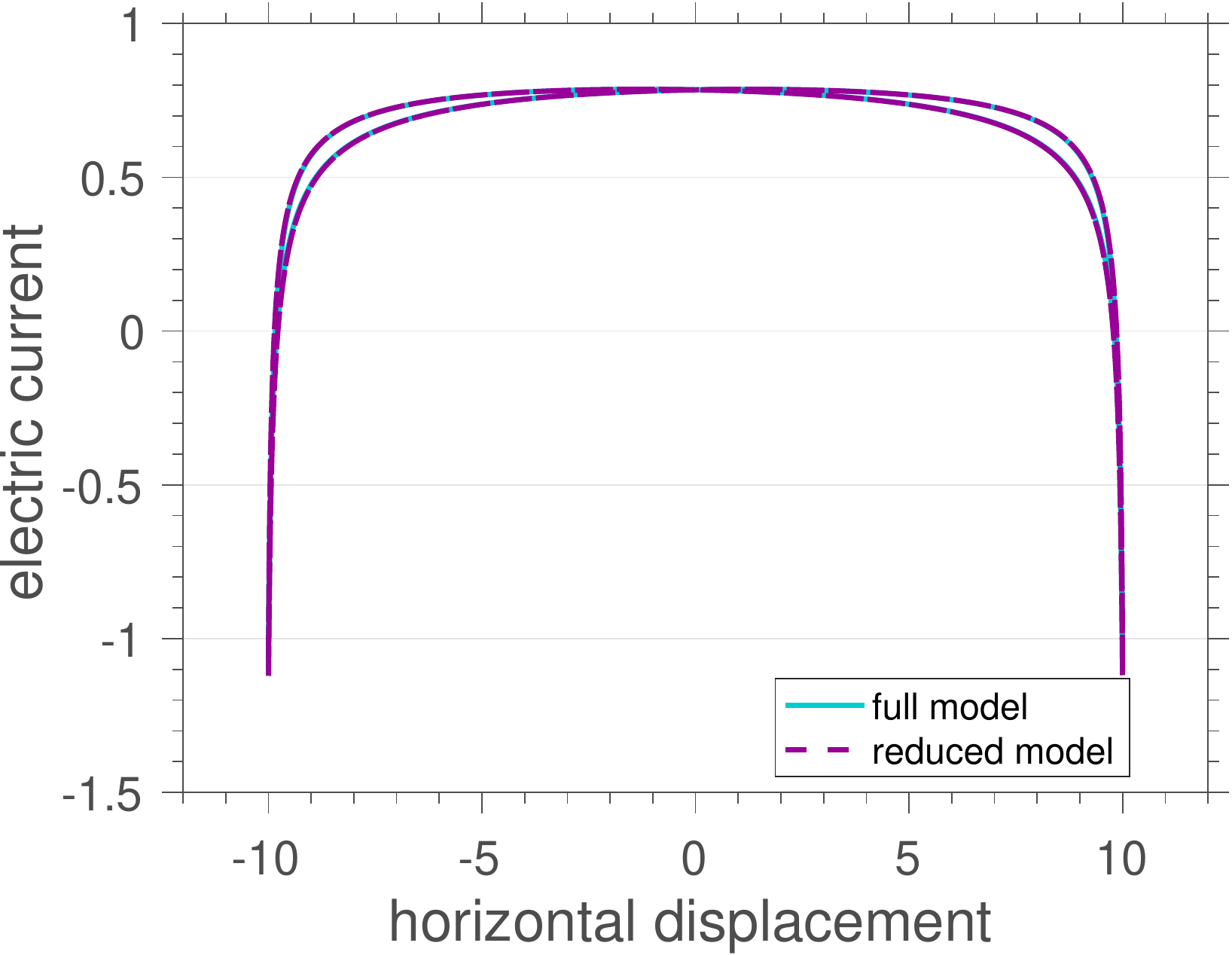}
\includegraphics[scale=0.25]{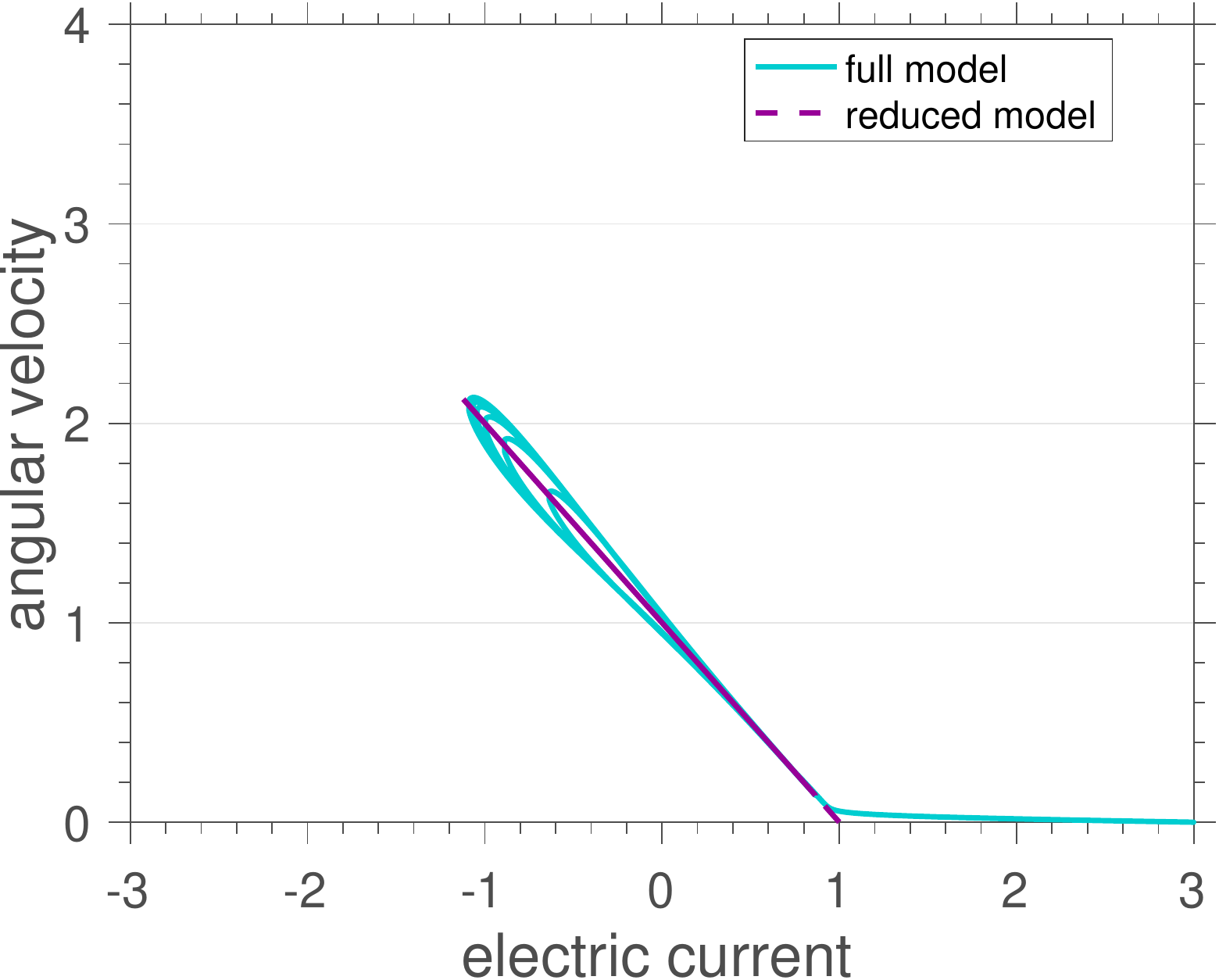}
\includegraphics[scale=0.25]{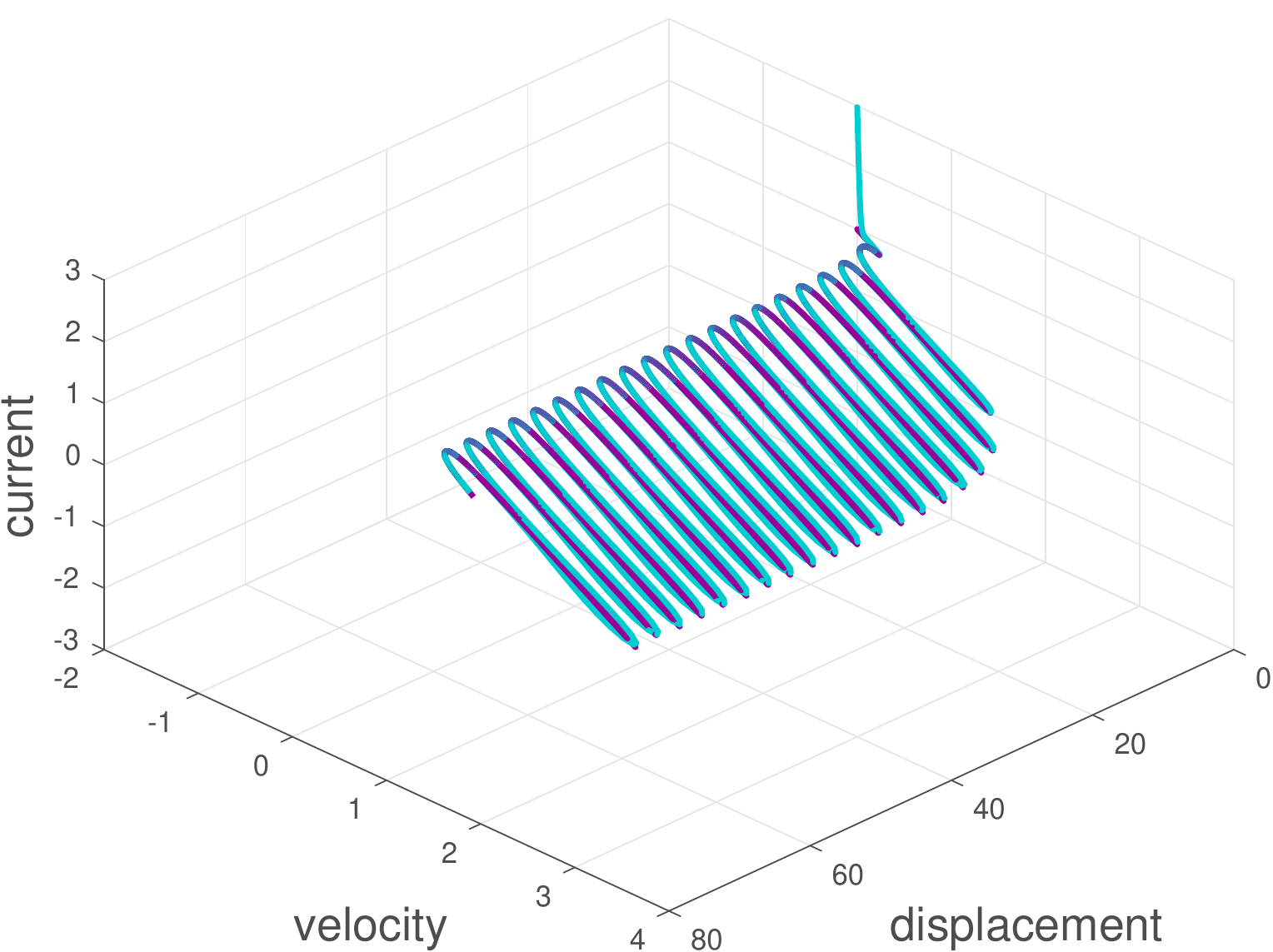}\\
\includegraphics[scale=0.25]{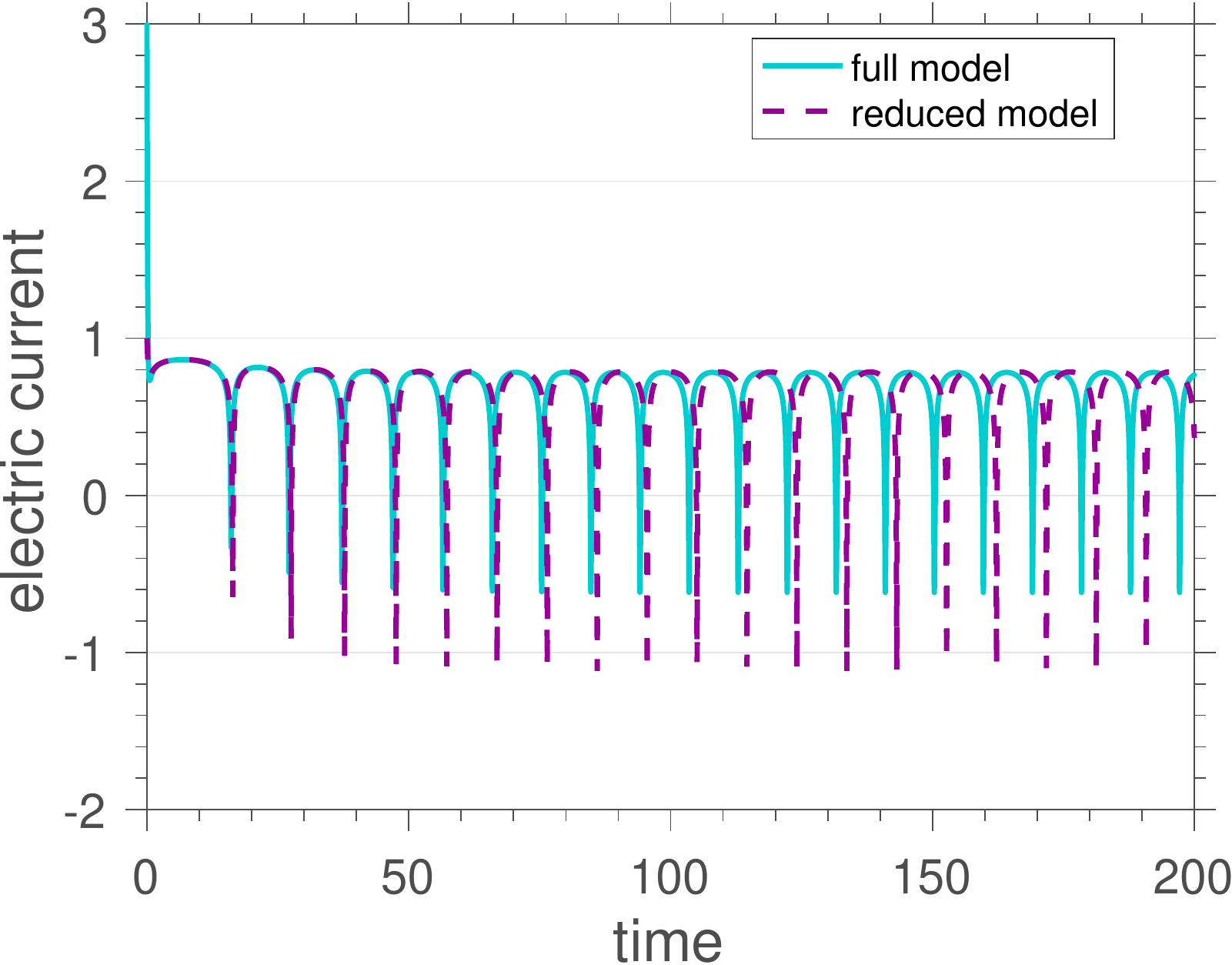}
\includegraphics[scale=0.25]{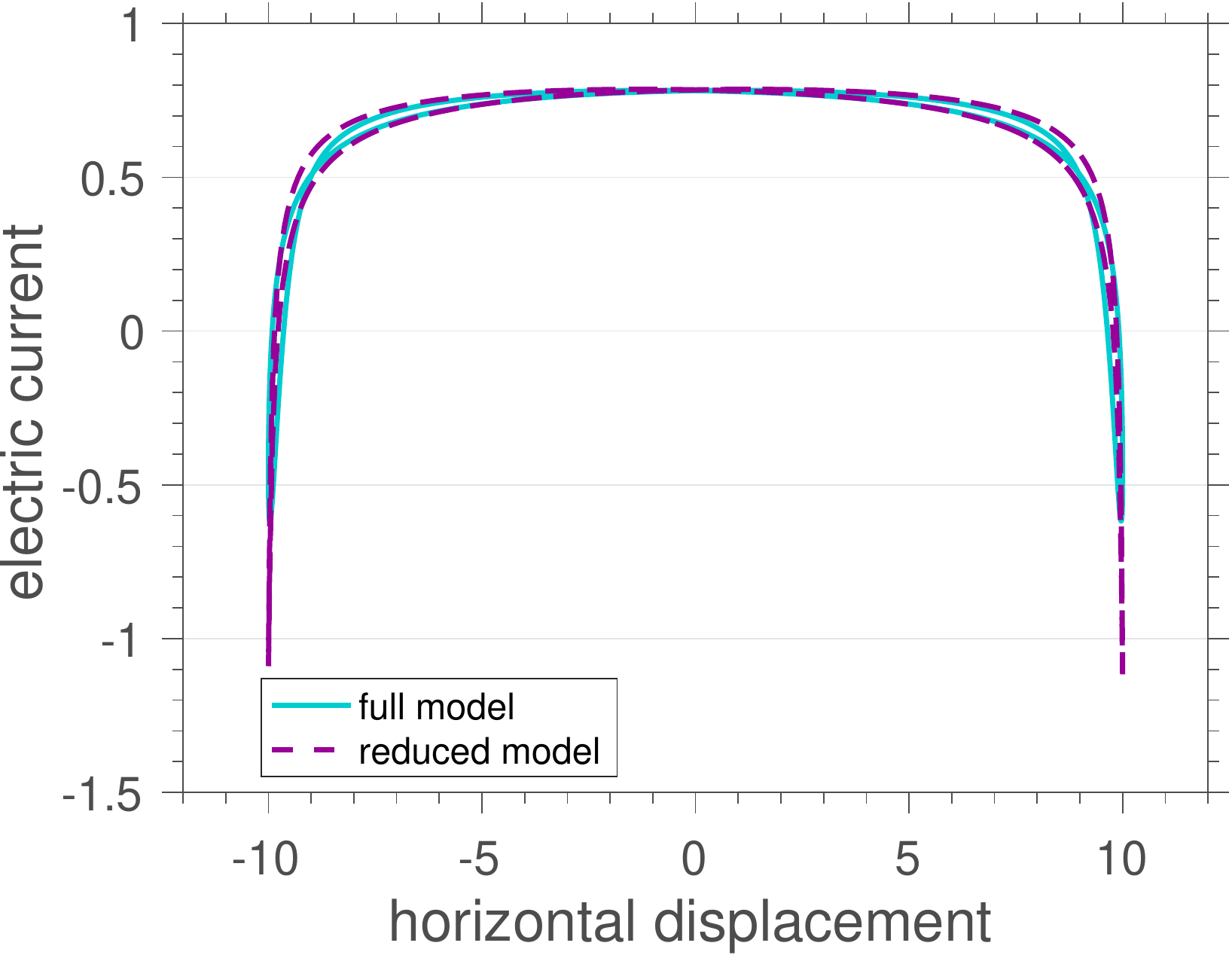}
\includegraphics[scale=0.25]{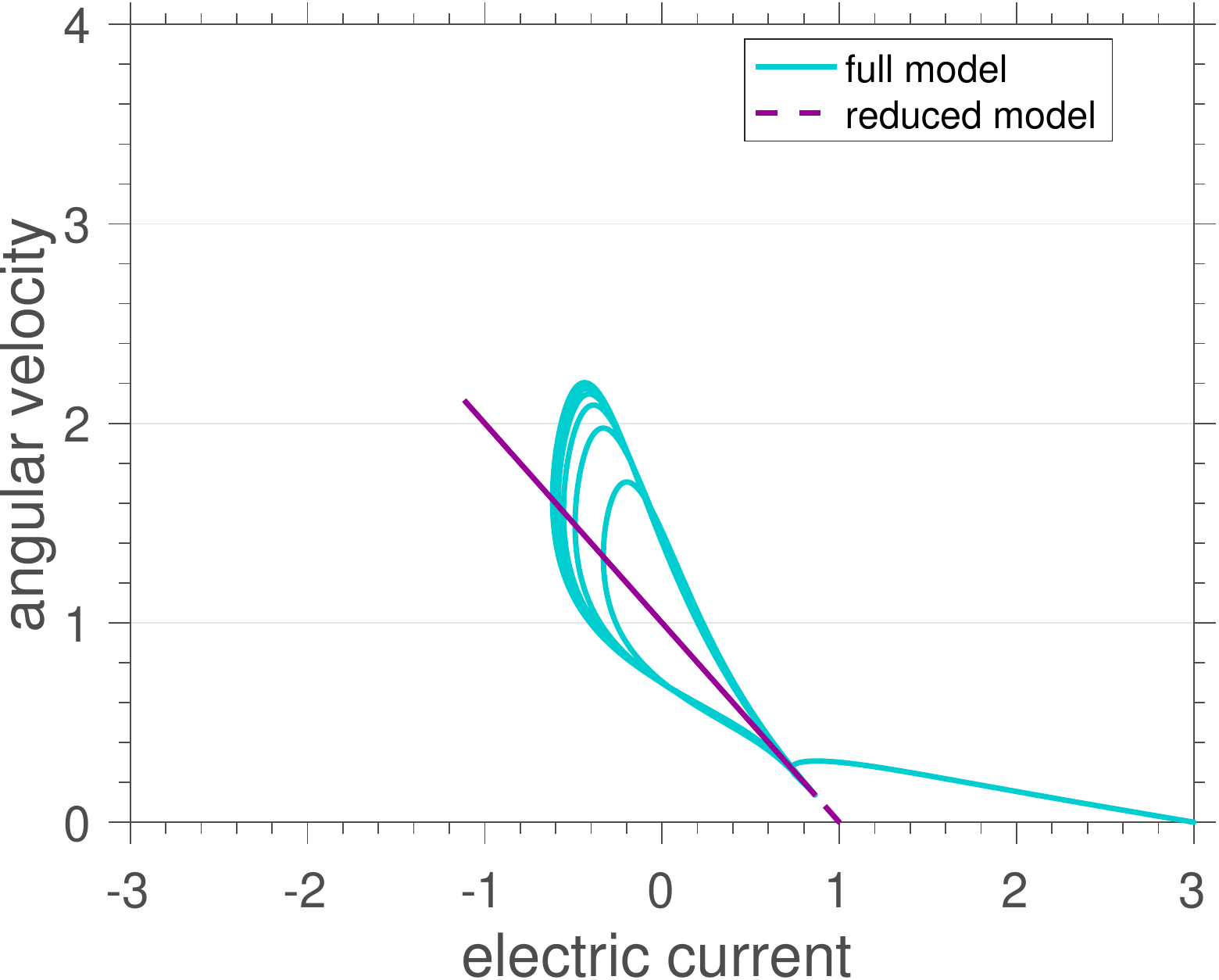}
\includegraphics[scale=0.25]{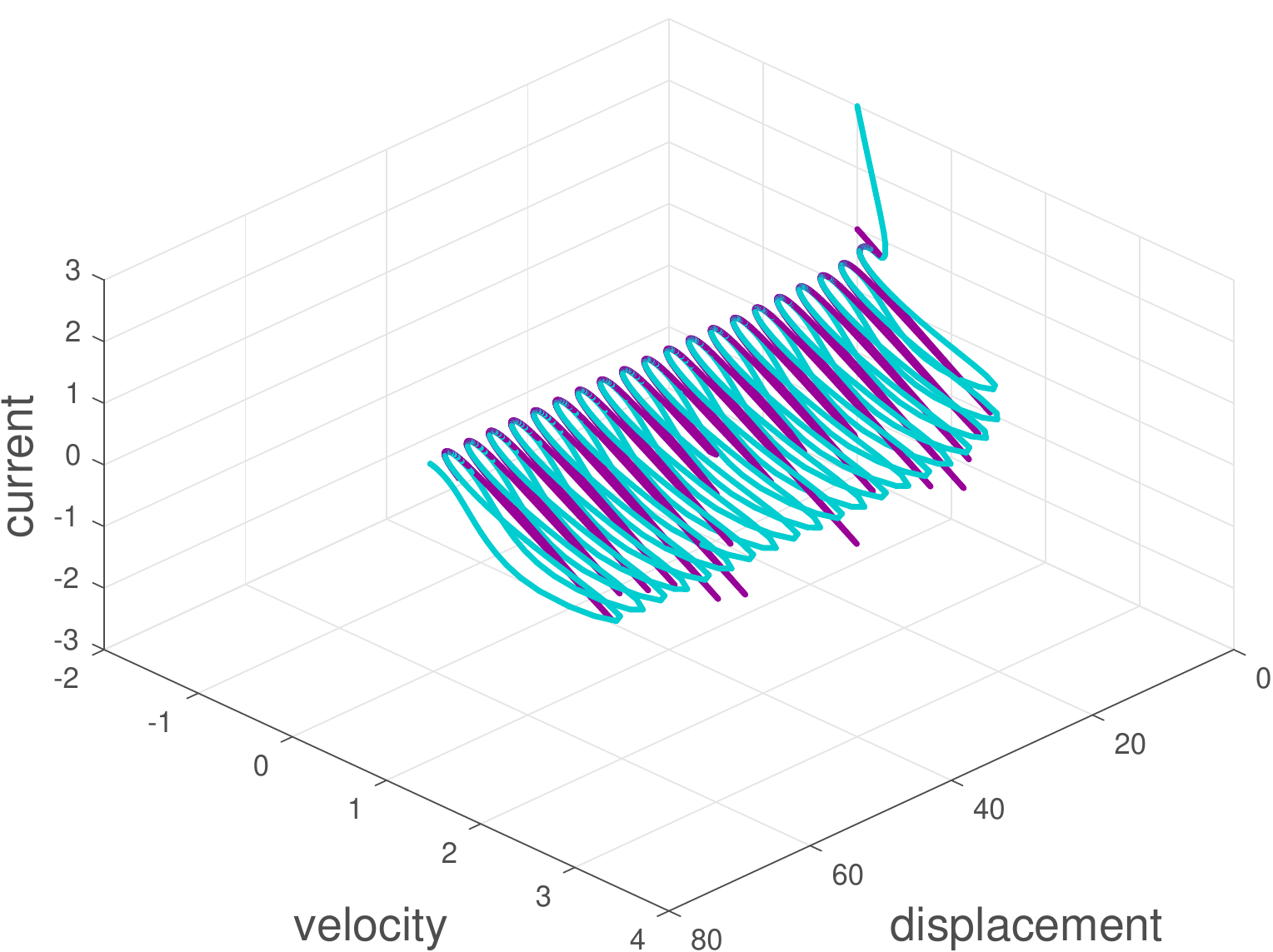}\\
\includegraphics[scale=0.25]{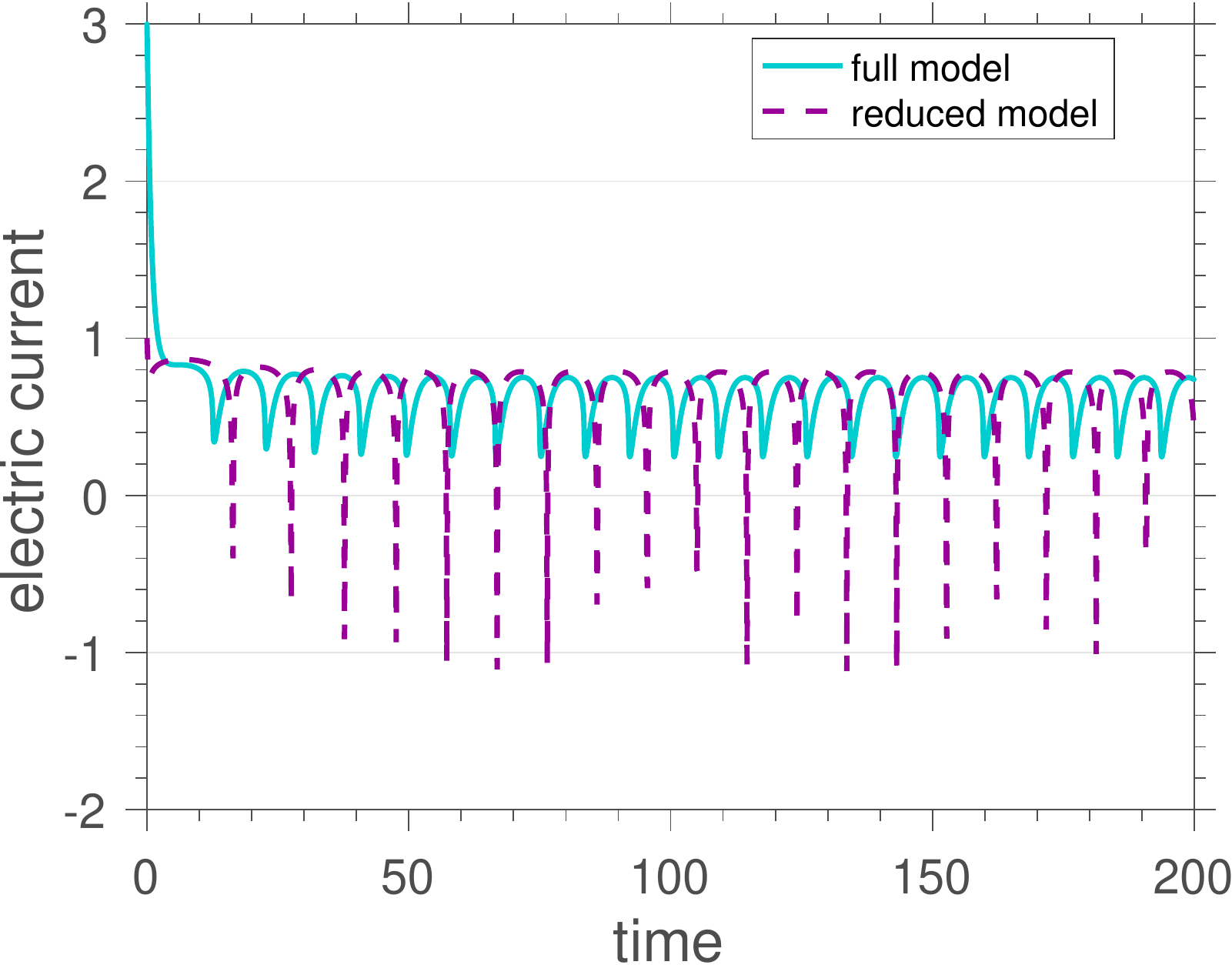}
\includegraphics[scale=0.25]{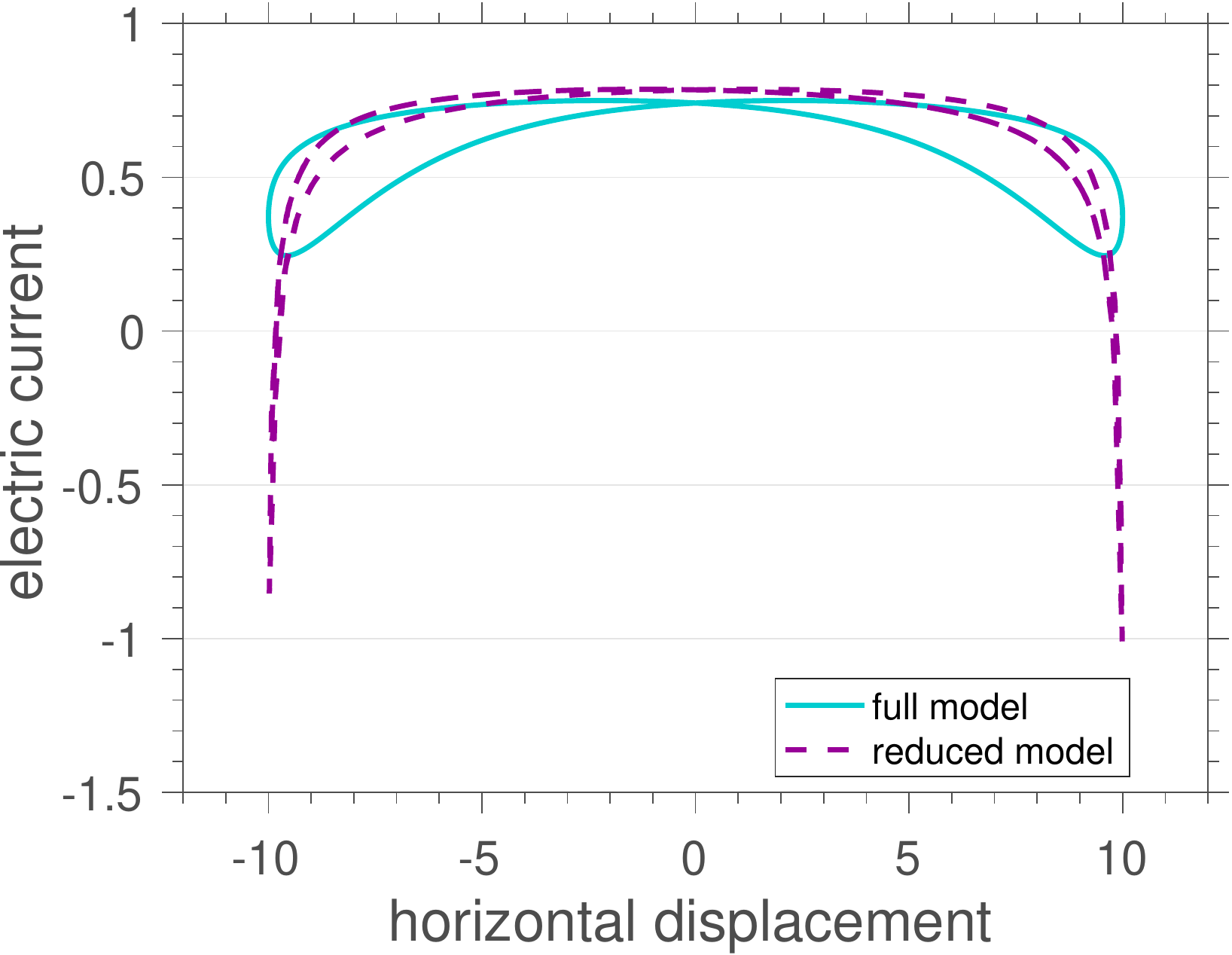}
\includegraphics[scale=0.25]{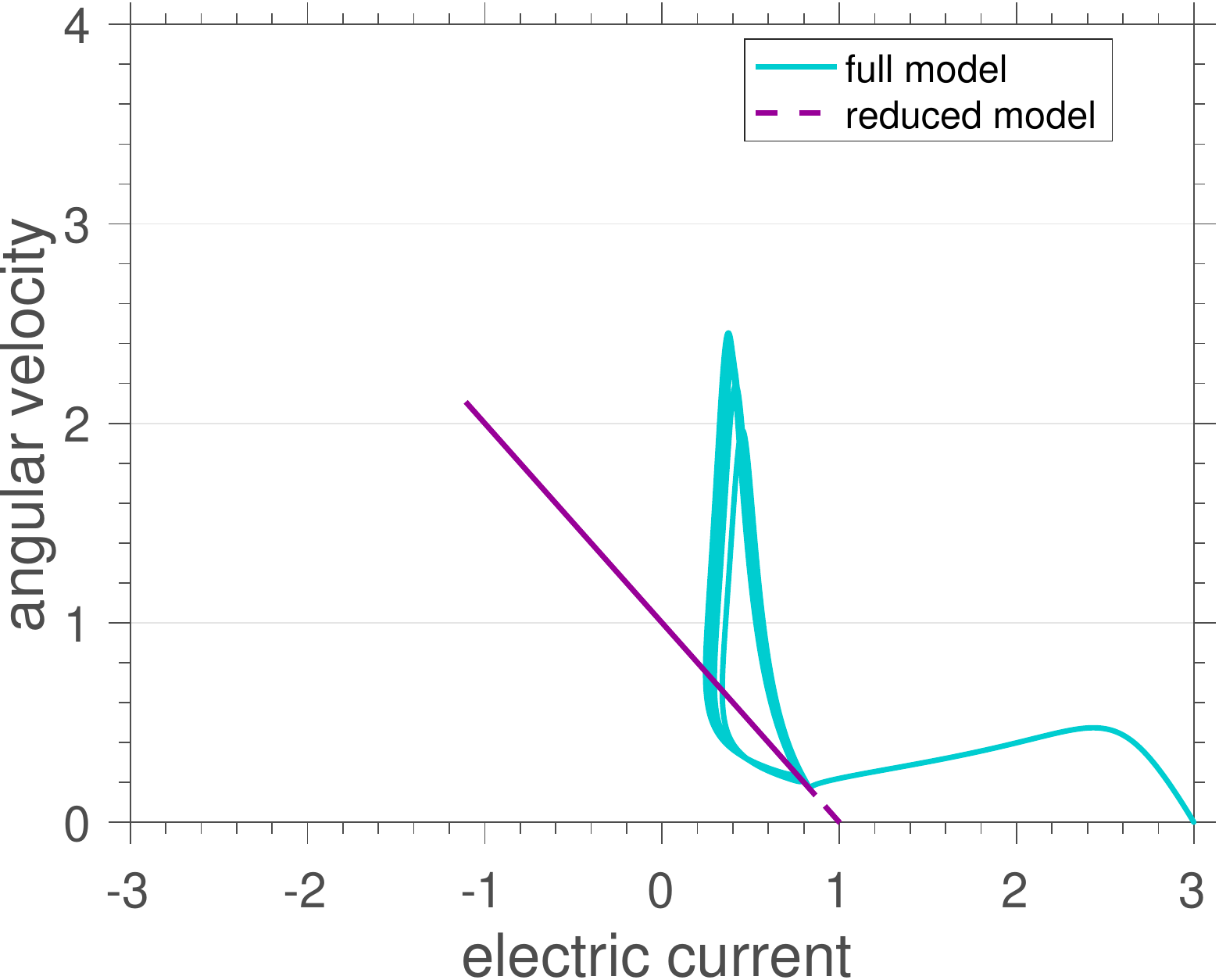}
\includegraphics[scale=0.25]{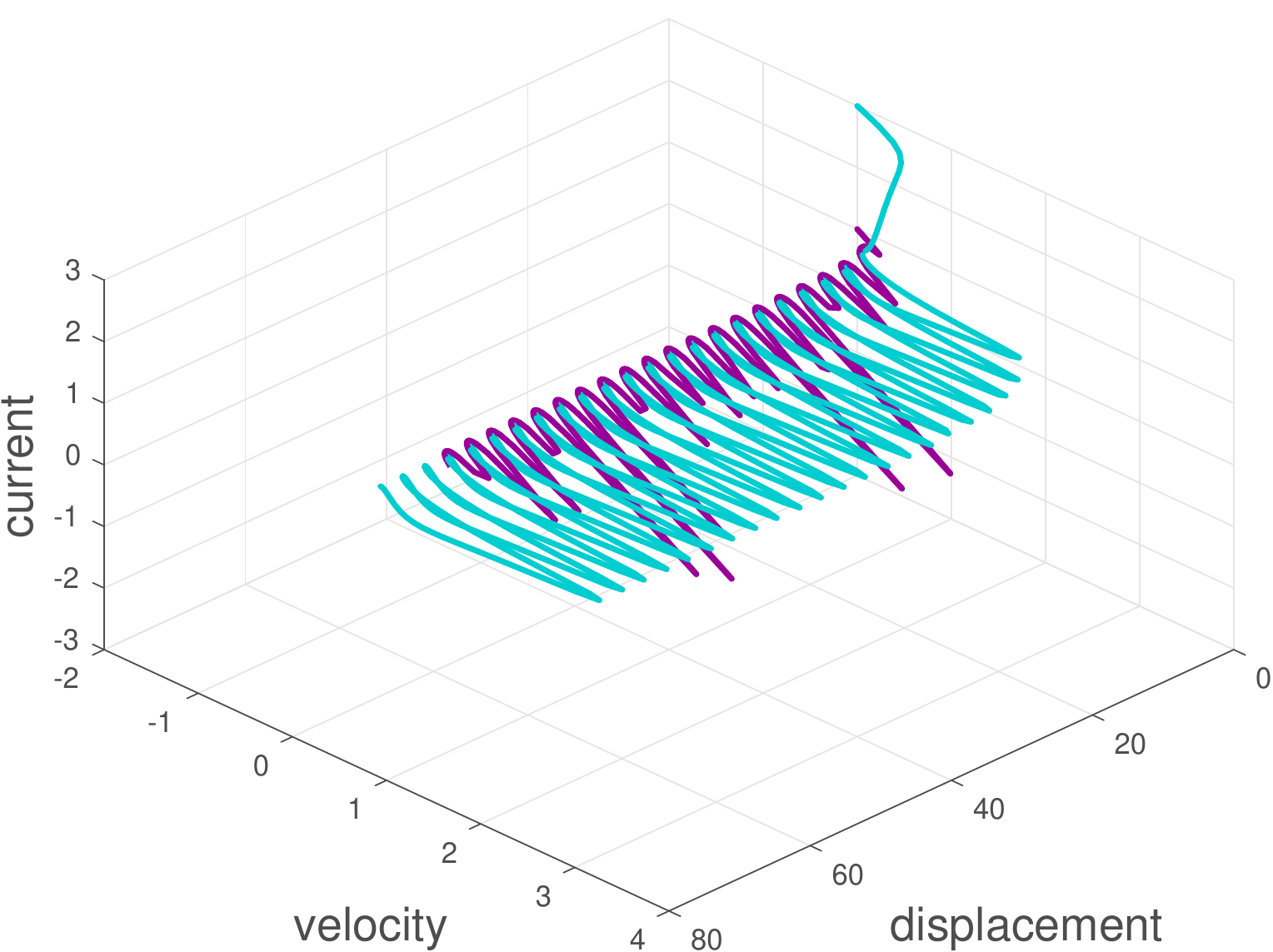}\\
\includegraphics[scale=0.25]{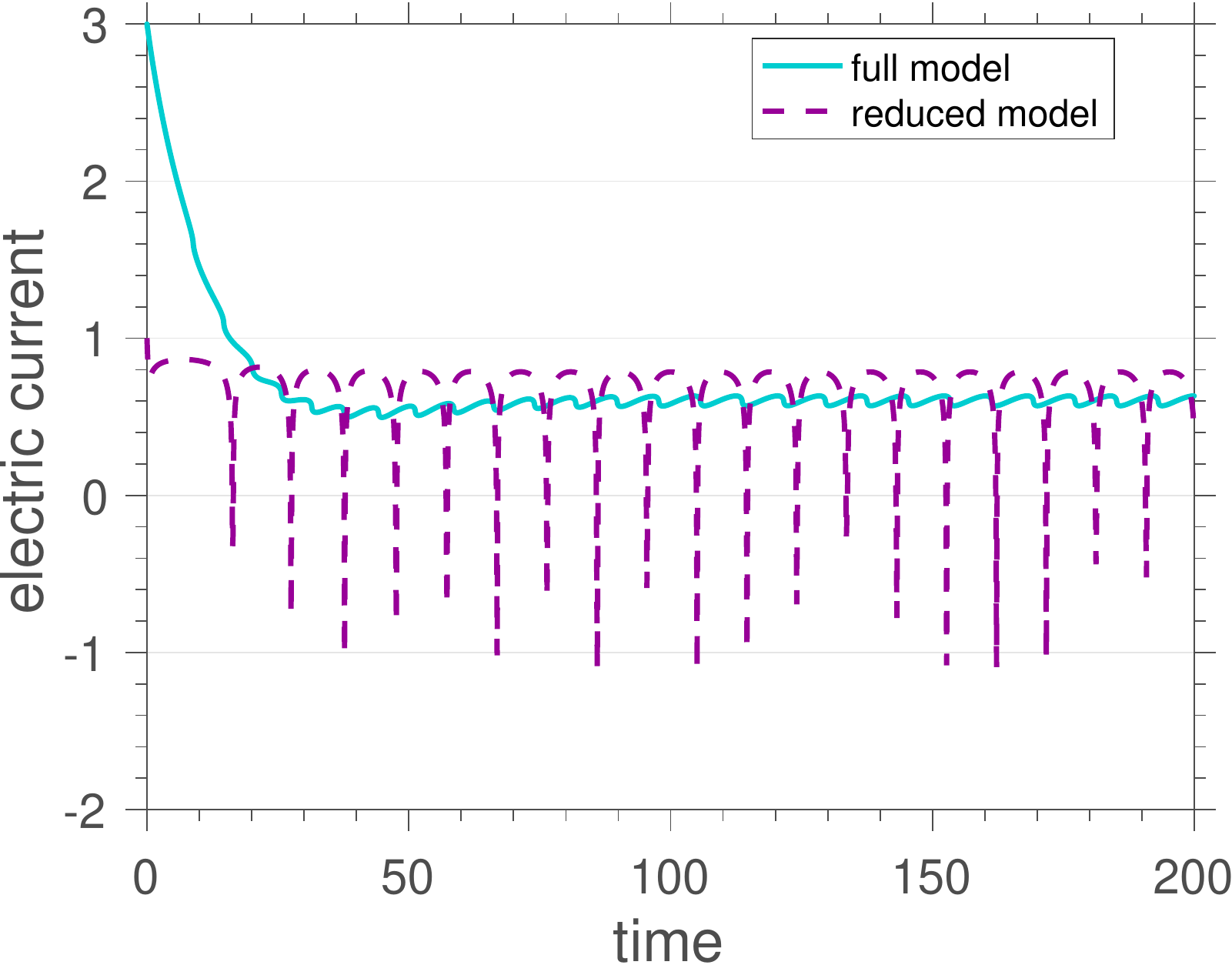}
\includegraphics[scale=0.25]{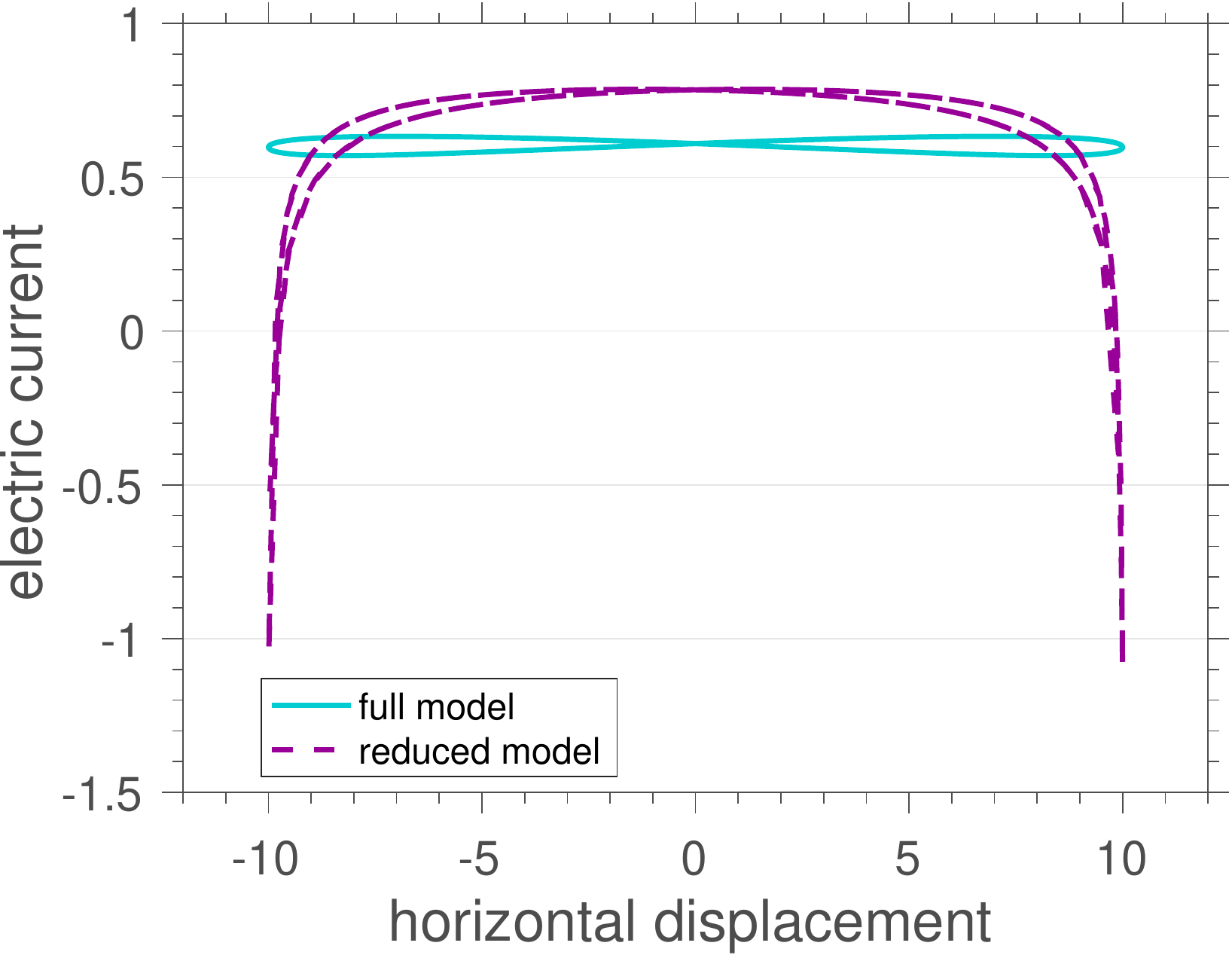}
\includegraphics[scale=0.25]{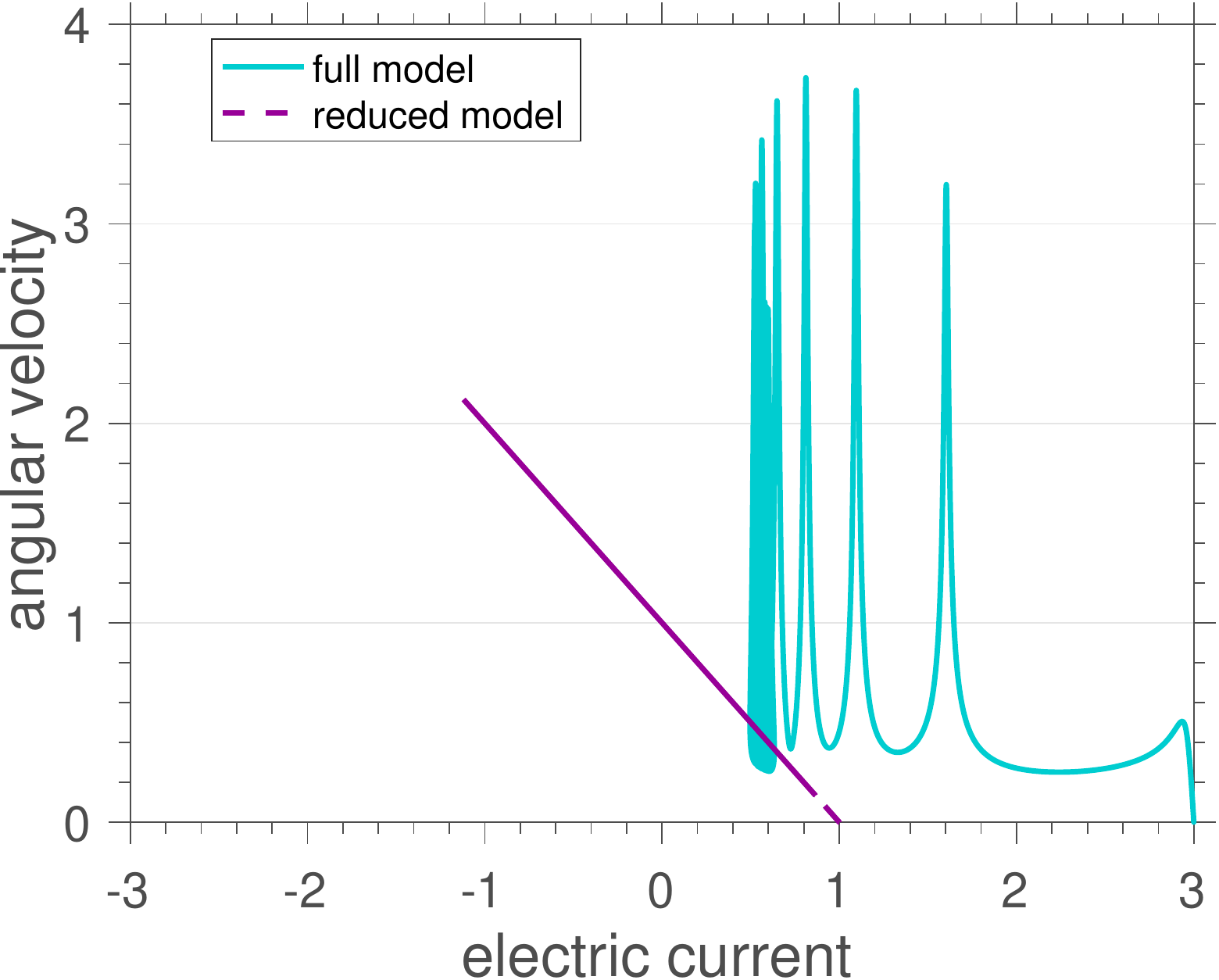}
\includegraphics[scale=0.25]{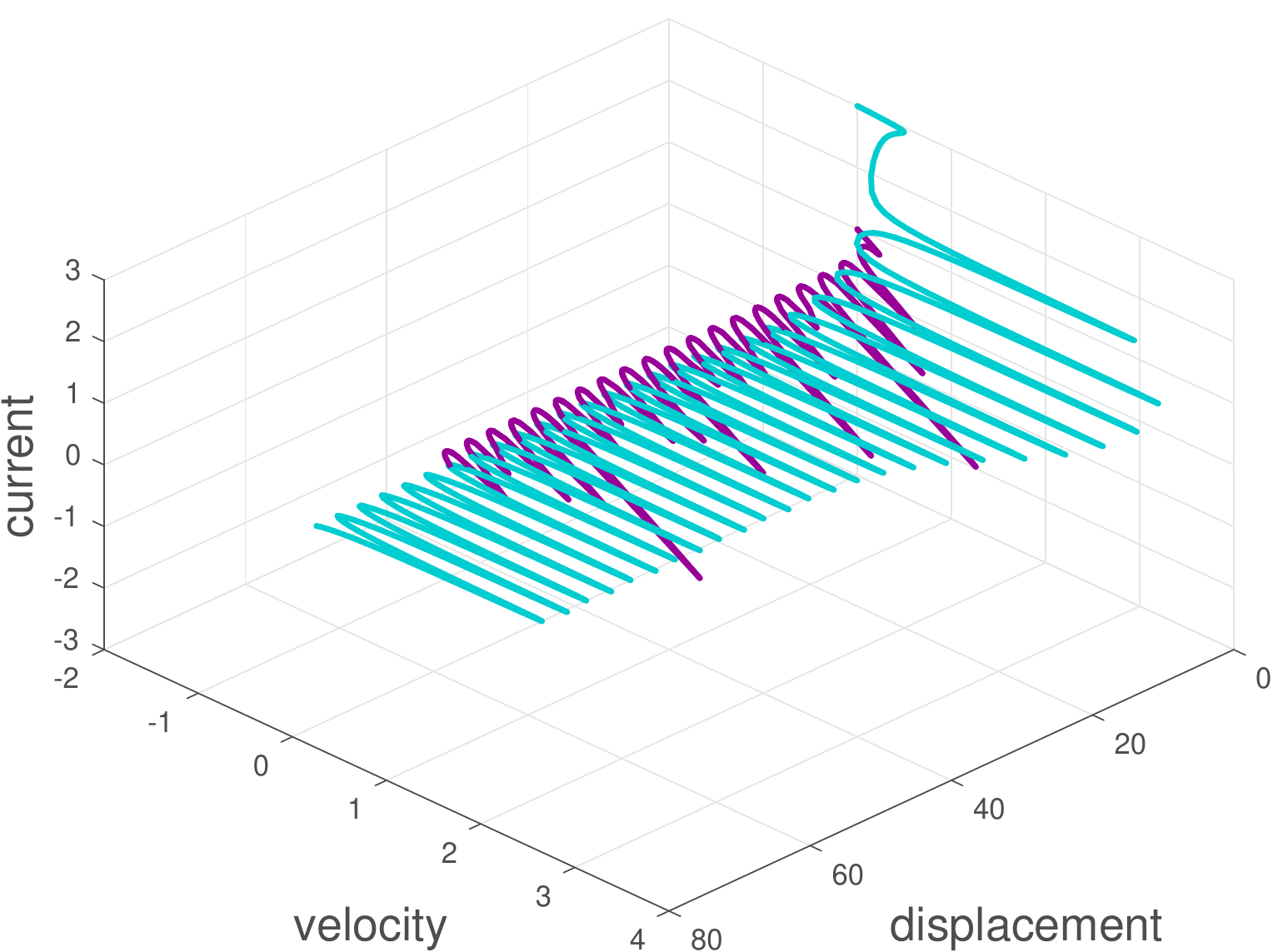}
\caption{Comparison of the evolution of the two models as a function of $\ell$ value: $\ell=0.01$ (first line); $\ell=0.1$ (second line); $\ell=1$ (third line); $\ell=10$ (fourth line). First column: $\dot{q}$ time series; Second column: projection in $\dot{q} \times x$ plane; Third column: projection in $\dot{\theta} \times \dot{q}$ plane; Fourth column: phase-space trajectory. Dimensionless parameters: $b=1$, $\nu=1$, $d=10$, $(\theta_0, \dot{\theta}_0, \dot{q}_0) = (0,0,3 \, \nu)$.}
\label{ell_effect_fig}
\end{figure*}

To access the validity limit of the quasi-steady-state approximation in the electromechanical system, Figure~\ref{ell_effect_fig} presents the comparison of several observables for the two dynamics, varying the value of the dimensionless inductance value $\ell \in \lbrace 0.01, \, 0.1, \, 1, \, 10 \rbrace$ across the lines. These dynamic observations correspond to: $\dot{q}$ time series (first column); projection in $\dot{q} \times x$ plane (second column); projection in $\dot{\theta} \times \dot{q}$ plane (third column); phase-space trajectory (fourth column). Animations of these simulations are available in Supplementary Material 03 - 06.

Note that for small (dimensionless) inductance values ($\ell = 0.01$ or $\ell = 0.1$), the two dynamics have an excellent correlation so that the reduced dynamic perfectly reproduces the main characteristics of the full dynamics, observing the quantitative limitations punctuated in the two previous sections. In the case where $\ell = 1$, although some qualitative aspects of the full dynamics can still be observed in the reduced one, the quantitative errors are already high enough so that the correlation between the two behaviors is not so high. In the limit for enormous inductance values ($\ell = 10$ in this example), one has a quasi-stationary behavior in full-order dynamics that is so out of phase with its reduced-order counterpart that, even from a qualitative perspective, the approximation is not very informative. 

This degeneration of the approximation is evident when looking at the projection in $\dot{\theta} \times \dot{q}$ plane (third column of Figure~\ref{ell_effect_fig}), remembering that it is not only the similarity in shape of the attractors that establishes a good correlation but how these trajectories are traversed. For small $\ell$, the ``airfoil'' and the plane (corresponding to the attractors of the two dynamics) are very similar in shape, which acts in favor of good correlation. However, as $\ell$ grows, the full-order dynamics ``airfoil'' opens and tilts, becoming more and more discrepant with the reduced-order dynamics affine manifold. In the beginning, this behavior, despite reducing, does not entirely compromise the correlation, but in the limit, for $\ell \gg 1$, it makes the two dynamics quite distinct. 

According to the numerical studies presented here, a reasonable limit to separate the region where the reduced-order dynamics effectively represent the full-order dynamics from that region where the approximation degenerates is around $\ell = 0.1$. However, from the practical point of view, it is important to note that the condition $\ell \gg 0.1$ rarely is seen in a typical electromechanical system since it requires very small values for the rotational inertia $J$ or the electrical resistance $R$; alternatively, large values for the inductance $L$ or the coupling coefficient $G$. For instance, when the rotational inertia is tiny, significant velocity fluctuations may occur, leading to induce fluctuations in the electromagnetic induction, making the inductance effects relevant to the dynamics.

\begin{figure*}
\centering
\includegraphics[scale=0.25]{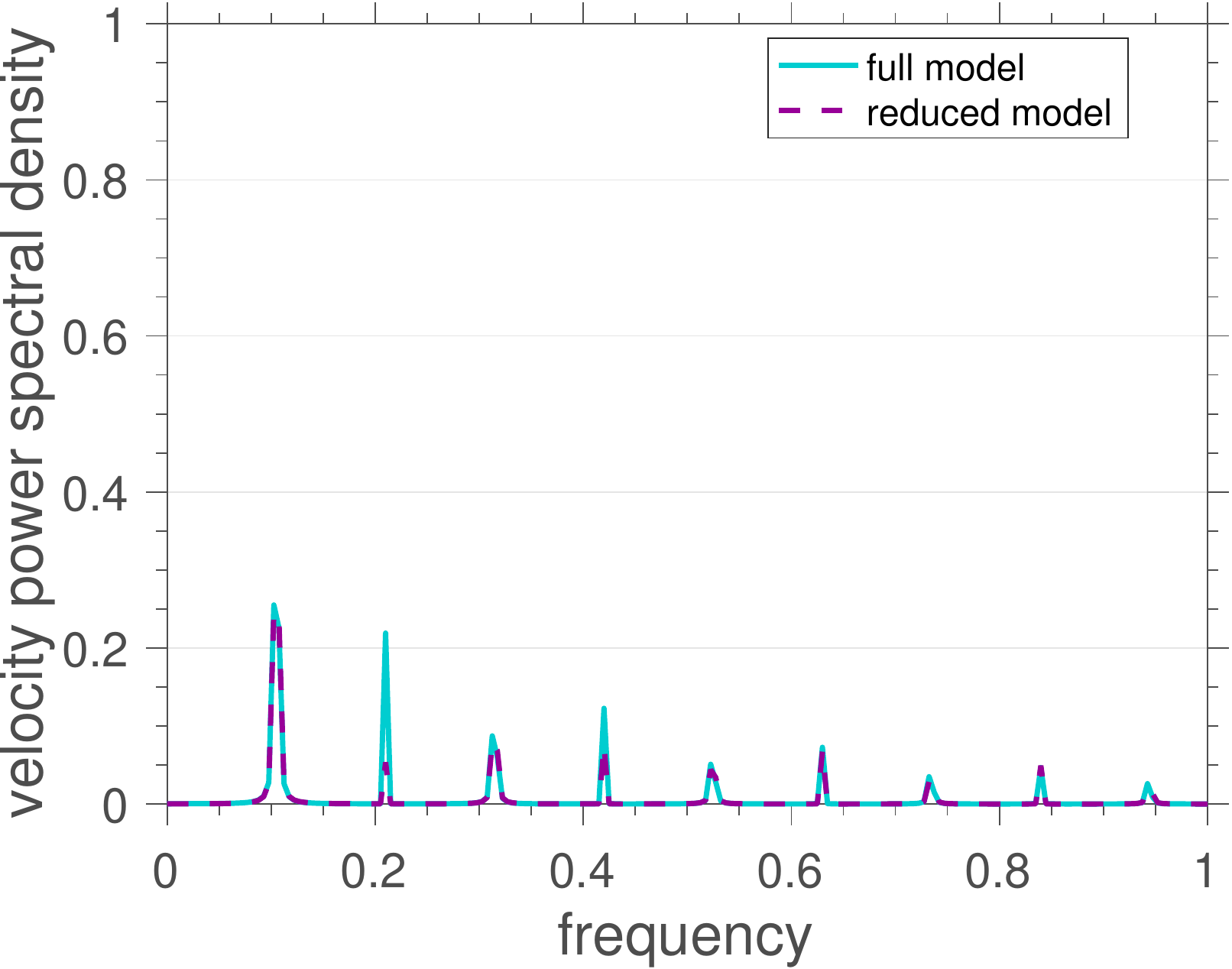}
\includegraphics[scale=0.25]{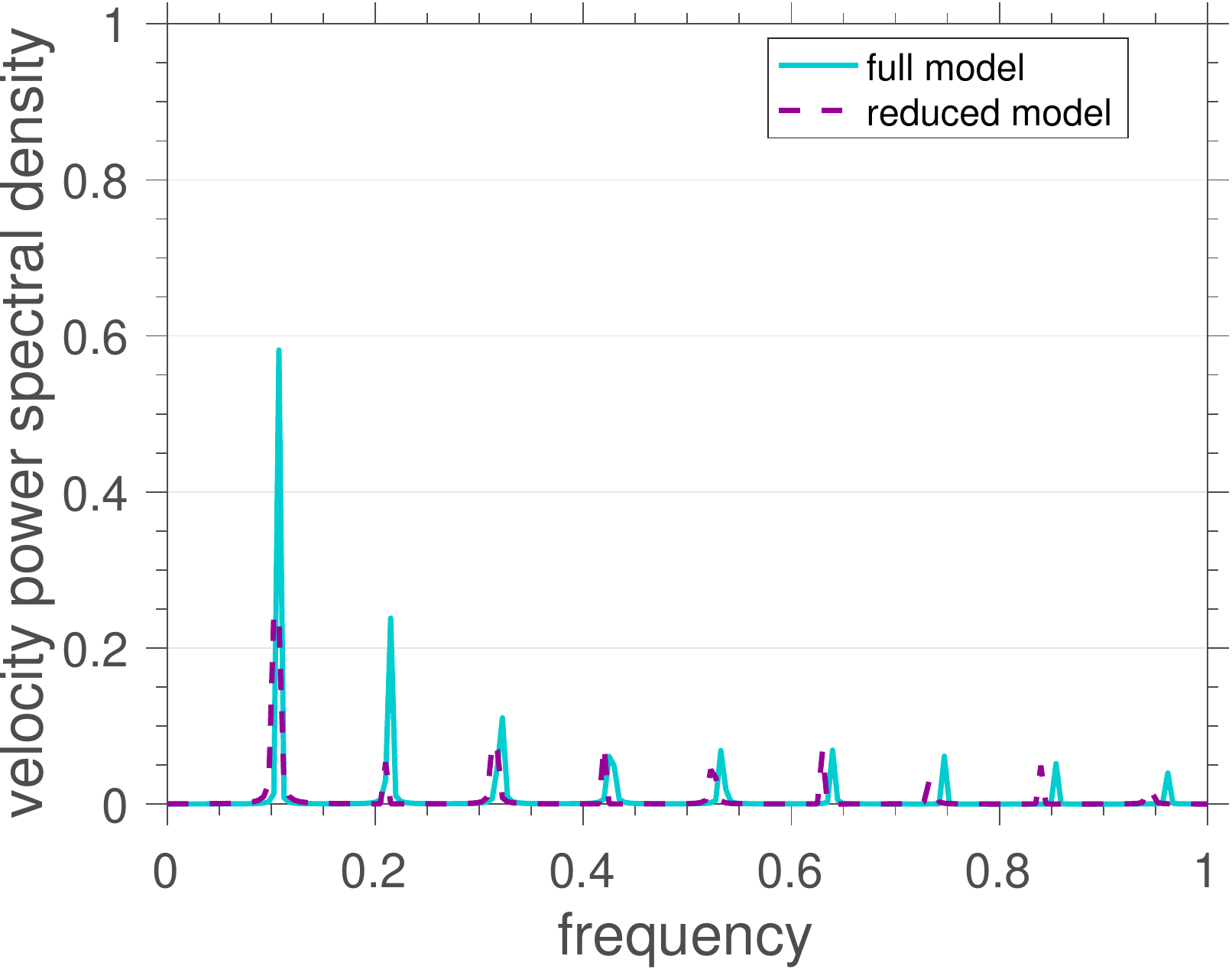}
\includegraphics[scale=0.25]{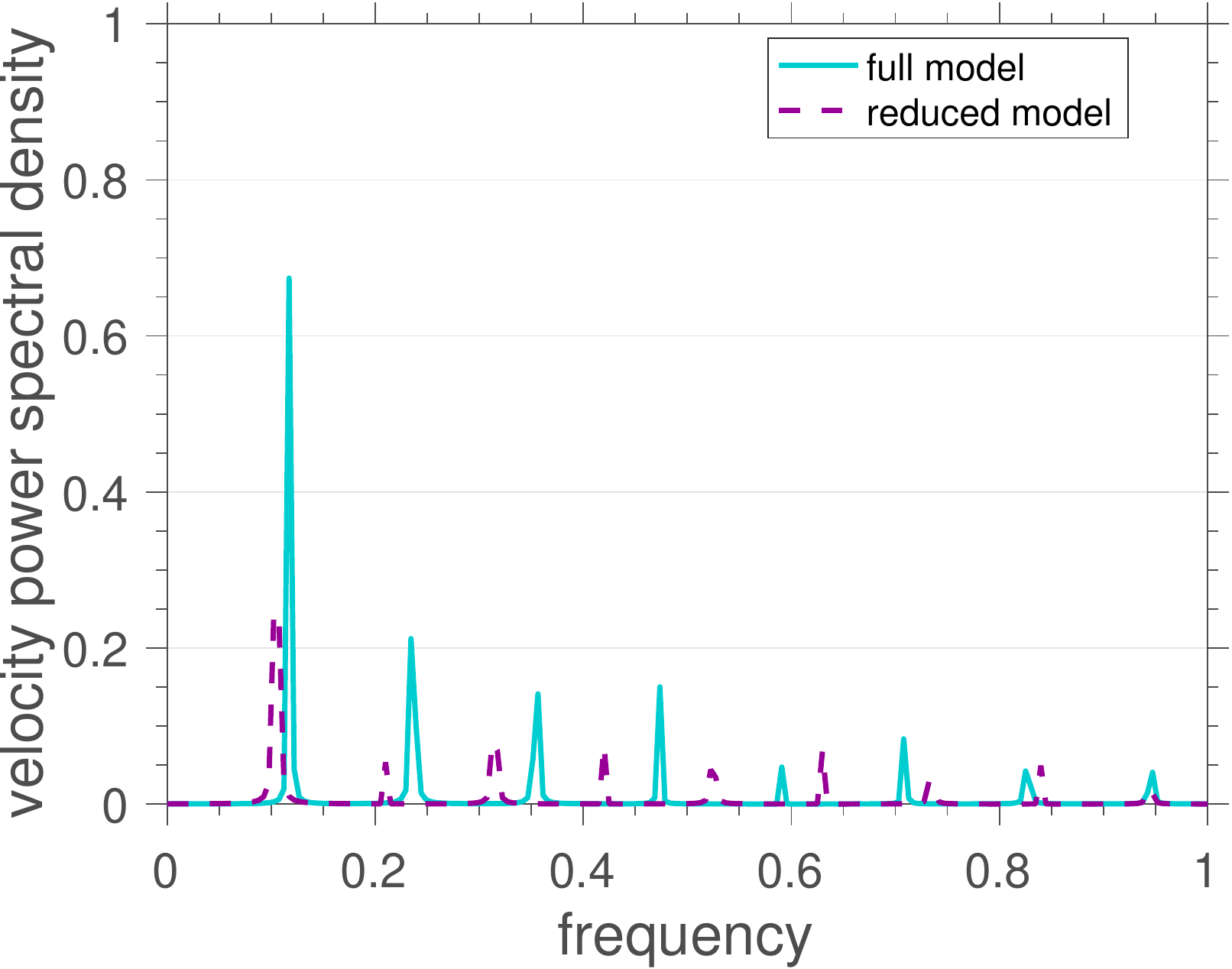}
\includegraphics[scale=0.25]{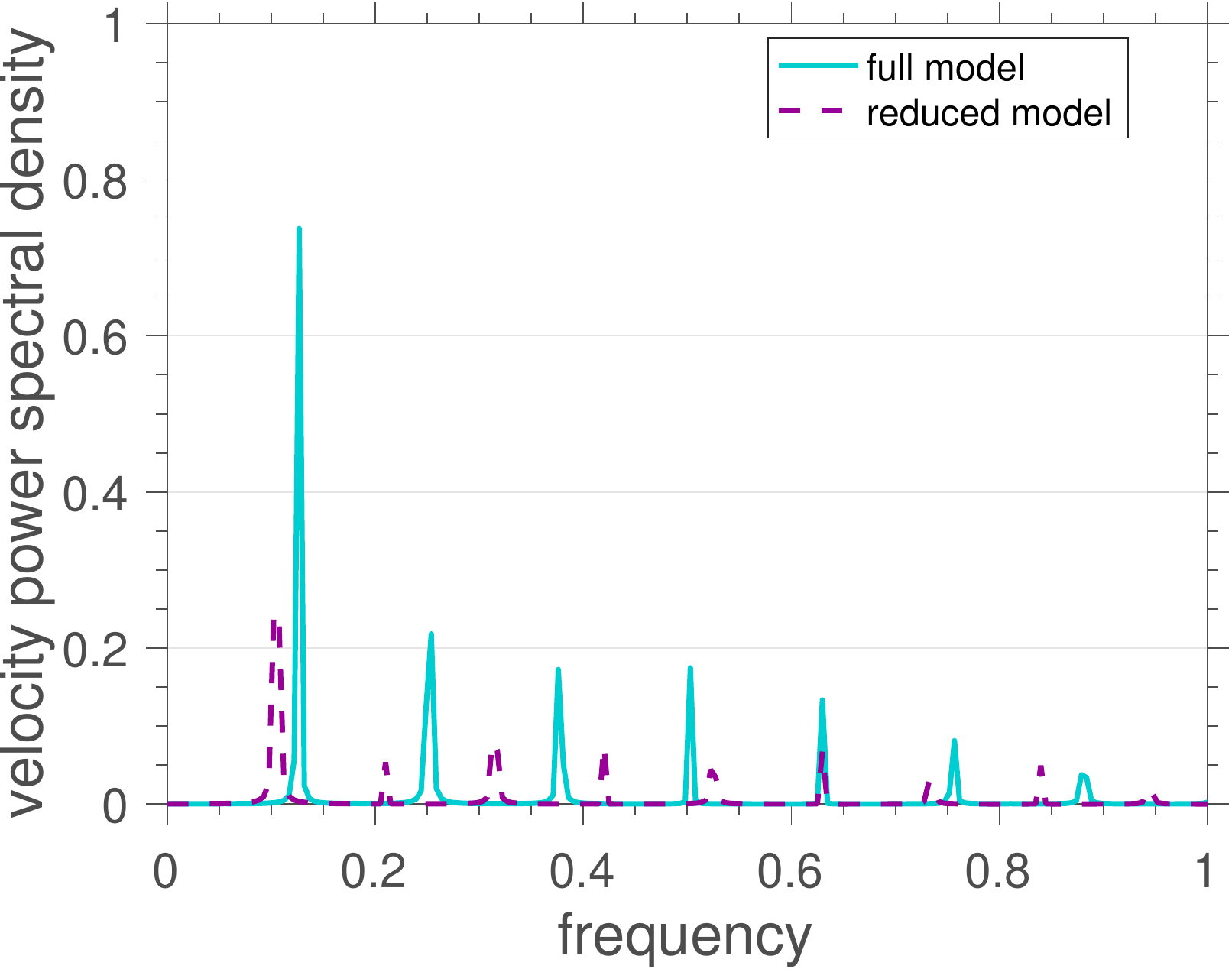}
\caption{Comparison of the angular velocity power spectral density of the two models as a function of $\ell$ value: $\ell=0.01$ (outer left); $\ell=0.1$ (inner left); $\ell=1$ (inner right); $\ell=10$ (outer right). Dimensionless parameters: $b=1$, $\nu=1$, $d=10$, $(\theta_0, \dot{\theta}_0, \dot{q}_0) = (0,0,3 \, \nu)$.}
\label{psd_ell_effect_fig}
\end{figure*}

To better understand this behavior change, it is necessary to look at the system's response in the frequency domain. In this sense, Figure~\ref{psd_ell_effect_fig} shows the power spectral density function of the angular velocity time series for several values of the dimensionless inductance $\ell$. It can be noted that as $\ell \to \infty$ the frequency value that corresponds to the first peak of the full-order model moves to the right of its counterpart in the reduced model, which incurs a reduced-order dynamic that is delayed to the original dynamics. This fact was well observed by Lima et al. \cite{lima2018_1,Lima2019p552}, which also shows that this effect occurs when increasing the value of the parameters $\nu$ or $d$.

\begin{figure*}
\centering
\includegraphics[scale=0.35]{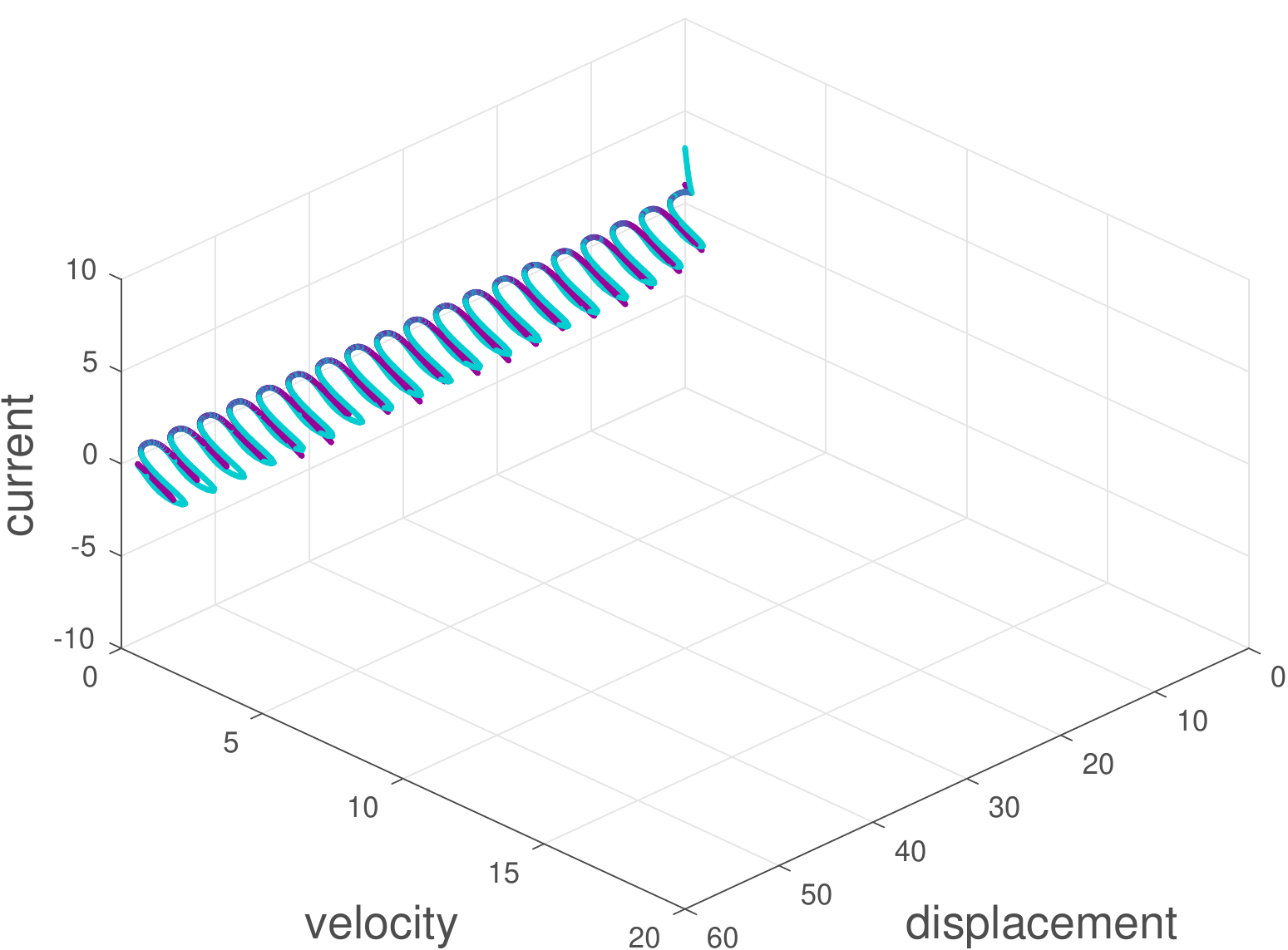}
\includegraphics[scale=0.35]{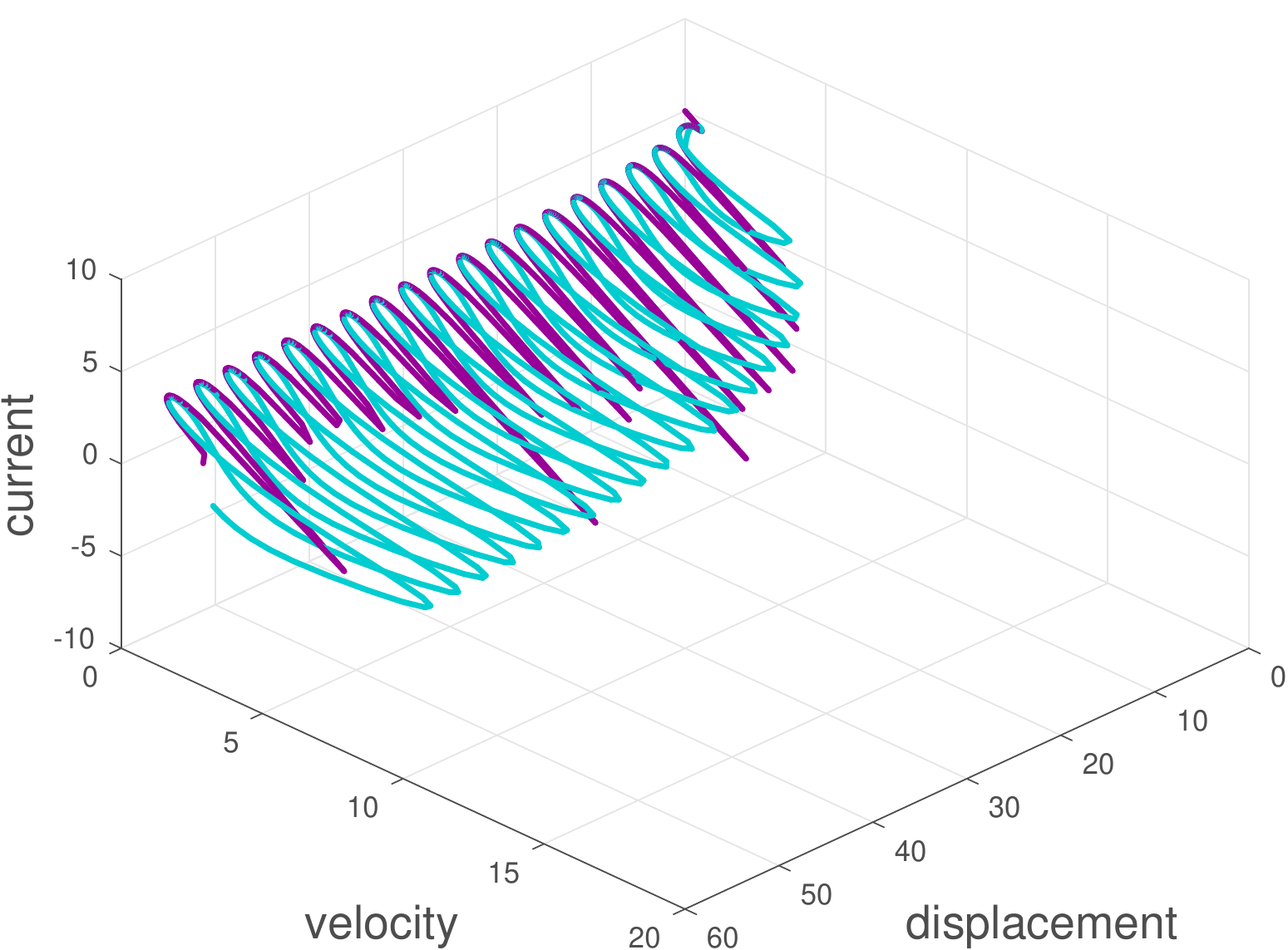}
\includegraphics[scale=0.35]{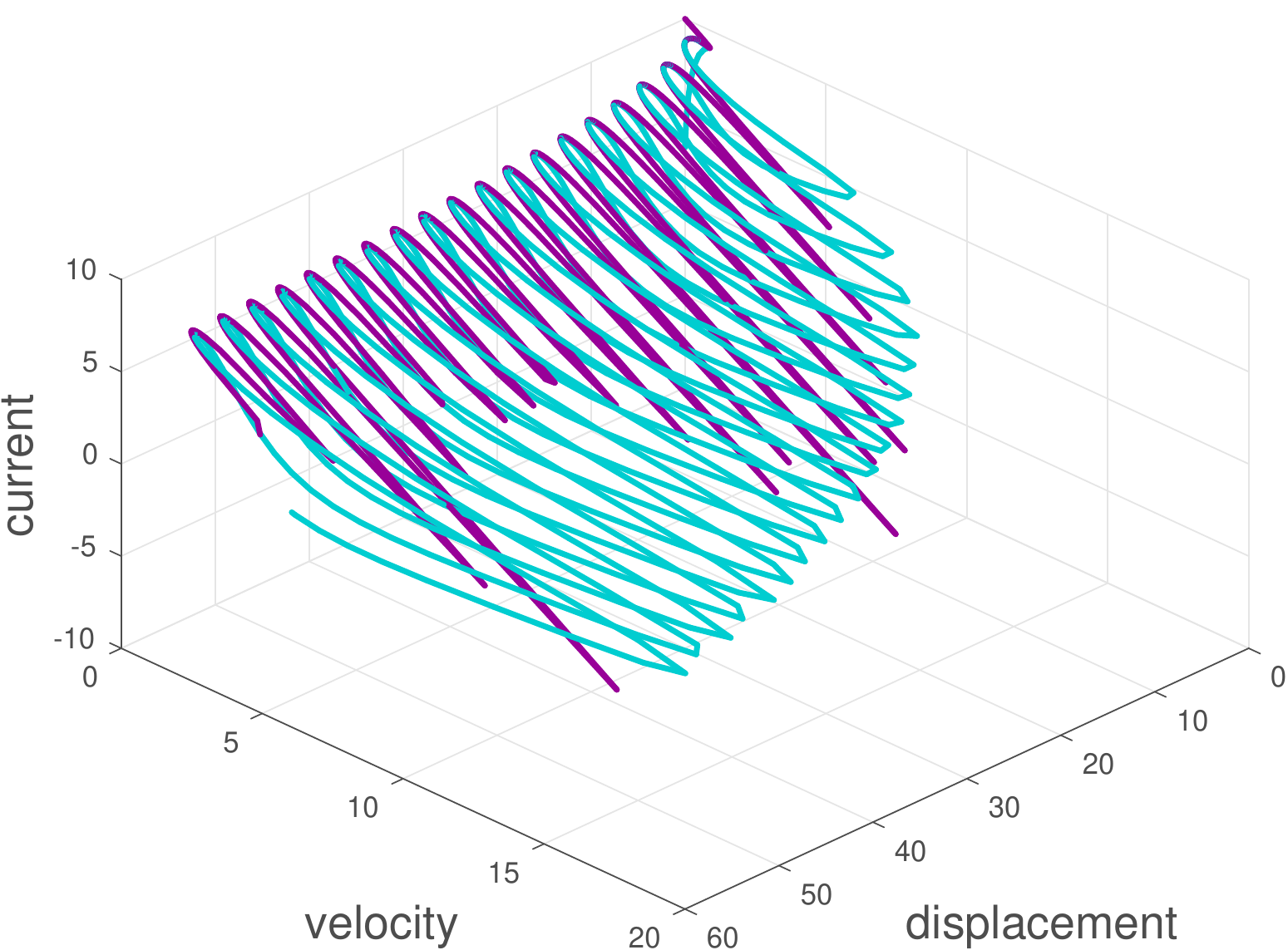}
\caption{Comparison between the trajectories of the two models in phase-space as a function of $\nu$ value: $\nu=1$ (first column); $\nu=5$ (second column); $\nu=10$ (third column). Dimensionless parameters: $\ell=0.05$, $b=1$, $d=10$, $(\theta_0, \dot{\theta}_0, \dot{q}_0) = (0,0,3 \, \nu)$.}
\label{nu_effect_fig}
\end{figure*}

\begin{figure*}
\centering
\includegraphics[scale=0.35]{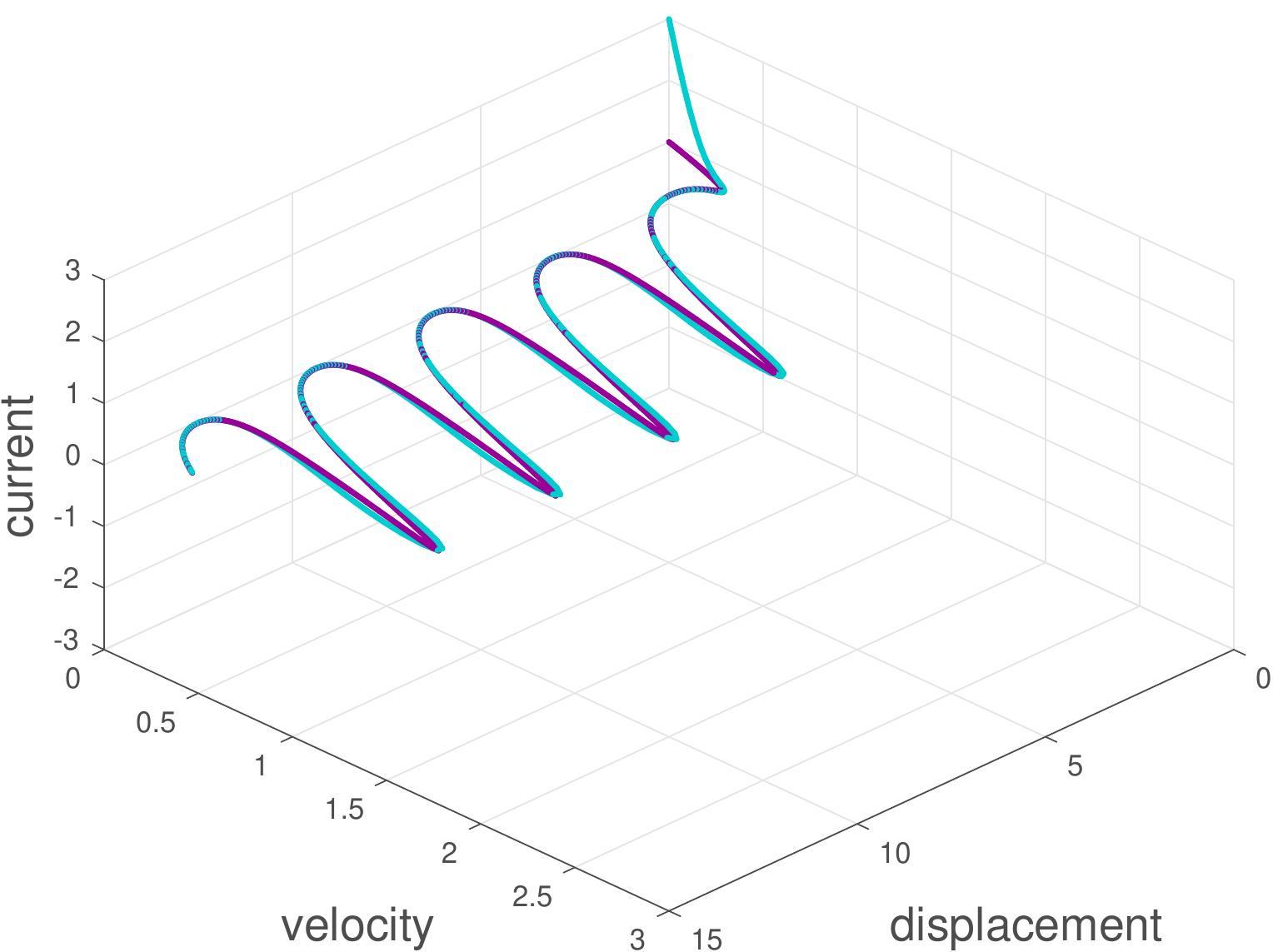}
\includegraphics[scale=0.35]{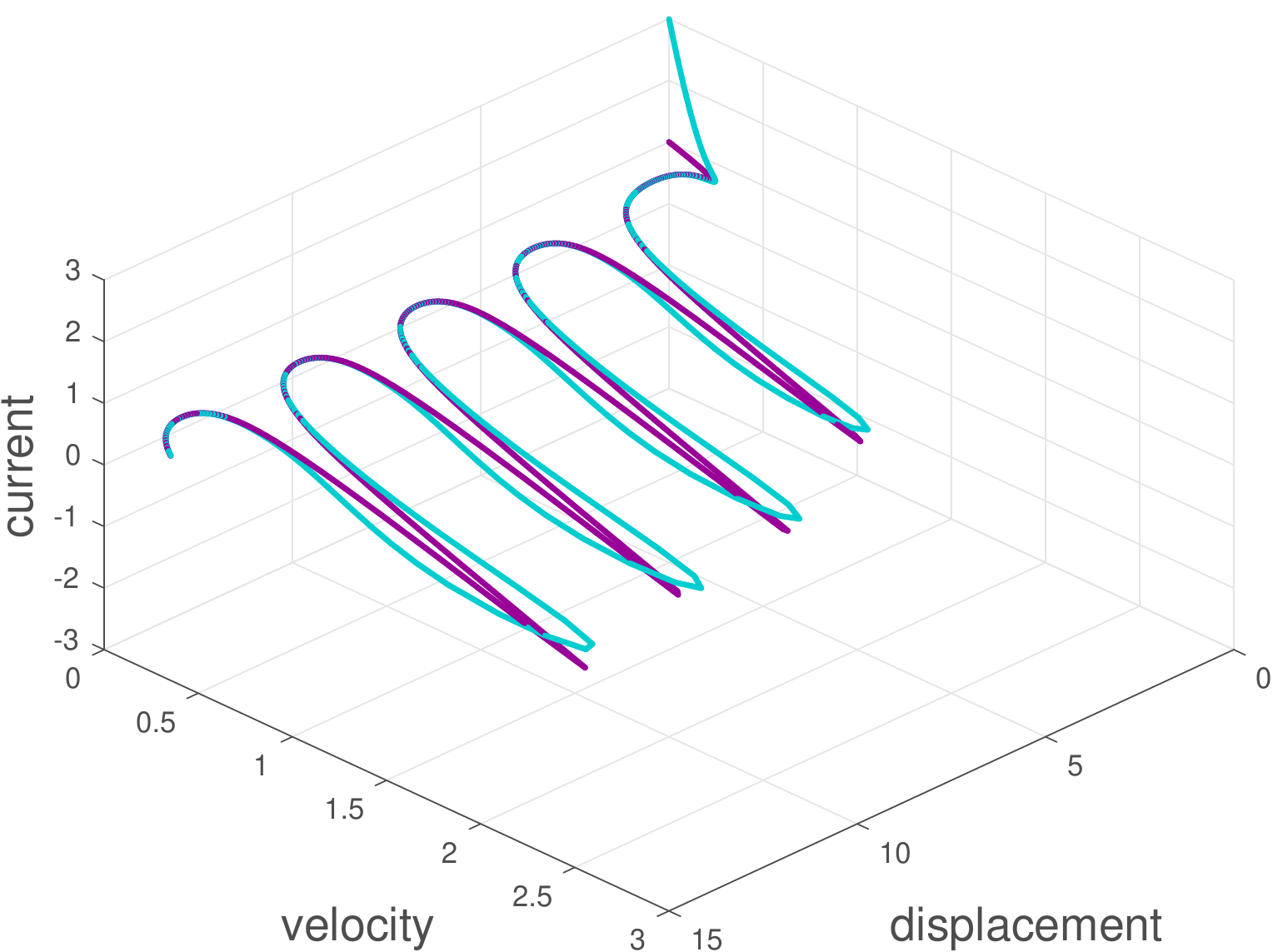}
\includegraphics[scale=0.35]{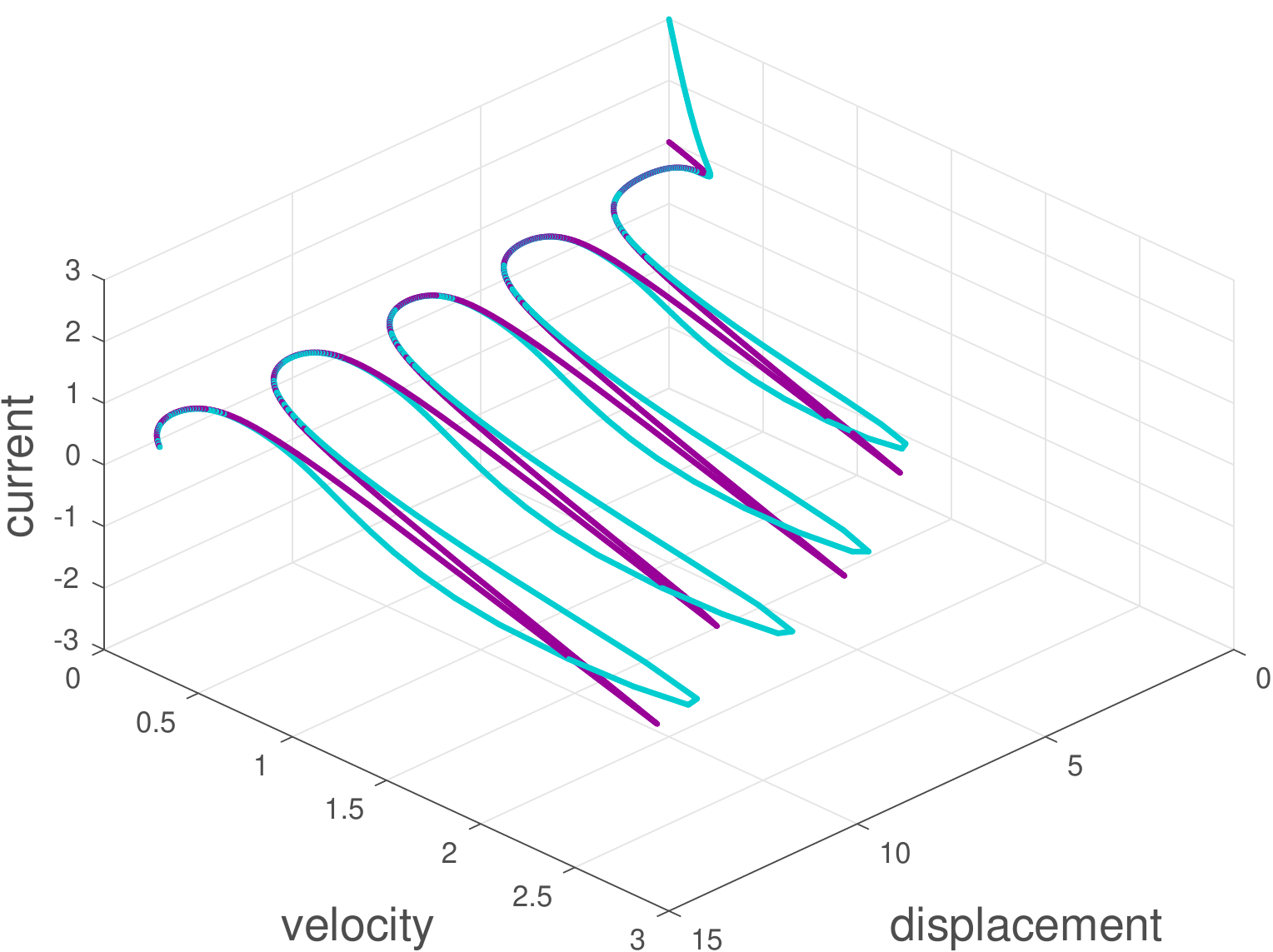}
\caption{Comparison between the trajectories of the two models in phase-space as a function of $d$ value: $d=5$ (first column); $d=10$ (second column); $d=15$ (third column). Dimensionless parameters: $\ell=0.05$, $\nu=1$, $b=1$, $(\theta_0, \dot{\theta}_0, \dot{q}_0) = (0,0,3 \, \nu)$.}
\label{d_effect_fig}
\end{figure*}

Thus, the validity limits of the approximation are also explored for different values of the dimensionless parameters $\nu$ and $d$, for which the phase-space trajectories evolution are shown in the Figures~\ref{nu_effect_fig} and Figure~\ref{d_effect_fig}, respectively. 

As the value of $\nu$ increases, in addition to the amplitude of the attractors increasing, it can be seen in Figures~\ref{nu_effect_fig} that the ``airfoil'' moves away from the plane associated with the reduced-order dynamics. Although bad from a quantitative point of view, it can be observed that qualitatively the approximation maintains a correlation with the original dynamics, even for moderately high values of $\nu$. A similar analysis for the parameter $d$, in Figure~\ref{d_effect_fig}, reaches the same conclusion (with small values of $\ell$ in both cases).

\subsection{Different modeling perspectives in physics and engineering: a philosophical point of view}

In this last part, the authors want to present, from a philosophical point of view, their particular opinion about the different perspectives in which the modeling of a physical system can be tackled in physics and engineering.

Nature has its laws responsible for crafting the mechanisms by which natural (and even artificial) processes occur. Understanding the essence of such laws and how they shape the evolutionary mechanisms of the universe is the fundamental objective of physics while knowing that, given the enormous underlying complexity, a complete understanding is impossible. In fact, as knowledge advances, ignorance also grows, as the boundary between known and unknown varies over time, increasing whenever the state-of-art is expanded \cite{Gleiser2015}. In this sense, physics increasingly seeks to deepen the understanding of a phenomenon until the knowledge is considered deep enough there are only a few (or perhaps none) non-incremental questions left to be answered.

To be able to make ``quantifications'' related to the mechanisms associated with a phenomenon, it is necessary to have a representation in the mathematical language (which plays the role of a \emph{lingua franca} between physicists and nature) of the known fundamental laws related to it. These representations are helpful and necessary, but it should be noted that they are not necessarily the actual form (the purest and ultimate essence) of the natural laws. They express physicists' perception of these fundamental laws, deep enough not to be refuted in the laboratory but incomplete by construction. As much as one master a particular language, he/she does not know all the words and grammar rules, not even the native speakers. No doubt nature is fluent in mathematics, but is it her mother tongue? Even if it is, she may not be able to express some ``sentences'' in that language. As a result of this limitation, combined with considerations (hypotheses) that disregard certain secondary aspects of a phenomenon in favor of simplicity, any description of natural processes derived from the mathematical representation of fundamental laws will be an approximation (an emulation) of the reality. Such mimics of reality are called \emph{models}.

It is clear then that any physical-mathematical model is not the reality. In the best scenario, it is a good imitation of reality. As a model is only a caricature of reality, all of them are erroneous in the formal logical sense. Nevertheless, this strictly logical view of models is an impoverished vision of their utility. Even though all of them are ``wrong'', most of them can be very useful, and they are the best rational tool one has for making quantitative inferences about the physical reality of things.

Both physicists and engineers make frequent use of models. They are tools for both, being more or less detailed according to need, convenience, or processing capacity. However, there is a profound difference between the two when thinking about what model structure to adopt. In its purest face, physics seeks a much deeper understanding of a phenomenon, demanding a more complete (and consequently complex) model, closer to including all the fundamental laws that whistle in that context. The model must have all the ingredients to represent reality reliably, no more, no less. On the other hand, due to its applied nature, often dealing with time and cost constraints, engineering frequently cannot afford to increase the complexity of a model beyond what is necessary to answer a certain question or solve a problem, no matter that information about the nature of the phenomenon is lost. The model must be enough to solve the problem, no matter that it is not the most trustworthy to reality. Sometimes these roles are reversed, engineers use more sophisticated models than physicists on the same problem, but typically the opposite is more usual. There is no conflict of interest in these two cases in the authors' view. They are distinct worldviews that are entirely compatible with the primary objectives of the two areas of knowledge. Ultimately, the complexity of the model will be dictated by the analysis goals.

This duality of views on modeling is at the heart of the topic discussed in this paper. There is no doubt that the complete model describes the behavior of the electromechanical system closer to reality in its fine detail. However, as exemplified in the sections above, a significant loss of information inherent in the reduced model only occurs in a (fast) temporal boundary layer close to the initial steps of the dynamics, not being important when one wants to analyze, for example, the asymptotic behavior of the system after a long time. It is hard to believe that, with the typical precision with which experiments are conducted in engineering laboratories, for the steady-state dynamics, the discrepancy between reduced model predictions and observations will be much more significant than the case where full-order model predictions are confronted with measurements. Therefore, from an engineering point of view, if knowing the dynamics in the first instants of time is not essential, there is no discredit in using the reduced-order model since it produces a valuable description of reality. In summary, it is possible to choose both models (full and reduced-order) from a rational perspective.


\section{Final remarks}
\label{concl_remaks}

The present work addresses the question of the validity of the quasi-steady-state assumption to reduce the order of a dynamical system that models the behavior of an electromechanical system, presenting a comprehensive discussion on the topic. Two electromechanical models (full and reduced) are described in detail, where the reduced model is obtained from the full-order model by discarding the inductive term in the electrical equation. This procedure is justified in light of physical analysis, with equal mathematical support, involving comparing scales of representative time of mechanical and electrical dynamics. To the best of the authors' knowledge, the mathematical justification for the model reduction in the format presented here is not found in the classic books that address the subject and the analogy between the simplification of the electromechanical system with its counterpart in chemical kinetics.

Based on the simulation results presented in this paper, from a qualitative perspective, one can conclude that the reduced-order model is an excellent exploratory tool because it faithfully reproduces the behavior of the original system. The reduced-order model can also be considered good for short time analysis intervals from the quantitative viewpoint. In this case, the reduced-order results differ very little from the original dynamics. These numerical results also show that the practical limit where simplification loses validity is (typically) when the ratio between the electrical and mechanical time scales is of the order of 10\%.

The discussion carried out in this paper is limited to showing that the reduced model can reproduce well characteristics of the dynamic behavior of the full-order model. The authors did not assert the validity of these models when confronted with data from laboratory measurements. Such a test can provide the final word on the validity of models as predictive tools. Unfortunately, the authors do not have at their disposal laboratory facilities with instrumentation capacity to analyze the electromechanical system studied here, which did not make it possible to compare the models' predictions with actual observations. It would be fascinating to carry out such an experiment, so this is the primary recommendation for future works on this topic.

One of the anonymous reviewers pointed out the possibility of obtaining an original analogy between the electromechanical dynamics investigated here and aeroelastic systems with lock-in effects \cite{NAPRSTEK2019p106103,NAPRSTEK2020p103441}. Exploring such a possibility in detail is undoubtedly an exciting direction to continue this work. Besides that, it would also be nice to investigate how perturbations in the initial conditions and forcing propagate to the electromechanical system response and bifurcation effects that exist in such a system, which these perturbations can induce. The authors will address these topics in future works.

In science, the final word is always open so that even an established theme can be discussed again from the ground up if reasoned questions arise. Indeed, the discussion raised in \cite{lima2018_1,Lima2019p552} is legitimate, most of the comments being interesting. The points of disagreement the authors of this work have concerning these papers were punctuated throughout this manuscript, the main one being the argument against the effectiveness of simplifying the model by assuming a quasi-steady state for the dynamics, which, as shown here, preserves the main qualitative characteristics of the original dynamics.

Inspired by this discussion, the authors would like to open another front for reflection on the modeling of this electromechanical system. The full-order model discussed here does not include dry friction effects present in the contact between the cart and the ground and in the pin that slides in the slot attached to the cart. Understanding how these friction effects affect the actual system dynamics and measuring how impoverished the predictions that do not consider them are, is an exciting line of investigation to pursue.

\section*{Dedication}

This paper is dedicated to the memory and legacy of Prof. Ali H. Nayfeh (1933 - 2017) \cite{Balachandran2017,Rega2020p1,Younis2017p1535} and Prof. Dean T. Mook (1935 - 2020) \cite{Hajj2020p1173}, who have inspired generations of theoretical and applied mechanists over the past 50 years, including the authors of this work.

\section*{Acknowledgements}

The authors thank Prof. Samuel da Silva (UNESP) for the fruitful discussions on the topic addressed in this paper and for the careful reader of the manuscript. They also thank the anonymous reviewers for their careful reading of the manuscript and their suggestions that helped improve the paper's final version.

\section*{Funding}

This research received financial support from the Brazilian agencies Coordena\c{c}\~{a}o de Aperfei\c{c}oamento de Pessoal de N\'{\i}vel Superior - Brasil (CAPES) - Finance Code 001, the Brazilian National Council for Scientific and Technological Development (CNPq), grants 307371/2017-4 and 309799/2021-0, and the Carlos Chagas Filho Research Foundation of Rio de Janeiro State (FAPERJ) under the following grants: 211.304/2015, 210.021/2018, 210.167/2019, 211.037/2019 and 201.294/2021.

\section*{Code availability and animations}

The simulations of this paper used a Matlab code dubbed \textbf{ElectroM - ElectroMechanical Dynamic Code}. To facilitate the reproduction of the results, this code is available for free on GitHub \cite{ElectroM}. Animations of the electromechanical dynamics are available on the supplementary material and on a YouTube playlist \cite{eletromech_video3}.

\section*{Compliance with ethical standards}

\section*{Conflict of Interest }
The authors declares that they have no conflict of interest.

\bibliographystyle{spbasic}      

\bibliography{references}

\end{document}